\documentclass[11pt,a4paper]{article}

\usepackage{amsmath}
\usepackage{mathtools}
\usepackage{nccmath}
\usepackage{amsthm}
\usepackage{amssymb}
\usepackage{amsfonts}
\usepackage{anyfontsize}
\usepackage{mathrsfs}
\usepackage{bbm}
\usepackage{graphicx}
\usepackage{caption}
\usepackage{bookmark,hyperref}
\usepackage{12many}
\usepackage{parskip}
\usepackage{setspace}
\usepackage{multirow}
\usepackage{arydshln}
\usepackage{anyfontsize}
\usepackage{changepage}
\usepackage{todonotes}
\usepackage{braket}
\usepackage{dirtytalk}
\usepackage{MnSymbol}

\newcommand{\spinfourfour}{\text{Spin$(4,4)$}}
\newcommand{\spintwotwo}{\text{Spin$(2,2)$}}
\newcommand{\sofourfour}{\text{SO$(4,4)$}}
\newcommand{\sotwotwo}{\text{SO$(2,2)$}}

\newcommand{\sofour}{\text{SO$(4)$}}

\newcommand{\spinfour}{\text{Spin$(4)$}}
\newcommand{\sotwo}{\text{SO$(2)$}}
\newcommand{\sutwo}{\text{SU$(2)$}}
\newcommand{\sltwo}{\text{SL$(2)$}}

\newcommand{\psymbol}[2]{\genfrac{}{}{0pt}{1}{#1}{#2}}

\DeclareMathOperator{\Tr}{Tr}

\numberwithin{equation}{section}
\usepackage{fullpage}

\begin{document}

\begin{titlepage}
 \thispagestyle{empty}
 \begin{flushright}
 \hfill{Imperial-TP-2023-CH-01 }\\
 \end{flushright}

 \vspace{30pt}

 \begin{center}
     
  {\fontsize{20}{24} \bf {Freely acting orbifolds of type IIB\\
  \vspace{20pt} string theory on $T^5$}}

     \vspace{30pt}
{\fontsize{13}{16}\selectfont {George Gkountoumis$^1$, Chris Hull$^2$, Koen Stemerdink$^1$ and Stefan Vandoren$^1$}} \\[10mm]

{\small\it
${}^1$ Institute for Theoretical Physics {and} Center for Extreme Matter and Emergent Phenomena \\
Utrecht University, 3508 TD Utrecht, The Netherlands \\[3mm]

${}^2$ 
{{The Blackett Laboratory}},
{{Imperial College London}}\\
{{Prince Consort Road}}, 
{{London SW7 2AZ, U.K.}}\\[3mm]}

\vspace{1cm}

{\bf Abstract}

\vspace{0.3cm}

\begin{adjustwidth}{12pt}{12pt}
We study freely acting orbifolds of type IIB string theory on $T^5$ that spontaneously break supersymmetry from $\mathcal{N}=8$ to $\mathcal{N}=6,4,2$ or 0 in five dimensions. We focus on  orbifolds that are a $\mathbb{Z}_p$ quotient by a T-duality acting on $T^4$ and a shift on the remaining $S^1$. Modular invariant partition functions are constructed and 
 detailed examples of both symmetric and asymmetric orbifolds are presented, including new examples of five-dimensional non-supersymmetric string theories with no tachyons. The orbifolds we consider arise at special points in the moduli space of string theory compactifications with a duality twist. The
supergravity limit of these are Scherk-Schwarz reductions which generate gauged supergravities with   positive definite classical potentials on the moduli space in five dimensions.
Both symmetric and asymmetric freely acting orbifolds give a landscape of Minkowski vacua. For gauged supergravities to belong to this landscape, we find a number of constraints and conditions. Firstly, the scalar potential should lead to a massive spectrum with masses that obey quantization conditions arising from a string theory orbifold, which we discuss in detail. Secondly, we find constraints on the massless sector, e.g. in the examples  of orbifolds preserving sixteen supersymmetries in five dimensions that we consider, only an odd number of vector multiplets arises. Lastly, we present new examples of candidate asymmetric orbifolds with modular invariant partition functions, but with non-integral coefficients in the $q\bar{q}$-expansion in the twisted sectors.

\end{adjustwidth}

\end{center}

\vspace{5pt}

\newcommand\blfootnote[1]{%
  \begingroup
  \renewcommand\thefootnote{}\footnote{#1}%
  \addtocounter{footnote}{-1}%
  \endgroup}

\blfootnote{g.gkountoumis@uu.nl \quad c.hull@imperial.ac.uk \quad koenstemerdink@gmail.com \quad s.j.g.vandoren@uu.nl}

\noindent

\end{titlepage}

\begin{spacing}{1.15}
\tableofcontents
\end{spacing}

\section{Introduction}

Spontaneous breaking of supersymmetry in supergravity can be realized via Scherk-Schwarz reductions on a circle with a twist by a supergravity duality symmetry \cite{Scherk:1978ta,Scherk:1979zr,Cremmer:1979uq}. The string theory uplifts of these are typically reductions of string theories with a duality twist \cite{Dabholkar:2002sy} which can in some cases be 
viewed as compactifications on torus bundles over a circle but in general can give
 reductions on non-geometric T-folds or U-folds \cite{Hull:2004in}. For the case in which the duality twist is a T-duality, these become freely acting orbifolds at special points of the moduli space \cite{Dabholkar:2002sy}. (For some early references on such orbifolds see e.g. \cite{Rohm:1983aq,Kounnas:1988ye,Ferrara:1987es,Ferrara:1988jx,Kiritsis:1997ca}.) Our interest here lies mostly in partial supersymmetry breaking, which can easily be accommodated with the Scherk-Schwarz mechanism and its stringy uplift. This yields a variety of string theories in lower dimensions with various amounts of supersymmetry. Complete breaking of supersymmetry without the presence of tachyons can be achieved as well, and we will present some new examples of such string theories. Typically they suffer from known problems of generating either a cosmological constant that is too large, or extra dimensions that are too large. This renders these models not realistic for phenomenology, though still interesting for present purposes, as we shall discuss. 

Scherk-Schwarz reductions over a circle $S^1$ from $D+1$ to $D$ dimensions arise when the fields in $D+1$ dimensions pick up a monodromy around the circle contained in the continuous duality symmetry group $G$ of the $D+1$ dimensional supergravity. The ansatz for the fields is
\begin{equation}
    \psi(x^\mu,y)=g(y)[\psi(x^\mu)]\ ,\qquad g(y)=\exp \Big(\frac{My}{2\pi R}\Big)\ ,
\end{equation}
where $g\in G$, $y$ is the coordinate on $S^1$ with radius $R$ and $M$ is an element of the Lie algebra of $G$. The fields then pick up a monodromy matrix $e^{M}$ going around the $S^1$.

To fully specify the reduction, we need to also choose a spin structure on the circle; this will be discussed in more detail later on. Scherk-Schwarz reductions of supergravity theories yield gauged supergravities in $D$ dimensions, and when the (classical) moduli space in $D+1$ dimensions is a coset $G/K$ with $K$ compact, the scalar potential on the moduli space in $D$ dimensions is positive definite. If furthermore the monodromy matrix is $G$-conjugate to a rotation matrix in $K$, there exist stable or at least marginally stable Minkowski vacua that may preserve some of the supersymmetry \cite{Dabholkar:2002sy}. The quadratic fluctuations around the vacuum then yield the mass spectrum with masses determined by the parameters in the monodromy matrix. 

Scherk-Schwarz reductions can   be generalised to string theory \cite{Dabholkar:2002sy,Hull:1998vy}.
In string theory  the global symmetry group $G$ of the low-energy effective action is broken to the discrete U-duality group $G(\mathbb{Z})$ which is a symmetry of the non-perturbative string theory \cite{Hull:1994ys} 
and the monodromy is required to be in $G(\mathbb{Z})$ \cite{Dabholkar:2002sy,Hull:1998vy} giving non-linear quantization conditions on the mass matrix. Now, if the quantized monodromy is conjugate to a rotation, the construction is an orbifold \cite{Dabholkar:2002sy}. Each choice of a monodromy matrix corresponds to a choice of orbifold and yields a particular quantization condition on the masses. From the point of view of the string landscape, it is therefore important to study the mass spectra allowed by these quantization conditions. When the scalar potential in supergravity leads to a mass spectrum that does not obey these quantization conditions, it belongs to the swampland. One of the aims of our paper is to study this question in detail for a large class of freely acting orbifolds.

 For the purposes of this paper, we will consider monodromies contained in the T-duality subgroup of $G(\mathbb{Z})$. Such supergravity vacua can be realized in string theory as freely acting orbifolds. 
 We focus on the case of type IIB string theory on $T^5$ and 
 the orbifold actions we discuss here involve a T-duality action on $T^4$ together with a shift along the remaining $S^1$. The six-dimensional supergravity theory obtained by compactification on $T^4$ has symmetries given by
 $G=\text{Spin}(5,5)$ and $K=[\text{Spin}(5)\times \text{Spin}(5)]/ \mathbb{Z}_2$. The U-duality symmetry is then $G(\mathbb{Z})=\text{Spin}(5,5;\mathbb{Z})$ and the T-duality subgroup in which the monodromy lies is $\text{Spin}(4,4;\mathbb{Z})$.
In addition to the $\text{Spin}(5,5;\mathbb{Z})$ U-duality, the string theory compactified on $T^4$ also has a local symmetry given by the double cover $\tilde K$ of $K$, $\tilde K=[\text{Spin}(5)\times \text{Spin}(5)]$ \cite{hull2007generalised}: while the local symmetry acts through $K$ on the bosons, the double cover is 
  needed for the action on the fermions, requiring a generalized spin structure \cite{hull2007generalised}. We will work in a gauge in which the local symmetry is fixed, so that the U-duality transformations act on the fermions through compensating $\tilde K$ transformations.

  If the monodromy acts as a diffeomorphism on $T^4$, this corresponds to a symmetric orbifold of the IIB string, so that the compactification geometry is a $T^4$ bundle over $S^1$ \cite{Hull:2004in}. If the monodromy acts as a T-duality on $T^4$, this constructs a T-fold background which is realized as an asymmetric orbifold at a special point in the moduli space \cite{Dabholkar:2002sy,Hull:2020byc}. Asymmetric orbifolds are interesting for moduli stabilization because more of the moduli  are fixed than in the symmetric case. Furthermore, by combining the T-duality action on $T^4$ with a shift on $S^1$ one ensures that no extra moduli arise from the twisted sector. (See e.g. \cite{dine1997new,Angelantonj:2006ut,Anastasopoulos:2009kj,Condeescu:2013yma} for some references.) The original references for symmetric and asymmetric orbifolds are \cite{dixon1985strings,dixon1986strings} and \cite{Narain:1986qm,narain1991asymmetric} respectively. 
  These references mainly focus on the lattice approach, which  is particularly useful when constructing partition functions, as we will discuss. However, for making contact with the Scherk-Schwarz reductions the language of monodromies and duality twists is more useful. These reductions correspond to orbifolds by symmetries which are freely acting and these have some important differences with the non-freely acting ones. In this paper, we will use these two complementary approaches and discuss their relationship.

  The T-duality transformations on $T^4$ can be realised as diffeomorphisms of a doubled torus $T^8$, so that the T-fold construction described above can be viewed as a bundle of the doubled torus $T^8$ over a circle \cite{Hull:2004in}. More general constructions are possible in which the dependence on the momentum on the circle introduced by the monodromy is generalised to include dependence on the string winding number on the circle. String theory on the circle  can be formulated  by introducing a dual coordinate conjugate to the winding number so that the circle is promoted to a doubled circle.  Then the general construction has separate monodromies which are introduced on both the circle and the dual circle; these two monodromies  are given by two commuting T-duality transformations \cite{Dabholkar:2005ve}. This gives a construction that is not even locally geometric and can be understood in terms of a doubled torus bundle over the doubled circle \cite{Hull:2007jy,Hull:2009sg}. (Such constructions are sometimes referred to as having \say{R-flux}.) The corresponding orbifolds involve a shift on both the circle and the dual circle, introducing phases depending on both momentum and winding \cite{Dabholkar:2005ve}. Further possibilities arise with shifts on several circles, see e.g. \cite{dabholkar1999string}. In this paper we will focus on the cases in which there is a shift on just one circle and 
  no shift on the dual circle.

  We will carefully discuss the orbifold action on the fermions. The theory has further discrete symmetries such as $(-1)^{F_s}$ where $F_s$
is the spacetime fermion number, together with refinements of this into chiral fermion numbers. We will see that we need to consider monodromies such as ${\mathcal M}(-1)^{F_s}$ where
${\mathcal M}$ is in $\text{Spin}(4,4;\mathbb{Z})$. Note that for a monodromy $(-1)^{F_s}$
(with ${\mathcal M}=1$), the orbifold by this together with a shift on the circle amounts to choosing the anti-periodic spin structure for the circle. This corresponds to a non-supersymmetric string compactification which can be made free of tachyons. We will discuss explicit examples of such compactifications in this paper.

The choice of $T^5$ was motivated by our earlier study of black holes in type IIB string theory with a duality twist \cite{Hull:2020byc}, in which supersymmetry is (partially) broken in the vacuum already in the absence of the black hole. In that paper, we studied in detail the supergravity arising from the reduction with a duality twist. In this paper we analyse the string vacua in their own right, and postpone the further discussion of D-branes and black holes to a future study \cite{GHSV}. 
The compactifications we study here are closely related to those of \cite{Hull:2017llx,Gautier:2019qiq}
which involved a similar construction with $K3$ instead of $T^4$, and their heterotic duals.

It turns out that the landscape of freely acting orbifolds is rather rich  and connects to   questions that arise  in the string landscape programme. One issue is mentioned above, namely finding the  quantization conditions on the masses. Moreover, for the massless sector, there are constraints on the moduli spaces that may appear. One example of such a constraint  is that we always find an odd number of abelian massless vectors in the class of orbifolds we consider, independent of the number of supersymmetries. For orbifolds preserving 16 supersymmetries, this implies that only an odd number of abelian vector multiplets arises. This is much in line with the analysis in higher dimensions \cite{montero2021cobordism,kim2020four}. 

We also study the case of $\mathcal{N}=2$ (eight supersymmetries), and find that only some  of the magic square supergravities \cite{Gunaydin:1983rk,Gunaydin:1983bi}  arise in the orbifold landscape, while the others do not arise in this way. 
It is important to stress  that we have not studied all possible duality twists here, but only those with monodromy matrices that are conjugate to $\text{Spin}(4)\times \text{Spin}(4)$ rotations in  the T-duality group. 
These are the ones that lead to stable Minkowski vacua with spontaneously broken supersymmetry. 
Even in this class, there are surprises. As it was recently shown in \cite{bianchi2022perturbative}, asymmetric orbifolds of the type $(T^4\times S^1)/\mathbb{Z}_p$ with $\mathcal{N}=6$ (24 supersymmetries) only exist for $p=2,3$ but lead to problems with the partition function for $p=4,6$. These partition functions are modular invariant, but have non-integral coefficients in the $q\bar{q}$-expansion in the twisted sectors and  hence appear to be unphysical\footnote{The same integrality problem in fact also arises for the non-freely acting asymmetric orbifolds $T^4/\mathbb{Z}_{4,6}$.}. 
This integrality constraint was analyzed in the original papers \cite{Narain:1986qm,narain1991asymmetric}, and it was satisfied by all the models presented there. Later, models based on the idea of quasicrystalline compactification violating the integrality constraint were discussed in \cite{harvey1988quasicrystalline}. 
Here, we will review the aforementioned $\mathcal{N}=6$ models and find new examples with less supersymmetry where the integrality condition is violated.

The following two sections deal with the orbifold constructions in the closed string sector. In section \ref{sec:Orbifold constructions} we discuss the various possible orbifolds, the relation with  duality twists and the quantization conditions on the mass parameters. In section \ref{The orbifold action and partition function} we present the orbifold action on the world-sheet fields and the general formalism for constructing modular invariant orbifold partition functions. In section \ref{closed string spectrum} we present the details of some examples, including asymmetric $\mathbb{Z}_2$ orbifolds that preserve $\mathcal{N}=6$ or $\mathcal{N}=2$ supersymmetry, as well as a symmetric $\mathbb{Z}_4$ orbifold and an asymmetric $\mathbb{Z}_2$ orbifold that preserve $\mathcal{N}=4$ supersymmetry. As an example with completely broken supersymmetry we  discuss a symmetric $\mathbb{Z}_3$, $\mathcal{N}=0$ orbifold, and find the conditions for the absence of tachyons that appear in the twisted sectors. In section \ref{Supergravity} we relate the freely acting orbifolds with the supergravity Scherk-Schwarz reductions presented in \cite{Hull:2020byc}, we analyse the moduli spaces of massless scalars that can appear and find complete agreement of the low-lying massive spectrum. Furthermore, we discuss the supertrace formulae for the $\mathcal{N}=0$ supergravity models, and show that the first non-vanishing supertrace is Str$M^8>0$, leading to negative corrections to the scalar potential. This is consistent with earlier predictions from string theory \cite{Rohm:1983aq}, and similar to the supergravity results in four dimensions obtained in \cite{Cremmer:1979uq,ferrara1979mass}. Finally, in section \ref{sec:bad examples} we give some explicit examples of swampland models and end with some conclusions and our outlook in section \ref{conclusion}.

\section{Orbifolds and duality twists}
\label{sec:Orbifold constructions}
\subsection{The orbifolds}
The orbifolds we are interested in have target spaces of the form 
\begin{equation}\label{background}
\mathbb{R}^{1,4}\times  S^1\times T^4\ , \end{equation}
identified under the action of a $\mathbb{Z}_p$ symmetry.
This requires being at a point in the moduli space in which the fields on $T^4$ are invariant under the action of 
$\mathbb{Z}_p$. We then orbifold by a $\mathbb{Z}_p$ given by this $T^4$ symmetry combined with a shift by $2\pi \mathcal{R}/p$ on the circle $S^1$ with radius $\mathcal{R}$ which makes the orbifolds freely acting. Freely acting orbifolds have no fixed points and at generic points in the moduli space, all states coming from the twisted sectors are massive. Furthermore, as compared to non-freely acting orbifolds, supersymmetry is spontaneously broken instead of being explicitly broken, manifested by the fact that gravitini  become massive instead of being projected out. 

The symmetric orbifolds arise when the $\mathbb{Z}_p$ action on $T^4$ is a geometric discrete symmetry of $T^4$, generated by a diffeomorphism on $T^4$, i.e. by an element in GL(4;$\mathbb{Z}$);  this generator is the monodromy of the corresponding duality twist. The requirement that the monodromy is in the discrete group GL(4;$\mathbb{Z}$)
restricts the rank $p$ of the group $\mathbb{Z}_p$. All possible values of $p$ have been classified in \cite{Erler:1992ki} and for  $T^4$ these are $p=2,3,4,5,6,8,10,12$ and $24$.  Not all values of $p$ yield supersymmetric string theories, and it is worth mentioning that symmetric orbifolds of $T^4$ of rank $p=5,8,10,12$ and 24 break all supersymmetry.

For the asymmetric orbifolds, the $\mathbb{Z}_p$ group   acts as a T-duality transformation on the $T^4$ CFT.
The T-duality group for superstrings on $T^4$ is Spin$(4,4;\mathbb{Z})$, a discrete subgroup
of the double cover Spin$(4,4)$ of $\text{SO}(4,4)$, as the D-brane states transform as a spinor representation of $\text{Spin}(4,4)$ \cite{Hull:1994ys}. Then, in this case, the
 monodromy matrix lies in Spin$(4,4;\mathbb{Z})$.\footnote{When the monodromy matrix is not in the T-duality subgroup of the U-duality group, one has generalized orbifolds that quotient by a non-perturbative symmetry \cite{Dabholkar:2002sy}. Such orbifolds don't have a CFT description on the worldsheet and we will not consider them in this paper.} The background fields, namely the  torus metric $G$ and the two-form $B$-field, can be combined  into a matrix $E=G+B$. T-duality transforms $E$ to a new background $E'$ through a fractional linear transformation\footnote{For details on how T-duality acts on the background fields we refer to the classic review \cite{giveon1994target} and e.g. \cite{tan2015t,satoh2017lie}. }. Consistency of the asymmetric orbifold then requires that the $\mathbb{Z}_p$ transformation is a symmetry under which $E'=E$. This can be achieved only for special values of the  moduli, which are therefore stabilized in these non-geometric constructions. Requiring modular invariance puts severe constraints on which asymmetric orbifolds are allowed and in particular restricts the values of $p$. Nevertheless, asymmetric orbifolds allow for more possibilities compared to the symmetric ones (see also \cite{harvey1988quasicrystalline,nibbelink2017t}). We will return to these issues when we discuss specific examples.

\subsection{Moduli space and quantization conditions}
\label{sec:quantization conditions}

For a coset space $G/K$, the points are cosets: $[g]=gK$, where $g\in G$. The stabilizer of the origin $[\mathbbm{1}]=K$ is a subgroup $K_0\subset G$, so that the stabilizer of a point $[g]$ is the conjugate subgroup $K_g=gK_0g^{-1}$. We are interested in those points $[\bar g]$ for 
which there is a $\mathbb{Z}_p$ symmetry preserving $[\bar g]$  which is also in the duality group $G(\mathbb{Z})$, so that 
$G(\mathbb{Z})\cap K_{\bar g}$ is non-trivial and contains the $\mathbb{Z}_p$. 
This is a hard condition to solve in general, and $G(\mathbb{Z})\cap K_{\bar{g}}$ is non-trivial only for special points.
Such a fixed point is necessarily at a minimum of the associated Scherk-Schwarz potential and gives a stable Minkowski vacuum \cite{Dabholkar:2002sy}.

If the generator of the $\mathbb{Z}_p$ symmetry is $\mathcal{M}\in G(\mathbb{Z})$, then 
\begin{equation}\label{conjugational}
\mathcal{M} = {\bar g} \tilde{\mathcal{M}} {\bar g}^{-1} \,; \qquad\quad \tilde{\mathcal{M}}\in K_0 \subset G \,, \qquad {\bar g}\in G \,.
\end{equation}
At a point in the moduli space fixed under the action of such a $\mathbb{Z}_p$ symmetry, the corresponding background field configuration on $T^4\times S^1$ is preserved by the $\mathbb{Z}_p$ symmetry. 
If 
 $   {\mathcal M}=e^{M'}\ $,
then
\begin{equation}
    \tilde{\mathcal M}=e^{M}\ ,
\end{equation}
where 
\begin{equation}
    M={\bar g}^{-1} M'{\bar g}
\end{equation}
will be referred to as the mass-matrix: it is the matrix of masses that appear in the effective action.  In this way, a monodromy  $ {\mathcal M}\in G(\mathbb{Z})$ satisfying $ {\mathcal M}^p=1$ defines a matrix $\tilde{\mathcal{M}}\in K_0$ that also satisfies
$\tilde{\mathcal{M}}^p=1$ and so generates a  $\mathbb{Z}_p$  subgroup of $K$.
The action on fermions is through an element of the double cover $\tilde K$ of $K$ and this in general can generate a  $\mathbb{Z}_{2p}$  subgroup of $\tilde K$.

Monodromies ${\mathcal{M}}, {\mathcal{M}}' $ that are $G(\mathbb{Z})$-conjugate, i.e.
\begin{equation}
    {\mathcal{M}}' =g{\mathcal{M}}g^{-1}\,, \qquad { g}\in G (\mathbb{Z})\,,
\end{equation}
define the same theory. Then the distinct orbifolds correspond to conjugacy classes of monodromies \cite{Hull:1998vy}. This conjugation changes the fixed point $\bar g \to \bar g'= g \bar g $ but leaves 
$\tilde{\mathcal{M}}$ unchanged.
Changing $\bar g$ to $\bar g k$ with $k\in K_0$ gives another representative of the coset $[\bar g]$ and transforms 
$\tilde{\mathcal{M}}$ to $k\tilde{\mathcal{M}}k^{-1}$. Then the monodromy only defines the rotation matrix $\tilde{\mathcal{M}}$ and the mass matrix $M$ up to conjugation by an element of $K_0$.

\subsection{Type IIB string theory compactified on $T^4$ }

For  type IIB string theory compactified on $T^4$, 
 $G=\text{Spin}(5,5)$ and $K=[\text{Spin}(5)\times \text{Spin}(5)]/ \mathbb{Z}_2$ with double cover $\tilde K=[\text{Spin}(5)\times \text{Spin}(5)]$.
 We will focus on the T-duality subgroup $\text{Spin}(4,4)\subset \text{Spin}(5,5)$ which is a perturbative symmetry that can be realised in the world-sheet formulation.
In the classical supergravity theory, the 1/2-BPS 0-branes in $6D$ obtained by compactification on $T^4$ are in the 16-dimensional spinor representation of $\text{Spin}(5,5)$.
In the quantum string theory, the charges of these 0-branes are quantized and take values in a 16-dimensional charge lattice. The U-duality group $G(\mathbb{Z})=\text{Spin}(5,5;\mathbb{Z})$ is the discrete subgroup of $\text{Spin}(5,5)$
preserving this lattice. Under the subgroup $\text{Spin}(4,4)\subset \text{Spin}(5,5)$, the 16 0-brane charges decompose into an 8-dimensional vector representation $8_v$ of $\text{Spin}(4,4)$
corresponding to the NS-NS 0-branes from 4  momentum modes and 4 winding modes on $T^4$, together with an 8-dimensional chiral spinor representation $8_s$ of $\text{Spin}(4,4)$
corresponding to the R-R 0-branes arising from D1-branes and D3-branes wrapping the $T^4$.
Then the charges transform as the $8_v+8_s $ representation of $\text{Spin}(4,4)$, and $\text{Spin}(4,4;\mathbb{Z})$ is the discrete subgroup of  $\text{Spin}(4,4)$ preserving the charge lattice.
 
 At a point $[g]$ in the moduli space that is fixed under a $\mathbb{Z}_p$ symmetry, we compactify on a further $S^1$ and orbifold by the $\mathbb{Z}_p$ symmetry of the $T^4$ combined with a shift 
 on the $S^1$. The orbifold group is then generated by $ \mathcal{M}$ combined with a shift of the $S^1 $ coordinate 
 by $2\pi\mathcal{R}/p$.
This corresponds to a compactification on $T^4$ followed by a compactification on $S^1 $ with a duality twist by the T-duality transformation $\mathcal{M}\in G(\mathbb{Z})$, so that there is a monodromy
$\mathcal{M}$ on the $S^1$.
Then $\mathcal{M}\in\text{Spin}(4,4;\mathbb{Z})$ is required to satisfy
\begin{equation}\label{conjugation}
\mathcal{M} = g \tilde{\mathcal{M}} g^{-1} \,; \qquad\quad \tilde{\mathcal{M}}\in   [(\spinfour\times\spinfour )
/ \mathbb{Z}_2]_0
\subset\text{Spin}(4,4) \,, \qquad g\in\text{Spin}(4,4) \,.
\end{equation}

Note that $\text{Spin}(4,4)$ is the double cover of $\text{SO}(4,4)$ and $[\spinfour\times\spinfour]
/ \mathbb{Z}_2$ is a double cover of the $\text{SO}(4)\times \text{SO}(4)$ subgroup of $\text{SO}(4,4)$. This means that it is the double cover of an asymmetric rotation. In the world-sheet theory, one $\text{SO}(4)$ factor acts as a rotation on the left-movers and the other acts as a rotation on the right-movers. The restriction to $\text{Spin}(4,4;\mathbb{Z})$ ensures that the periodicity condition of the bosonic coordinates on $T^4$ is preserved. The group $\tilde K=\spinfour\times\spinfour
 $ is a quadruple cover of the $\text{SO}(4)\times \text{SO}(4)$.

The elements $ \pm \mathbbm{1} $ of $\spinfour$ both project to the 
identity $ \mathbbm{1} $ of $\text{SO}(4)$, while the four elements
$ (\mathbbm{1} , \mathbbm{1})$, $ (\mathbbm{1} ,- \mathbbm{1})$, $ (-\mathbbm{1} , \mathbbm{1})$, $ (-\mathbbm{1} ,- \mathbbm{1})$ of $\spinfour\times\spinfour$ all project to the 
identity $ (\mathbbm{1} , \mathbbm{1})$ of $\text{SO}(4)\times \text{SO}(4)$, exhibiting the quadruple cover. Then
\begin{equation}
    \text{SO}(4)\times \text{SO}(4)\cong
 \frac {   \spinfour\times\spinfour}
{
\mathbb{Z}_2 \times \mathbb{Z}_2
}\,,
\label{equation 2.7 ref}
\end{equation}
with the subgroup $\mathbb{Z}_2 \times \mathbb{Z}_2$ consisting of
 the elements
$ (\mathbbm{1} , \mathbbm{1})$, $ (\mathbbm{1} ,- \mathbbm{1})$, $ (-\mathbbm{1} , \mathbbm{1})$, $ (-\mathbbm{1} ,- \mathbbm{1})$.
The duality group $K$ is the double cover of \eqref{equation 2.7 ref}, given by
\begin{equation}
   K=
 \frac {   \spinfour\times\spinfour}
{
\mathbb{Z}_2 
}\,,
\end{equation}
with the $\mathbb{Z}_2 $ generated by $ (-\mathbbm{1} ,- \mathbbm{1})$.

All fermions in the theory transform as representations of $\spinfour\times\spinfour$
while the bosons transform as representations of $\text{SO}(4)\times \text{SO}(4)$.
The element $ (-\mathbbm{1} ,- \mathbbm{1})$ of $\spinfour\times\spinfour$ leaves
all bosons invariant but multiplies each spacetime fermion by $-1$. This means that it acts as 
$(-1)^{F_s}$, where ${F_s}$ is the spacetime fermion number.
Next, $ (-\mathbbm{1} , \mathbbm{1})$ acts as $-1$ on fermions transforming under the $\spinfour_L$ subgroup of 
$\spinfour_L\times\spinfour_R$ and so acts as $(-1)^{F_L}$, where $F_L$ is the corresponding fermion number. Similarly, $ (\mathbbm{1} ,- \mathbbm{1})$ 
acts as $-1$ on fermions transforming under the $\spinfour_R$ subgroup of 
$\spinfour_L\times\spinfour_R$ and so acts as $(-1)^{F_R}$. Note that $F_s$ should not be confused with the world-sheet fermion number and $F_L,F_R$ should not be confused with the left and right-moving world-sheet fermion numbers, although they are of course related via the GSO projection.
Thus, there are four possible lifts to $\spinfour\times\spinfour$
of the identity in $\text{SO}(4)\times \text{SO}(4)$, given by $\mathbbm{1},(-1)^{F_s}$,$(-1)^{F_L}$,$(-1)^{F_R}$.
Of these, $(-1)^{F_s}$ is represented trivially in $\text{Spin}(4,4;\mathbb{Z})$
by $\tilde {\mathcal{M}}= \mathbbm{1}$, while $(-1)^{F_L}$,$(-1)^{F_R}$ are represented in 
$\text{Spin}(4,4;\mathbb{Z})$ by $\tilde {\mathcal{M}}= -\mathbbm{1}$.

We can now specify the orbifolds we will be considering here.
We choose a $ \mathcal{M} \in \text{Spin}(4,4;\mathbb{Z})$ satisfying $ \mathcal{M}^p=1$, for some $p$, and take the monodromy to be
$ \mathcal{M}$.
This then determines an element $\tilde{\mathcal{M}}\in   [(\spinfour\times\spinfour )
/ \mathbb{Z}_2]_0$ via \eqref{conjugation}. To define the transformation of the fermions, we then choose a 
  lift of  $\tilde{\mathcal{M}}$ to the double cover 
  $\hat{\mathcal{M}}\in   \spinfour\times\spinfour $.
  It is also possible to include a fermionic twist, in which case 
  the monodromy becomes $ \mathcal{M}(-1)^{F_s}$, and similarly for $F_{L/R}$.
Note that if $p$ is odd, $ (\mathcal{M}(-1)^{F_s})^p=(-1)^{F_s}$, so that this generates a 
$\mathbb{Z}_{2p}$ symmetry, while for even $p$ it generates a $\mathbb{Z}_{p}$ symmetry. 

Practically, it is often useful to define the orbifold by a choice of matrix $\hat{\mathcal{M}}\in   \spinfour\times\spinfour $ which gives the orbifold action on all the fields. This matrix is only determined up to $\spinfour\times\spinfour$ conjugation, and by conjugation we can bring it to a standard form in a convenient maximal torus $\text{SO}(2)^4$
\begin{equation}
    \hat{\mathcal{M}}= ({\mathcal{M}}_L,{\mathcal{M}}_R)\in \spinfour _L\times\spinfour _R \,,
    \label{mhat}
\end{equation}
with 
\begin{equation}
    {\mathcal{M}}_L=\begin{pmatrix}
    R(m_1) & 0  \\
    0 & R(m_3) 
    \end{pmatrix} \,,\qquad {\mathcal{M}}_R=\begin{pmatrix}
    R(m_2) & 0  \\
    0 & R(m_4) 
    \end{pmatrix} 
     \label{mhat2}\,,
\end{equation}
where
we use the notation $R(x)=\begin{psmallmatrix}\cos x & \,\,\,\,-\sin x \\ \sin x & \,\,\,\,\cos x \end{psmallmatrix}$ for a two by two rotation matrix. Here each matrix acts in the $(2,0)+(0,2)$ representation of $\spinfour\cong\text{SU}(2)\times \text{SU}(2)$. The matrices $\mathcal{M}_{L/R}$ act on spinors on $T^4$ such as the internal part of the R-vacua,  as we discuss in the next section, see \eqref{transformation of ramond vacua}.
Thus, the monodromy is specified by four angles $m_i$, and then the key step is determining what angles are allowed, i.e. for which choices of the angles $m_i$ there is a monodromy matrix $ \mathcal{M} \in \text{Spin}(4,4;\mathbb{Z})$.

 These $m_i$ are the same parameters that were used in the supergravity analysis of \cite{Hull:2020byc}, so they can be used to make contact with the results that were obtained there.  Each of the mass parameters $m_i$ gives the mass of exactly two of the eight gravitini, so the amount of preserved supersymmetry in the orbifold can be tuned by choosing the $m_i$ to be zero or non-zero.
Our orbifolds preserve $\mathcal{N}=2r$ supersymmetry, where $r$ is the number of $m_i$ that is zero (or a multiple of $2\pi$).
For $\mathcal{N}=4$ supersymmetry with two of the $m_i$ non-zero, there are two possibilities depending on whether the twist is chiral with both of the non-zero $m_i$ in either ${\mathcal{M}}_L$ or in ${\mathcal{M}}_R$, or non-chiral with one of the non-zero $m_i$ in ${\mathcal{M}}_L$ and one in ${\mathcal{M}}_R$. A chiral twist (e.g. with $m_2=m_4=0$) leads to a (1,1) supergravity theory, as the massive multiplets are (1,1) massive supermultiplets in the terminology of \cite{Hull:2000cf}. Regarding the non-chiral twists, we refer to the twist with $m_3=m_4=0$ as (0,2) theory as the massive multiplets are (0,2) massive supermultiplets, and we refer to the twist with $m_1=m_2=0$ as a (2,0) theory as the massive multiplets are (2,0) massive supermultiplets. More details on the supergravity aspects are discussed in section \ref{Supergravity} and in \cite{Hull:2020byc}.

Using the isomorphism $\spinfour\cong\text{SU}(2)\times \text{SU}(2)$ (see appendix \ref{app: group theory} and in particular \eqref{embedding SO4}) the matrix $\hat{\mathcal{M}}\in   \spinfour\times\spinfour $ projects onto a matrix 
\begin{equation}
    {\mathcal{M}}_\theta= ({\mathcal{N}}_L,{\mathcal{N}}_R)\in  \text{SO}(4)_L\times \text{SO}(4)_R \ ,
\end{equation}
 which (by conjugation) can be brought to the standard form 
\begin{equation}
    {\mathcal{N}}_L=\begin{pmatrix}
    R(\theta_L) & 0  \\
    0 & R(\theta'_L) 
    \end{pmatrix} \,,\qquad
    {\mathcal{N}}_R=\begin{pmatrix}
    R(\theta_R) & 0  \\
    0 & R(\theta'_R) 
    \end{pmatrix} \ ,
\end{equation}
for four angles $\theta_L,\theta '_L,\theta_R,\theta'_R$ which are related to the $m_i$ by 
\begin{equation}
\begin{alignedat}{4}
\theta_L &=  {m_1+m_3}  \,, &\qquad\quad \theta_R &= {m_2+m_4}  \,, \\[3pt]
\theta '_L &= {m_1-m_3}  \,, &\qquad\quad \theta'_R &= {m_2-m_4}  \,.
\end{alignedat}
\label{theta's}
\end{equation}
Then e.g. $m_1= \frac 1 2 (\theta_L+\theta'_L)$ so that taking $\theta'_L\to \theta'_L+2\pi$ takes $m_1 $ to $m_1+\pi$ and $R(m_1)$ to $-R(m_1)$.
Note that $(m_1,m_2,m_3,m_4)$, $(m_1+\pi,m_2,m_3+\pi,m_4)$, $(m_1,m_2+\pi,m_3,m_4+\pi)$, $(m_1+\pi,m_2+\pi,m_3+\pi,m_4+\pi)$ all give the same angles 
 $\theta_L,\theta '_L,\theta_R,\theta'_R$ (mod $2\pi$), exhibiting the quadruple cover.

\subsection{An ansatz}

Solving in general for all possible integer valued T-duality elements that are conjugate to a rotation is a  difficult problem. Here we consider a special case for which results are known in the literature. 
For the case $G= \text{SL}(2,\mathbb{R})$, $K=\text{SO}(2)$ all
monodromies satisfying
\begin{equation}\label{conjugationsl2}
\mathcal{M} = g \tilde{\mathcal{M}} g^{-1} \,; \qquad\quad
\mathcal{M}\in\text{SL}(2,\mathbb{Z})\,,\qquad
\tilde{\mathcal{M}}\in\sotwo \,, \qquad g\in\sltwo \,
\end{equation}
are given in \cite{Dabholkar:2002sy,DeWolfe:1998eu}.
The $\tilde{\mathcal{M}}\in\sotwo$ for which this is possible are rotations by angles 
\begin{equation}\label{quantalpha}
\alpha\in\big\{0,\pm\tfrac{\pi}{3},\pm\tfrac{\pi}{2},\pm\tfrac{2\pi}{3},\pi\big\} \quad {\rm mod}\,\, 2\pi\ .
\end{equation}
Such rotations then generate a $\mathbb{Z}_2$, $\mathbb{Z}_3$, $\mathbb{Z}_4$ or $\mathbb{Z}_6$ subgroup of $\sltwo$ generated by $\tilde{\mathcal{M}}\in\sotwo$ and these in turn yield a $\mathbb{Z}_2$, $\mathbb{Z}_3$, $\mathbb{Z}_4$ or $\mathbb{Z}_6$ subgroup of $\text{SL}(2,\mathbb{Z})$
generated by $\mathcal{M}$.

The group $\text{SO}(4,4)$ has a subgroup $\text{SO}(2,2)\times \text{SO}(2,2)$ as well as a subgroup $\text{SO}(4)\times \text{SO}(4)$, and the group theory regarding these two subgroups is very similar.
Now, $\spinfourfour$ has a subgroup that is a double cover of this $\text{SO}(2,2)\times \text{SO}(2,2)$
given by 
\begin{equation}
  \frac {\spintwotwo\times\spintwotwo}   {\mathbb{Z}_2}\subset \spinfourfour\,.
\end{equation}
As
\begin{equation}\label{sl2subgroups}
\sltwo^2 \cong \spintwotwo\,,
\end{equation}
we see that $\spinfourfour$ has a subgroup
\begin{equation}\label{sl2subgroups monodromy}
\frac {\sltwo^4}   {\mathbb{Z}_2}
 \subset \spinfourfour \,.
\end{equation}
The quotient by ${\mathbb{Z}_2}$ reflects the fact that the element $(-\mathbbm{1},-\mathbbm{1},-\mathbbm{1},-\mathbbm{1})\in {\sltwo^4} $ maps to the identity in $\spinfourfour$.
We will then restrict our monodromy to lie in this subgroup
and use the known results for $\sltwo$. We   stress that these are not   all possible $[\spinfour\times\spinfour]
/ \mathbb{Z}_2$ rotations that can be conjugated to integer valued elements of the T-duality group $\text{Spin}(4,4;\mathbb{Z})$, but these are the ones that we focus on for the purposes of this work.

Restricting   the  subgroup $\sltwo^4  /\mathbb{Z}_2 \subset \spinfourfour$  to a subgroup of 
$\text{Spin}(4,4;\mathbb{Z})$, restricts us to a discrete subgroup of $\sltwo^4$ and we need to check what subgroup arises in this way.
The $8_v $ of $\text{Spin}(4,4)$ restricts to a $(4,1)+(1,4)$ of $\spintwotwo\times\spintwotwo $ and as the vector representation $4$ of $\spintwotwo$
is the $(2,2) $ representation of $\sltwo^2$, the $8_v$ is in the
$$(2,2,1,1)+(1,1,2,2)$$
representation of $\sltwo^4$. The $8_s$ representation of Spin(4,4) is in the 
$$(1,2,2,1)+(2,1,1,2)$$
representation of $\sltwo^4$. Thus the 16 charges are in the 
\begin{equation}(2,2,1,1)+(1,1,2,2)+(1,2,2,1)+(2,1,1,2)\,.
\label{chargereps}
\end{equation}
Choosing 16 basis vectors $e_i$ for the 16-dimensional charge lattice, the allowed charges are $n^i e_i$ where $n^i$ is a 16-vector of integers, $n^i\in \mathbb{Z}^{16}$.
The group $\sltwo^4$ acts on the integers $n^i$ in the representation (\ref{chargereps})
and will preserve the lattice if restricted to the discrete subgroup 
$\text{SL}(2,\mathbb{Z})^4$.

Then the monodromy $\mathcal{M}\in\text{SL}(2,\mathbb{Z})^4/\mathbb{Z}_2$ 
is conjugate to a rotation in $\text{SO}(2)^4$ specified (up to conjugation) by four angles $\alpha _i\in \big\{0,\pm\tfrac{\pi}{3},\pm\tfrac{\pi}{2},\pm\tfrac{2\pi}{3},\pi\big\}$ mod $2\pi$, where $i=1,\ldots,4$. 
The element $(-\mathbbm{1},-\mathbbm{1},-\mathbbm{1},-\mathbbm{1})\in {\sltwo^4} $ with trivial monodromy corresponds to
$ \alpha_1=\alpha_2=  \alpha_3=\alpha_4 $; this will play a role in the ${\cal N}=6$ supersymmetric cases discussed below.

The angles $\alpha _i$ are related to the angles $\theta_i$ by (for details see appendix \ref{Ap:embedding})
\begin{equation}
\begin{alignedat}{4}
\theta_L &=  {\alpha_1+\alpha_3}  \,, &\qquad\quad \theta_R &= {\alpha_1-\alpha_3}  \,, \\[3pt]
\theta '_L &= {\alpha_2+\alpha_4}  \,, &\qquad\quad \theta'_R &= {\alpha_2-\alpha_4}  \,.
\end{alignedat}
\label{thetaalph's}
\end{equation}
This is similar in form to (\ref{theta's}) and so we see again that the $\text{SO}(2)^4$ parameterised by the $\alpha_i$ provides a quadruple cover of the $\text{SO}(2)^4$ parameterised by the $\theta_i$.
Comparing \eqref{thetaalph's} with (\ref{theta's}), we see that
\begin{equation}
\begin{alignedat}{4}
{m_1+m_3} &=  {\alpha_1+\alpha_3}  \,, &\qquad\quad {m_2+m_4} &= {\alpha_1-\alpha_3}  \,, \\[3pt]
{m_1-m_3}  &= {\alpha_2+\alpha_4}   \,, &\qquad\quad {m_2-m_4} &= {\alpha_2-\alpha_4}  \,.
\end{alignedat}
\label{malph}
\end{equation}
All the equations relating angles hold modulo $2\pi$.

The equations (\ref{malph}) then determine the mass parameters $m_i$, giving one solution as
\begin{equation}
\begin{aligned}\label{m's in terms of alpha's}
m_1&=\tfrac{1}{2}(\alpha_1+\alpha_2+\alpha_3+\alpha_4) \,,\qquad\quad
&m_2=\tfrac{1}{2}(\alpha_1+\alpha_2-\alpha_3-\alpha_4) \,,\\[4pt]
m_3&=\tfrac{1}{2}(\alpha_1-\alpha_2+\alpha_3-\alpha_4) \,,\qquad\quad
&m_4=\tfrac{1}{2}(\alpha_1-\alpha_2-\alpha_3+\alpha_4) \,.
\end{aligned}
\end{equation}
For a given set of $\alpha_i$, the complete set of solutions is given by the 
 $(m_1,m_2,m_3,m_4)$ in (\ref{m's in terms of alpha's}), together with $(m_1+\pi,m_2,m_3+\pi,m_4)$, 
  $(m_1,m_2+\pi,m_3,m_4+\pi)$, $(m_1+\pi,m_2+\pi,m_3+\pi,m_4+\pi)$. All these yield the same angles 
 $\theta_L,\theta '_L,\theta_R,\theta'_R$, giving a quadruple cover.
Then for a given set of $\alpha_i$, there are four possible choices of $\hat {\mathcal{M}} $. Denoting the canonical $\hat {\mathcal{M}} $ (given by (\ref{mhat}),(\ref{mhat2}) with the $m$'s given by (\ref{m's in terms of alpha's})) by $\hat {\mathcal{M}}(m) $, the four choices are
$\hat {\mathcal{M}}(m) $, $\hat {\mathcal{M}} (m)(-1)^{F_s}$, $\hat {\mathcal{M}} (m)(-1)^{F_L}$ and $\hat {\mathcal{M}}(m)(-1)^{F_R} $.

The results just stated come from requiring agreement of the two different parameterisations of $\mathcal{M}_\theta$ in terms of the $m$'s or the $\alpha$'s respectively. Half of this ambiguity is lifted by requiring agreement for the two different parameterisations of $\tilde {\mathcal{M}}$. Then  a given set of $\alpha$'s determines a  
$\tilde {\mathcal{M}}$, which we denote $\tilde {\mathcal{M}}(\alpha)$ and is conjugate to a monodromy $ {\mathcal{M}}(\alpha)$.
For this,
there remain two possible choices of $\hat {\mathcal{M}} $, 
which are $\hat {\mathcal{M}}(m) $ and $\hat {\mathcal{M}} (m)(-1)^{F_s}$, corresponding to a choice of generalized spin structure (see the discussion of fermionic monodromies below). We choose the generalized spin structure so that 
the monodromy ${\mathcal{M}}(\alpha)$ gives the twist $\hat {\mathcal{M}}(m) $
and ${\mathcal{M}}(\alpha)(-1)^{F_s}$ gives the twist $\hat {\mathcal{M}}(m) (-1)^{F_s}$.

It will be useful to note that (\ref{m's in terms of alpha's}) can be inverted to give 
\begin{equation}
\begin{aligned}
\alpha_1&=\tfrac{1}{2}(m_1+m_2+m_3+m_4) \,,\qquad\quad
&\alpha_2=\tfrac{1}{2}(m_1+m_2-m_3-m_4) \,,\\[4pt]
\alpha_3&=\tfrac{1}{2}(m_1-m_2+m_3-m_4) \,,\qquad\quad
&\alpha_4=\tfrac{1}{2}(m_1-m_2-m_3+m_4) \,.
\end{aligned}
\label{alpha's in terms of m's}
\end{equation}

As the allowed values for each of the $\alpha_i$ are $\big\{0,\pm\tfrac{\pi}{3},\pm\tfrac{\pi}{2},\pm\tfrac{2\pi}{3},\pi\big\}$ mod $2\pi$, the allowed values of the $m_i$ can be found by taking linear combinations of these. It will often be useful to rewrite these parameters as
\begin{equation}\label{miN}
m_i = \frac{2\pi N_i}{p} \,.
\end{equation}
Here the $N_i$ are integers, and $p$ is the smallest positive integer such that all four $m_i$ can be written like this. This relation defines the quantization condition on the mass parameters and guarantees that
the integer $p$ is the order of the monodromy matrix, $\hat {\cal M}^p=1$.

As follows from \eqref{m's in terms of alpha's}, the quantization of the $\alpha$'s allows for the values $p\in\big\{2,3,4,6,8,12,24\big\}$. Notice that the values $p=5$ and $p=10$ found in \cite{Erler:1992ki} and mentioned in the beginning of this section do not appear in our list. This is because the duality twists from our ansatz lie in a particular $\text{SL}(2)^4$ subgroup given by \eqref{sl2subgroups monodromy}, and the values $p=5,10$ do not arise from this subgroup.

 In \autoref{allowed orbifolds} we list which values of $p$ yield orbifolds that preserve a certain amount of supersymmetry. Recall that our orbifolds preserve $\mathcal{N}=2r$ supersymmetry where $r$ is the number of $m_i$ that is zero (mod $2\pi$).  Notice that fixing $p$ does not fix the orbifold. For a given $p$, there can be more than one possibility. An example is $\mathcal{N}=4\,(0,2)$, with $p=4$, for which there can be both a symmetric and an asymmetric orbifold.

\renewcommand{\arraystretch}{1.4}
\begin{table}[h!]
\centering
\begin{tabular}{|c|c|c|}
\hline
\;Preserved supersymmetry\; & \;(A)symmetric\; & \;Possible $\mathbb{Z}_p$ orbifold ranks\; \\ \hline\hline
$\mathcal{N}=6$ & A & $p=2,3$ \\ \hline
 \multirow{2}{*}{$\mathcal{N}=4\;(0,2)$} & S & $p=2,3,4,6$ \\
& A & $p=3,4,6,12$ \\ \hline
$\mathcal{N}=4\;(1,1)$ & A & $p=2,3,4,6,12$ \\ \hline
$\mathcal{N}=2$ & A & $p=2,3,4,6,12$ \\ \hline
 \multirow{2}{*}{$\mathcal{N}=0$} & S & $p=2,3,4,6,8,12,24$ \\
& A & $p=3,4,6,8,12,24$ \\ \hline
\end{tabular}
\captionsetup{width=.83\linewidth}
\caption{\textit{The class of orbifolds studied in this paper, indicating the amount of preserved supersymmetry and the rank of the orbifold. We also indicate which orbifolds are symmetric (S, $m_1=m_2$ and $m_3=m_4$) or asymmetric (A). The $\mathcal{N}=4\,(0,2)$ is defined from having $m_3=m_4=0$ whereas the $(1,1)$ is defined from $m_2=m_4=0$, up to trivial permutations exchanging $m_{1,2}$ with $m_{3,4}$. Notice the absence of $p=2$ for asymmetric $\mathcal{N}=4\,(0,2)$ orbifolds. In this case we have $m_{1,2}=\pm \pi$ but $+\pi$ is the same as $-\pi$, and so this is a symmetric orbifold in disguise.}}
\label{allowed orbifolds}
\end{table}
\renewcommand{\arraystretch}{1}

\subsection{Special cases and examples}

The action of the group $\text{SL}(2)^4$ on $T^4$ we are considering splits it into the product of   $T^2\times T^2$ with the $\text{SL}(2)^2$ parameterised by $\alpha_1,\alpha _3$ acting on one $T^2$ and the   $\text{SL}(2)^2$ parameterised by $\alpha_2,\alpha _4$ acting on the other $T^2$. 

\subsubsection*{Fermionic monodromies $(-1)^{F_s}$,$(-1)^{F_L}$,$(-1)^{F_R}$}
\label{fermionic monodromies}

Here we consider monodromies such that ${\mathcal{M}}_\theta=\mathbbm{1} $
(i.e. all angles $\theta_i=(\theta_L,\theta'_L,\theta_R,\theta'_R)$ are $0$ mod $2\pi$),
so that the NS-NS sector is invariant and the twist only acts on fermions and on the R-R sector. For $\theta_i=0$ mod $2\pi$, each of the $m_i$ and each of the $\alpha _i$ must be either $0$ or $\pi $ (mod $2\pi$).
Consider for instance the twist 
\begin{equation}
     m_1=m_3=\pi, \quad m_2=m_4=0 \ ,
\end{equation}
so that $\theta_L=2\pi, \theta'_L=0$ and $\theta_R=\theta'_R=0$. The $\alpha_i$ are given by $\alpha_1=\alpha_3=\pi,\quad \alpha_2=\alpha_4 = 0$.
This lifts to a monodromy on the double cover
\begin{equation}
    \hat{\mathcal{M}}= ({\mathcal{M}}_L,{\mathcal{M}}_R)=(- \mathbbm{1}, \mathbbm{1})\ ,
\end{equation}
so that $ \hat{\mathcal{M}}=(-1)^{F_L}$. Choosing instead $m_1=m_3=0$ and $m_2=m_4=\pi$, we get $\hat{\mathcal{M}}= ( \mathbbm{1}, -\mathbbm{1})$, so that $ \hat{\mathcal{M}}=(-1)^{F_R}$.

We now consider a monodromy with
\begin{equation}
m_1=m_2=m_3=m_4=\pi\,.
\end{equation}
Then
\begin{equation}
    \hat{\mathcal{M}}= ({\mathcal{M}}_L,{\mathcal{M}}_R)=(- \mathbbm{1}, -\mathbbm{1})\,,
\end{equation}
so that $ \hat{\mathcal{M}}=(-1)^{F_s}$. This has a trivial projection to $\tilde K$: $\tilde {\mathcal{M}}=\mathbbm{1}$.
This then corresponds to ${\mathcal{M}}=\mathbbm{1}$ so that orbifold with $m_1=m_2=m_3=m_4=\pi$
has
a monodromy ${\mathcal{M}}(-1)^{F_s}=(-1)^{F_s}$.

The orbifold of the type IIB string by $(-1)^{F_s}$ gives the non-supersymmetric type 0B string \cite{Dixon:1986iz,Seiberg:1986by}. Here we are combining the $(-1)^{F_s}$ with a half-shift on the $S^1$ to give an interesting $\mathbb{Z}_2$ orbifold breaking all supersymmetry, which can be made tachyon-free. We will discuss it further in subsection \ref{n=0 and supertraces}.

\subsubsection*{Symmetric orbifold of $T^2$}

Choosing
\begin{equation}
     \alpha_2= \alpha_3=\alpha_4 = 0 \qquad\Rightarrow\qquad
   \theta_L=\theta_R = \alpha_1; \qquad \theta'_L=\theta'_R=0\,.
\end{equation}
Then (\ref{m's in terms of alpha's}) gives
\begin{equation}
    m_1=m_2= m_3=m_4= \frac{\alpha_1}{2} \,.
\end{equation}

This is a symmetric orbifold with $\alpha_1$ parameterizing an $\text{SO}(2)$ acting as a rotation on one $T^2$. 
For $\alpha_1 =2\pi/n $, ${\mathcal{M}}^n=\mathbbm{1}$,
as ${\mathcal{M}}^n$ is a rotation through $2\pi$, but $\hat {{\mathcal{M}}}^n=-\mathbbm{1}$, as the double cover of the $2\pi $ rotation acts as $-1$ on spinors. Thus, this constitutes a $\mathbb{Z}_{p}$ orbifold with $p=2n$. From the allowed values of $\alpha_1$, we can read of the allowed values of $n$, which are $n=1,2,3,4,6$, such that $p=2,4,6,8,12$. As all $m_i\ne 0$, this orbifold breaks all supersymmetry.

\subsubsection*{Symmetric orbifold of $T^4$}

Choosing
\begin{equation}
    \alpha_3=\alpha_4 = 0 \qquad\Rightarrow\qquad \theta_L= \theta_R=  \alpha_1, \qquad \theta'_L= \theta'_R=  \alpha_2
     \,,
\end{equation}
and (\ref{m's in terms of alpha's}) gives
\begin{equation}
m_1=m_2=\tfrac 1 2 (\alpha_1+\alpha_2), \qquad m_3=m_4=\tfrac 1 2 (\alpha_1-\alpha_2)\,.
\end{equation}
Recall that the allowed values for each of the $\alpha_1,\alpha_2$ are $\big\{0,\pm\tfrac{\pi}{3},\pm\tfrac{\pi}{2},\pm\tfrac{2\pi}{3},\pi\big\}$ mod $2\pi$.
For most choices of $\alpha_{1,2}$, all $m_i\ne 0$, so this orbifold breaks all supersymmetry. When $\alpha_1=\alpha_2$, we have $\mathcal{N}=4$ supersymmetry of type $(0,2)$.

\subsubsection*{Example: a $\mathbb{Z}_{24}$ orbifold}
\label{a Z24 orbifold}
We now   give an explicit example of a $\mathbb{Z}_{24}$ orbifold arising as above, i.e. as a symmetric orbifold of $T^4$. We can see from \autoref{allowed orbifolds} that $\mathbb{Z}_{24}$ orbifolds always break all supersymmetry. The example that we choose is
\begin{equation}\label{parameters Z24}
    \alpha_1=\frac{\pi}{2},\quad \alpha_2=\frac{\pi}{3},\quad \alpha_3=\alpha_4 = 0 \qquad\Rightarrow\qquad m_1=m_2=\frac{5\pi}{12}, \quad m_3=m_4=\frac{\pi}{12} \,.
\end{equation}

The monodromies in the SL$(2,\mathbb{Z})$ subgroups \eqref{sl2subgroups monodromy} that the $\alpha$'s rotate in are known. They can be found e.g. in \cite{Dabholkar:2002sy}. The ones corresponding to $\alpha_1$ and $\alpha_2$ respectively read
\begin{equation}\label{sl2monodromies Z24}
    \begin{pmatrix}
    0 & -1 \\
    1 & 0
    \end{pmatrix} \,, \qquad
    \begin{pmatrix}
    0 & -1 \\
    1 & 1
    \end{pmatrix}\,.
\end{equation}
The first one is simply a rotation matrix over an angle $\alpha_1={\pi}/{2}$, and the second one is conjugate to a rotation matrix over an angle $\alpha_2={\pi}/{3}$ via a conjugation \`a la \eqref{conjugationsl2}:
\begin{equation}
    \begin{pmatrix}
    0 & -1 \\
    1 & 1
    \end{pmatrix} = \sqrt{\frac{2}{\sqrt{3}}}\begin{pmatrix}
    1 & 0 \\
    -\tfrac{1}{2} & \tfrac{\sqrt{3}}{2}
    \end{pmatrix} \cdot \begin{pmatrix}
    \cos \tfrac{\pi}{3} & -\sin \tfrac{\pi}{3}\\
    \sin \tfrac{\pi}{3} & \cos \tfrac{\pi}{3}
    \end{pmatrix} \cdot \left[ \sqrt{\frac{2}{\sqrt{3}}}\begin{pmatrix}
    1 & 0 \\
    -\tfrac{1}{2} & \tfrac{\sqrt{3}}{2}
    \end{pmatrix} \right]^{-1} \;.
\end{equation}
We can use appendix \ref{isomorphism so22} to map the $\sltwo$ matrices \eqref{sl2monodromies Z24} properly to $\sotwotwo$ matrices, which can then be combined into an $\sofourfour$ element. This yields the monodromy
\begin{equation}\label{monodromy Z24}
    \mathcal{M} =\begin{pmatrix}
    0 & -1 & 0 & 0 & 0 & 0 & 0 & 0 \\
    1 & 0 & 0 & 0 & 0 & 0 & 0 & 0 \\
    0 & 0 & 0 & -1 & 0 & 0 & 0 & 0 \\
    0 & 0 & 1 & 1 & 0 & 0 & 0 & 0 \\
    0 & 0 & 0 & 0 & 0 & -1 & 0 & 0 \\
    0 & 0 & 0 & 0 & 1 & 0 & 0 & 0 \\
    0 & 0 & 0 & 0 & 0 & 0 & 1 & -1 \\
    0 & 0 & 0 & 0 & 0 & 0 & 1 & 0
    \end{pmatrix} \in\sofourfour  \,.
\end{equation}
This monodromy is written in $\tau$-frame (using the language from appendix \ref{app: eta and tau frame}), meaning that the group $\sofourfour$ consists of matrices that preserve the metric
\begin{equation}
    \tau = \begin{pmatrix}
    0 & \mathbbm{1}_{4\times4} \\
    \mathbbm{1}_{4\times4} & 0 
    \end{pmatrix} \,.
\end{equation}
We see that, in this frame, the monodromy is integer-valued as it should be\footnote{This is the case because the T-duality group works on integer valued charges (winding and momentum numbers) in $\tau$-frame.}. 
Notice that the monodromy acts as a diffeomorphism on $T^4$, as it should, since we consider a symmetric orbifold. The geometric group GL($4,\mathbb{Z}$) is embedded in the T-duality group SO(4,4,$\mathbb{Z}$) as
\begin{equation}
   \mathcal{M}_{\text{SO}}=  \begin{pmatrix}
    g & 0 \\
    0 & g^{-t}
    \end{pmatrix}\,,\qquad g = \begin{pmatrix}
    0 & -1 & 0 & 0 \\
    1 & 0 & 0 & 0  \\
    0 & 0 & 0 & -1  \\
    0 & 0 & 1 & 1  \\
    \end{pmatrix}\,,\qquad g \in \text{GL}(4,\mathbb{Z})\,.
\end{equation}

Note that the monodromy in \eqref{monodromy Z24} generates an orbit of rank 12: $\mathcal{M}^{12}_{\text{SO}}=\mathbbm{1}$. 
The action on the fermions is through the matrix $\hat{\mathcal{M}}$ given by (\ref{mhat}),(\ref{mhat2}) with the $m$'s in \eqref{parameters Z24} and this generates an orbit of rank 24 as it satisfies $\hat{\mathcal{M}}^{24}=1$.

\subsubsection*{Asymmetric orbifold of $T^2$}

Choosing
\begin{equation}
    \alpha_2=\alpha_4 = 0 \qquad\Rightarrow\qquad \theta_L= \alpha_1+\alpha_3 ,\qquad
   \theta_R= \alpha_1-\alpha_3 ; \qquad \theta'_L=\theta'_R=0 \,,
\end{equation}
and (\ref{m's in terms of alpha's}) gives
\begin{equation}
  m_1=m_3=\tfrac{1}{2}(\alpha_1+\alpha_3), \qquad m_2=m_4=\tfrac{1}{2}(\alpha_1-\alpha_3)\,.
\end{equation}
As discussed previously,   $\alpha_1,\alpha_3\in\big\{0,\pm\tfrac{\pi}{3},\pm\tfrac{\pi}{2},\pm\tfrac{2\pi}{3},\pi\big\}$ mod $2\pi$.
For example, choosing
$\alpha_1=\pi/2,\alpha_3=\pi/3$ gives $m_1=m_3=5\pi/12$ and $m_2=m_4=\pi/12$ giving a $\mathbb{Z}_{24}$ orbifold. 
Again, as all $m_i\ne 0$, this orbifold breaks all supersymmetry. We can preserve some supersymmetry e.g. by choosing $\alpha_1=\alpha_3$, so that $m_2=m_4=0$ and there is $\mathcal{N}=4$ supersymmetry of type $(1,1)$.

\subsubsection*{Chiral orbifold of $T^4$}
Choosing
\begin{equation}
  \alpha_1=\alpha_3,\quad  \alpha_2=\alpha_4  \qquad\Rightarrow\qquad \theta_L= 2\alpha_1  ,\qquad \theta'_L= 2\alpha_2,\qquad 
   \theta_R=\theta'_R=0 \,,
   \label{chiral orbifold alphas}
\end{equation}
and (\ref{m's in terms of alpha's}) gives
\begin{equation}
  m_1=\alpha_1+\alpha_2,\qquad
  m_3=\alpha_1-\alpha_2, \qquad m_2=m_4=0\,.
  \label{chiral orbifold m's}
\end{equation}
This type of orbifold can preserve either $\mathcal{N}=6$ or $\mathcal{N}=4$ supersymmetry. We now look at these cases separately.

\subsubsection*{$\mathcal{N}=6$ supersymmetric orbifold}
\label{n=6 allows only p=2 or 3}

For $\mathcal{N}=6$ supersymmetry, precisely one of the $m_i$ should be non-zero. Choosing this to be $m_1\ne 0$ with $m_2=m_3=m_4=0$ requires (using \eqref{chiral orbifold alphas},\eqref{chiral orbifold m's})
\begin{equation}
  \alpha_1=\alpha_2=  \alpha_3=\alpha_4  = \tfrac 1 2  m_1
  \qquad\Rightarrow\qquad \theta_L= \theta'_L= 2\alpha_1 =m_1 ,\qquad  
   \theta_R=\theta'_R=0 \,,
\end{equation}
so that this is a chiral orbifold. Recall that
$\alpha_1=m_1/2\in\big\{0,\pm\tfrac{\pi}{3},\pm\tfrac{\pi}{2},\pm\tfrac{2\pi}{3},\pi\big\}$ mod $2\pi$.

The factor of two in the relation between $\alpha_i$ and $m_1$ is important.
The case $\alpha_1 =\pi$ gives $m_1=2\pi$ as well as $\theta_L= \theta'_L=2 \pi$, so that $\hat {\mathcal {M}}=\mathbbm{1}$ and $ \mathcal {M}=\mathbbm{1}$ so that 
the monodromy is trivial. This reflects the fact that
the monodromy is not in ${\sltwo^4}$ but in ${\sltwo^4}/   {\mathbb{Z}_2}$. Similarly, $\alpha _1=\pi/2$ gives a $\mathbb{Z}_2$ symmetry instead of the $\mathbb{Z}_4$ that might have been expected. 
Finally, $\alpha_1={\pi}/{3}$ gives a $\mathbb{Z}_3$ symmetry instead of a $\mathbb{Z}_6$ symmetry, while $\alpha_1={2\pi}/{3}$
also gives a $\mathbb{Z}_3$ symmetry. Thus, the only possible values of $p$ from our ansatz are 2 and 3.

\subsubsection*{$\mathcal{N}=4$ chiral supersymmetric orbifold}
\label{n=4 chiral orbis}

For an orbifold with $\mathcal{N}=4$ chiral supersymmetry we need to turn on two mass parameters. Requiring $m_1,m_3\neq 0$ and $m_2=m_4=0$ gives (using \eqref{chiral orbifold alphas},\eqref{chiral orbifold m's})
\begin{equation}
  \alpha_1=\alpha_3= \tfrac 1 2 (m_1+m_3), \quad  
  \alpha_2=\alpha_4  = \tfrac 1 2 (m_1-m_3)
  \quad\Rightarrow\quad \theta_L= m_1+m_3, \quad   \theta'_L= m_1-m_3 ,\quad  
   \theta_R=\theta'_R=0 \,.
\end{equation}
This choice of mass parameters leads to a (1,1) theory.

\subsubsection*{Non-chiral supersymmetric orbifolds of $T^4$}
We discuss two examples of this type, preserving $\mathcal{N}=4$ and $\mathcal{N}=2$ supersymmetry respectively.
\subsubsection*{$\mathcal{N}=4$ non-chiral supersymmetric orbifold}
\label{non-chiral N=4}
Requiring $m_1,m_2\neq 0$ and $m_3=m_4=0$ gives (using \eqref{alpha's in terms of m's})
\begin{equation}
  \alpha_1=\alpha_2= \tfrac 1 2 (m_1+m_2), \quad  
  \alpha_3=\alpha_4  = \tfrac 1 2 (m_1-m_2)
  \quad\Rightarrow\quad \theta_L=\theta'_L= m_1, \quad   
   \theta_R=\theta'_R=m_2 \,.
\end{equation}
This choice of mass parameters leads to a (0,2) theory. In general, this is an asymmetric orbifold, but for the special choice $m_1=m_2$, the orbifold is symmetric.

\subsubsection*{$\mathcal{N}=2$ supersymmetric orbifold}
\label{non-chiral N=2}
For $\mathcal{N}=2$, we turn on three mass parameters and the remaining one is zero. Requiring e.g. $m_4=0$ gives (using \eqref{alpha's in terms of m's})
\begin{equation}
\begin{aligned}
\alpha_1&=\tfrac{1}{2}(m_1+m_2+m_3) \,,\qquad\quad
\alpha_2&=\tfrac{1}{2}(m_1+m_2-m_3) \,,\\[4pt]
\alpha_3&=\tfrac{1}{2}(m_1-m_2+m_3) \,,\qquad\quad
\alpha_4&=\tfrac{1}{2}(m_1-m_2-m_3) \,.
\end{aligned}
\end{equation}
so that
\begin{equation}
 \theta_L= m_1+m_3, \quad   \theta'_L= m_1-m_3, \qquad   
   \theta_R=\theta'_R=m_2 \,.
\end{equation}
Notice that this is always an asymmetric orbifold. We will discuss some explicit examples of non-chiral orbifolds in section \ref{closed string spectrum}.

\section{The orbifold action on the world-sheet fields and the partition function}
\label{The orbifold action and partition function}

\subsection{Lattices and tori}

Our starting point was a torus compactification on a square torus $\mathbb{R}^4/\mathbb{Z}^4$ with periodic torus coordinates so that the
BPS 0-brane charge lattice was preserved by Spin$(4,4;\mathbb{Z})$.
The  moduli were packaged into the background metric and antisymmetric tensor gauge fields  on the torus and the fixed point under the action of the monodromy was at a point $[\bar g]$ in the moduli space.
Acting with a duality transformation $\bar g$ in $G$ moves the fixed point to the origin and diagonalizes the action of the monodromy on the fields. However, it also deforms the torus, so that it is no longer a square torus and the boundary conditions of the torus coordinates are changed, so that the left-moving coordinates take values on a torus $\mathbb{R}^4/\Lambda_L$ for some lattice $\Lambda_L$ and the right-moving coordinates take values on a torus $\mathbb{R}^4/\Lambda_R$ for some lattice $\Lambda_R$. 
The left-moving momenta $p_L$ take values in the lattice dual to $\Lambda_L$ and the right-moving momenta $p_R$ take values in the lattice dual to $\Lambda_R$. Then the vectors $(p_L,p_R)$ build an 8-dimensional even, self-dual Lorentzian lattice, which is known as the Narain lattice \cite{narain1989new} and we will denote by $\Gamma_{4,4}$. The sublattices $\Lambda_L,\Lambda_R$ of $\Gamma_{4,4}$ are invariant under the action of the ${\mathbb{Z}} _p$ symmetry that is used in the orbifold. These sublattices are associated with root lattices of Lie algebras\footnote{For a thorough discussion on lattices we refer to \cite{lerchie1989lattices}.}.

For the ansatz of the last section, the monodromy is in 
\begin{equation}
  \frac {\spintwotwo\times\spintwotwo}   {\mathbb{Z}_2}\subset \spinfourfour \ .
\end{equation}
This means that the $T^4$ can be regarded as $T^2\times T^2$ with one $\spintwotwo$ factor in the monodromy acting on the first $T^2$ and the other acting on the second $T^2$.
In this case, the two four-dimensional lattices $\Lambda_L,\Lambda_R$ must each decompose into the sum of two 2-dimensional lattices: $\Lambda_L=\Lambda_1\oplus \Lambda_2$ and similarly for $\Lambda_R$.
Each 2-dimensional lattice must then be $A_2$  or $A_1\oplus A_1\cong D_2$\footnote{$D_n$ and $A_n$ denote the root lattices of $\text{SO}(2n)$ and $\text{SU}(n+1)$ respectively.}.
However, below we will also consider other 4-dimensional lattices $\Lambda_L,\Lambda_R$ that fall outside the scope of the ansatz of section \ref{sec:Orbifold constructions}. This allows further possible values of $p$, e.g. $p=5$, and in some cases it is necessary for modular invariance, as we will show in detail.

\subsection{The world-sheet fields}
\label{sec:the world-sheet fields}
We are now ready to discuss the action of the orbifold on the worldsheet variables. We split the bosonic coordinates as $X^M \rightarrow (\hat{X}^{\hat{\mu}},Y^m) \rightarrow (X^\mu, Z, Y^m)$, where $Y^m$ ($m=1,\ldots,4$) are the $T^4$ coordinates, $Z$ is the circle coordinate, $X^\mu$ ($\mu=0,\ldots,4$) are the $\mathbb{R}^{1,4}$ coordinates, and $\hat{X}^{\hat{\mu}}$ ($\hat{\mu} = 0,\ldots,5$) are the coordinates on $\mathbb{R}^{1,4}\times S^1$. We often work in complex coordinates on the torus, which we denote by $W^i = \tfrac{1}{\sqrt{2}}(Y^{2i-1}+iY^{2i})$ with $i=1,2$. On-shell, the worldsheet coordinates split into  left and right-moving parts as
\begin{equation}
W^i(\sigma^1,\sigma^2) = {W}_{L}^i(\sigma^1+\sigma^2) + W_{R}^i(\sigma^1-\sigma^2) \,.
\end{equation}
Here, $\sigma^1$ and $\sigma^2$ are the coordinates on the worldsheet which we always take to be of Lorentzian signature.
We denote the oscillators of all bosonic coordinates by $\tilde{\alpha}^M_n$ and $\alpha^M_n$ where the tilde indicates a left-mover, and we use different indices ($\hat{\mu}$, $\mu$, $z$, $m$ or $i$) according to the above decomposition. The fermionic modes of the superstring are denoted by $\tilde{b}^M_n$ and $b^M_n$ with a similar index structure. In the case of complex modes, we use a bar to denote the complex conjugate.

Now that we have set up our notation, we are ready to present the orbifold action. It is most easily stated in terms of the matrix $\tilde{\mathcal{M}}$ in \eqref{conjugation}, parametrized by the four mass parameters $m_i$. It works on the bosonic torus coordinates with asymmetric rotations
\begin{equation}\label{orbiaction2}
\begin{aligned}
{W}_{L}^1 \;&\rightarrow\; e^{i(m_1+m_3)}\: {W}_{L}^1 \,, \\
{W}_{L}^2 \;&\rightarrow\; e^{i(m_1-m_3)}\: {W}_{L}^2 \,, \\
W_{R}^1 \;&\rightarrow\; e^{i(m_2+m_4)}\: W_{R}^1 \,, \\
W_{R}^2 \;&\rightarrow\; e^{i(m_2-m_4)}\: W_{R}^2 \,,
\end{aligned}
\end{equation}
and with the same action on the fermionic torus coordinates. 
In addition, symmetric orbifolds correspond to $m_1=m_2$ and $m_3=m_4$. Furthermore, the rotations on the torus are accompanied by a shift along the circle coordinate
\begin{equation}\label{shift}
Z \;\rightarrow\; Z + 2\pi \mathcal{R} / p \,,
\end{equation}
which makes the orbifold freely acting. Here $\mathcal{R}$ is the circle radius ($Z\sim Z+2\pi \mathcal{R}$). Due to this shift, states that carry momentum in the $Z$-direction obtain a phase $e^{2\pi i n / p}$ under the orbifold action, where $n$ is the momentum number of the state. 

Let us now discuss the orbifold action on the Neveu-Schwarz (NS) and Ramond (R) vacua, which we denote by $\ket{0}_{{L}/{R}}$ and $|s_1,s_2,s_3,s_4\rangle_{{L}/{R}}$ respectively, where $s_{\kappa} = \pm \tfrac{1}{2}$. The subscript ${L}/{R}$ is used to distinguish the left and the right-moving vacua. We choose the GSO projection in such a way that both R-vacua   satisfy 
\begin{equation}
    \sum_{\kappa=1}^4 s_{\kappa} \,\in\, 2\mathbb{Z}\,.
\end{equation}
The NS-vacua are spacetime scalars and are invariant under the orbifold action. On the other hand the R-vacua are $10D$ spinors and we know how they transform under rotations, so in particular under the orbifold action. In general we have
\begin{equation}\label{spinorrotation}
|s_1,s_2,s_3,s_4\rangle \;\;\rightarrow\;\; \exp \left( 2\pi i \sum_{\kappa=1}^4 v_{\kappa} \, S_{\kappa} \right) |s_1,s_2,s_3,s_4\rangle = e^{2\pi i \,\vec{v} \cdot \vec{s}}\; |s_1,s_2,s_3,s_4\rangle \,,
\end{equation}
where the $S_{\kappa} = J_{2\kappa-1,2\kappa}$ are the Cartan generators of the little group SO$(8)$ with eigenvalues $s_{\kappa}$. The $v_{\kappa}$ denotes a rotation in the $2\kappa-1$ and $2\kappa$ directions over an angle $2\pi v_{\kappa}$. Whenever the rotation works asymmetrically on left and right-movers, the formula above applies to spinors in each sector individually. In this case we use $\tilde{u}_{\kappa}$ and $u_{\kappa}$ for the left and right-moving rotation parameters respectively. Using this notation, we read off from \eqref{orbiaction2} that our orbifold action is a rotation with
\begin{equation}
\begin{alignedat}{4}
\tilde{u}_3 &= \frac{m_1+m_3}{2\pi} \,, &\qquad\quad u_3 &= \frac{m_2+m_4}{2\pi} \,, \\[3pt]
\tilde{u}_4 &= \frac{m_1-m_3}{2\pi} \,, &\qquad\quad u_4 &= \frac{m_2-m_4}{2\pi} \,.
\end{alignedat}
\label{u's}
\end{equation}
and the other rotation parameters ($\tilde{u}_{1,2}$ and ${u}_{1,2}$) equal to zero. These rotation parameters are subject to the constraint
\begin{equation}
   u_{3,4}=\frac{n_{3,4}}{p}\,, \qquad n_{3,4}\in \mathbb{Z}\,,
\end{equation}
and similarly for $\tilde{u}_{3,4}$, as follows from the quantization condition on the mass parameters \eqref{miN}.
These parameters are related to the angles introduced in the last section by
\begin{equation}
\begin{alignedat}{4}
\tilde{u}_3 &= \frac{\theta_L}{2\pi} \,, &\qquad\quad u_3 &= \frac{\theta_R}{2\pi} \,, \\[3pt]
\tilde{u}_4 &= \frac{\theta'_L}{2\pi} \,, &\qquad\quad u_4 &= \frac{\theta'_R}{2\pi} \,.
\end{alignedat}
\end{equation}
As we can see from \eqref{spinorrotation}, the orbifold action on the R-vacua depends only on the values of $s_3,s_4$. We introduce the following notation for the possible values of these spins:
\begin{equation}
\begin{aligned}
|a_1\rangle_{L/R} &= \big|s_1,s_1,\tfrac{1}{2},\tfrac{1}{2}\big\rangle_{L/R} \,, \\
|a_2\rangle_{L/R} &= \big|s_1,s_1,-\tfrac{1}{2},-\tfrac{1}{2}\big\rangle_{L/R} \,, \\
|a_3\rangle_{L/R} &= \big|s_1,-s_1,\tfrac{1}{2},-\tfrac{1}{2}\big\rangle_{L/R} \,, \\
|a_4\rangle_{L/R} &= \big|s_1,-s_1,-\tfrac{1}{2},\tfrac{1}{2}\big\rangle_{L/R} \,.
\end{aligned}
\end{equation}
Here the relative sign between $s_1$ and $s_2$ is fixed by the GSO projection. By using \eqref{spinorrotation} and \eqref{u's} we find that the orbifold action on each of these is
\begin{equation}
\begin{alignedat}{4}
|a_1\rangle_{L} \;&\rightarrow \; e^{im_1} \, |a_1\rangle_{L} \,, \qquad\qquad
&|a_1\rangle_{R} \;& \rightarrow\; e^{im_2} \, |a_1\rangle_{R} \,, \\
|a_2\rangle_{L} \;&\rightarrow \; e^{-im_1} \, |a_2\rangle_{L} \,, \qquad\qquad
&|a_2\rangle_{R} \;& \rightarrow\; e^{-im_2} \, |a_2\rangle_{R} \,, \\
|a_3\rangle_{L} \;&\rightarrow \; e^{im_3} \, |a_3\rangle_{L} \,, \qquad\qquad
&|a_3\rangle_{R} \;& \rightarrow\; e^{im_4} \, |a_3\rangle_{R} \,, \\
|a_4\rangle_{L} \;&\rightarrow \; e^{-im_3} \, |a_4\rangle_{L} \,, \qquad\qquad
&|a_4\rangle_{R} \;& \rightarrow\; e^{-im_4} \, |a_4\rangle_{R} \,.
\end{alignedat}
\label{transformation of ramond vacua}
\end{equation}

Requiring that the orbifold action on the R-vacua is of order $p$ yields the additional conditions
\begin{equation}\label{psum_u_even}
   p \sum_{i} u_i \,\in\, 2\mathbb{Z} \qquad\text{and}\qquad p \sum_{i} \tilde{u}_i \,\in\, 2\mathbb{Z} \,.
\end{equation}
This also follows easily from the earlier analysis on the quantization conditions of the mass parameters, e.g. using \eqref{u's} together with \eqref{miN}, one finds $p(u_3+u_4)=pm_2/\pi=2N_2\in 2\mathbb{Z}$ and $p(\tilde{u}_3+\tilde{u}_4)=2N_1\in2\mathbb{Z}$. Since $pu_i\in\mathbb{Z}$, one also finds that $p(u_3-u_4)\in2\mathbb{Z}$. The same also holds for $\tilde u$. An instructive example is the $\mathbb{Z}_{24}$ orbifold (cf \ref{a Z24 orbifold}) with $m_1=m_2=5\pi/12$ and $m_3=m_4=\pi/12$, which corresponds to $u_3=1/4$ and $u_4=1/6$ (see also \autoref{tab breaking all}) and  leads to $p(u_3+u_4)=10$.

Furthermore, if $\pm {u}_3\pm {u}_4=0$ mod 2 for some choice of signs, half of the right-moving supersymmetries are preserved in the orbifold. Essentially, this means that either $m_2$ or $m_4 = 0$ mod $2\pi$ and two of the four gravitini coming from the NS-R sector remain massless. On the other hand, if the above condition is not met, all right-moving supersymmetries are broken. Exactly the same argument holds for $\tilde{u},\,m_{1,3}$ and the left-moving supersymmetries.

Finally, in order for strings to close in our geometry, they need to satisfy the boundary conditions
\begin{equation}\label{boundaryconditions}
\begin{alignedat}{4}
X^\mu(\sigma^1, \sigma^2+2\pi) &= X^\mu(\sigma^1, \sigma^2) \,, \qquad
&Z(\sigma^1, \sigma^2+2\pi) &= Z(\sigma^1, \sigma^2) + 2\pi \mathcal{R} \,(w + k/p) \,, \\[4pt]
W_{L}^1(\sigma^1, \sigma^2+2\pi) &= \big(e^{i(m_1+m_3)}\big)^k \:W_{L}^1(\sigma^1, \sigma^2) \,, \qquad &W_{R}^1(\sigma^1, \sigma^2+2\pi) &= \big(e^{i(m_2+m_4)}\big)^k \:W_{R}^1(\sigma^1, \sigma^2) \,, \\[4pt]
W_{L}^2(\sigma^1, \sigma^2+2\pi) &= \big(e^{i(m_1-m_3)}\big)^k \:W_{L}^2(\sigma^1, \sigma^2) \,, \qquad &W_{R}^2(\sigma^1, \sigma^2+2\pi) &= \big(e^{i(m_2-m_4)}\big)^k \:W_{R}^2(\sigma^1, \sigma^2) \,.
\end{alignedat}
\end{equation}
Here $w \in \mathbb{Z}$ is the winding number along the $S^1$ (we omit winding modes on the torus here for simplicity of the formulae) and $k = 0,\ldots,p-1$ is an integer that distinguishes between the various sectors. We have the untwisted sector for $k=0$, and $p-1$ twisted sectors for the other values of $k$ in which case the string closes only under application of the orbifold action.

\subsection{The partition function}
\label{sec:Partition function}

Now that we have defined our orbifold action, we would like to construct the one-loop partition function. (We follow the conventions of \cite{Blumenhagen:2013fgp}; for some other references and recent examples, see e.g. \cite{aoki2004construction,nibbelink2021worldsheet}.) In general, the starting point for the partition function is
\begin{equation}
    Z(\tau,\bar{\tau})=\Tr\Big[q^{(L_0-\frac{c}{24})}\bar q^{(\bar{L}_0-\frac{\bar c}{24})}\Big]\ ,\qquad q=e^{2\pi i\tau}\ ,
\end{equation}
where $\tau=\tau_1+i\tau_2$ is the complex structure modulus of the torus. In an orbifold, we have twisted sectors and projectors in each of the sectors onto invariant states. Therefore, the trace over the Hilbert space decomposes according to
\begin{equation}
    Z(\tau,\bar \tau)=\frac{1}{p}\sum_{k,l=0}^{p-1}{Z}[k,l](\tau,\bar \tau)\ ,
\end{equation}
where, as mentioned before, $p$ is the orbifold rank and $k$ characterizes the various sectors. In addition, $l$ implements the orbifold projection in each sector\footnote{If we denote the orbifold group element by $g$, with $g^p=1$, then the projection operator takes the form $P=\frac{1}{p}(1+g+g^2+\cdots + g^{p-1})$. }. Furthermore, for our models the partition function will factorize into the following pieces (we omit writing the $\tau$ dependence for simplicity of the notation)
\begin{equation}
    {Z}[k,l]= {Z}_{\mathbb{R}^{1,4}}\,  {Z}_{S^1}[k,l]  {Z}_{T^4}[k,l]  {Z}_F[k,l]\,.
    \label{partition function first}
\end{equation}
Here ${Z}_{\mathbb{R}^{1,4}}$ is the contribution to the partition function from the non-compact bosons, ${Z}_{S^1}[k,l]$ and ${Z}_{T^4}[k,l]$ refer to the compact bosons on $S^1$ and $T^4$ respectively and ${Z}_F[k,l]$ is the fermionic contribution to the partition function.

In the remainder of this section, we construct the various parts of the partition function and discuss modular invariance of the full partition function. For symmetric orbifolds, showing modular invariance is rather easy, as the individual pieces in \eqref{partition function first} will have the same properties under the modular group (or be invariant), in such a way that the sum over $k$ and $l$ is modular invariant. For asymmetric orbifolds, one must take care of the possible phases that will arise in left and right-moving sectors under modular transformations, and show case by case that it all combines into a full modular invariant partition function $Z$.

First, we consider the bosonic piece of the partition function. The contribution from the three non-compact bosons (we work in lightcone gauge) to the partition function is
\begin{equation}
    {Z}_{\mathbb{R}^{1,4}}= \left(\sqrt{\tau_2}\,\eta\,\bar{\eta}\right)^{-3}\,.
\end{equation}
This term is invariant under both $\mathcal{T}$ and $\mathcal{S}$ modular transformations. (Modular functions and transformations are discussed in appendix \ref{ap B}.)

To compute the contribution to the partition function from the compact boson on $S^1$, recall that due to the shift along the circle coordinate, momentum states pick up a phase $e^{2\pi i n/p}$. In addition, the boundary condition of the circle coordinate \eqref{boundaryconditions} implies that in the twisted sectors fractional winding modes can appear. Combining these, we can write  
\begin{equation}
{Z}_{S^1}[k,l]=  \frac{1}{\eta\,\bar{\eta}}\sum_{n,w \in \mathbb{Z}}e^{\frac{2\pi i n}{p}l}\, q^{\frac{\alpha'}{4}P_{R}^2(k)}\, \bar{q}^{\frac{\alpha'}{4}P_{L}^2(k)}\,,
\end{equation}
where
\begin{equation}
     P_{{L}/{R}}(k)=\frac{n}{\mathcal{R}}\pm  \frac{\big(w+\frac{k}{p}\big)\mathcal{R}}{\alpha'}\,.
\end{equation}
${Z}_{S^1}[k,l]$ can be written in a manifestly modular invariant form by performing a Poisson resummation over the momentum number $n$. We find
\begin{equation}
     {Z}_{S^1}[k,l]=\frac{\mathcal{R}}{\sqrt{\alpha'}\sqrt{\tau_2}\,\eta\,\bar{\eta}}\sum_{n,w \in \mathbb{Z}}e^{-\frac{\pi \mathcal{R}^2}{\alpha'\tau_2}\left|n-\frac{l}{p}+\left(w+\frac{k}{p}\right)\tau\right|^2}\,.
\end{equation}
Note that the circle partition function consists of four building blocks: ${Z}_{S^1}[0,0]$, ${Z}_{S^1}[0,l], {Z}_{S^1}[k,0]$ and ${Z}_{S^1}[k,l]$. Of course, ${Z}_{S^1}[0,0]$ corresponds simply to a circle compactification and is invariant under both $\mathcal{T}$ and $\mathcal{S}$ modular transformations. The remaining blocks obey the following modular transformations
\begin{equation}
\begin{aligned}
   {Z}_{S^1}[k,l]\xrightarrow{\mathcal{T}}  \frac{\mathcal{R}}{\sqrt{\alpha'}\sqrt{\tau_2}\,\eta\,\bar{\eta}}&\sum_{n,w \in \mathbb{Z}}e^{-\frac{\pi \mathcal{R}^2}{\alpha'\tau_2}\left|(n+w)-\frac{l-k}{p}+\left(w+\frac{k}{p}\right)\tau\right|^2}= {Z}_{S^1}[k,l-k]\,,\\
{Z}_{S^1}[k,l]\xrightarrow{\mathcal{S}} \frac{\mathcal{R}}{\sqrt{\alpha'}\sqrt{\tau_2}\,\eta\,\bar{\eta}}&\sum_{n,w \in \mathbb{Z}}e^{-\frac{\pi \mathcal{R}^2}{\alpha'\tau_2}\left|w+\frac{k}{p}+\left(n+\frac{l}{p}\right)\tau\right|^2}={Z}_{S^1}[l,-k]\,.
\end{aligned}
\end{equation}
These transformation rules will be combined with similar ones from the $T^4$ and the fermions to ensure modular invariance of the full partition function.

Next, we discuss the contribution coming from the $T^4$. We consider left and right-movers separately, since the orbifold can act asymmetrically on the torus coordinates. For clarity of the partition function, it is convenient to parametrize the orbifold action by two twist vectors
$\tilde{u}=(0,0,\tilde{u}_3,\tilde{u}_4)$ and $u=(0,0,u_3,u_4)$, with $\Tilde{u}_i, u_i$ as given in \eqref{u's}. (For a discussion on twist vectors see e.g. \cite{ibanez2012string,font2005introduction}.) By this, we mean that
\begin{equation}\label{orbiaction}
\begin{aligned}
{W}_{L}^1 \;&\rightarrow\; e^{2\pi i\tilde{u}_3}\: {W}_{L}^1 \,, \\
{W}_{L}^2 \;&\rightarrow\; e^{{2\pi i\tilde{u}_4}}\: {W}_{L}^2 \,, \\
W_{R}^1 \;&\rightarrow\; e^{{2\pi i{u}_3}}\: W_{R}^1 \,, \\
W_{R}^2 \;&\rightarrow\; e^{2\pi i{u}_4}\: W_{R}^2 \,,
\end{aligned}
\end{equation}
and the coordinates of the non-compact dimensions are not rotated. Note that symmetric orbifolds correspond to $\tilde{u}=u$. Now, let us focus on the oscillator modes and postpone the discussion of the lattice sum over momenta and windings. The oscillator part of the $T^4$ partition function factorizes into left and right-moving pieces as
\begin{equation}
    Z_{T^4}[k,l] =  \widetilde{\mathcal{Z}}_{T^4}[\tilde{\theta}^k,\tilde{\theta}^l] \otimes \mathcal{Z}_{T^4}[\theta^k,\theta^l]\,.
\end{equation}
Here $\theta $ is the generator of the orbifold group. $\theta^k$ refers to twisted sectors where the torus coordinates obey boundary conditions of the form $W_{R}^i(\sigma^1,\sigma^2+2\pi)= \theta^k\, W_{R}^i(\sigma^1,\sigma^2)$  and  $\theta^l$ characterizes the orbifold action: $W_{R}^i \to \theta^l\,W_{R}^i$ ($\tilde{\theta}^k,\tilde{\theta}^l$ correspond to the left-movers). Similarly with the circle, ${Z}_{T^4}[0,0]$ corresponds simply to compactification on $T^4$ and is invariant under both $\mathcal{T}$ and $\mathcal{S}$ modular transformations.

First, we present the right-moving torus partition function. In the untwisted sector ($k=0$) it is given by 
\begin{equation}
   \mathcal{Z}_{T^4}[\mathbf{1},\theta^l]=q^{-\frac{2}{12}}\prod^4_{i=3} \prod_{n=1}^{\infty}(1-q^n\,e^{ 2\pi i  l u_i})^{-1} (1-q^n\,e^{-2\pi il u_i})^{-1} \,,
   \label{torus one}
\end{equation}
and can be rewritten in a more convenient form as 
\begin{equation}
   \mathcal{Z}_{T^4}[\mathbf{1},\theta^l]=\prod^4_{i=3}2 \sin(\pi lu_i)\, \frac{\eta} {\vartheta\Big[\psymbol{ \frac{1}{2}}{ -\frac{1}{2}+lu_i}\Big]}\ .
   \label{torus l}
\end{equation}
By performing successively $\mathcal{S}$ and $\mathcal{T}$ modular transformations, we find the partition function in a twisted sector labeled by $k$ ($k\neq 0$), which reads\footnote{Here we omit an irrelevant constant phase coming from the $\mathcal{T}$ transformation because it is always cancelled by left-moving contributions in both symmetric and asymmetric orbifolds.}
\begin{equation}
   \mathcal{Z}_{T^4}[\theta^k,\theta^l]=e^{-\pi i \sum\limits_{i=3}^4 \left(k u_i l u_i\right)}e^{\pi i \sum\limits_{i=3}^4 \left(k u_i-\frac{1}{2}\right)} {\chi} [\theta^k,\theta^l]\prod^4_{i=3}\,\frac{\eta}{\vartheta\Big[\psymbol{ \frac{1}{2}-ku_i}{ -\frac{1}{2}+lu_i}\Big]}\ ,
   \label{general torus partition}
\end{equation}
where
\begin{equation}
   \chi[\theta^k,\theta^l]= \prod^4_{i=3}2 \sin(\pi \text{\footnotesize{gcd}}(k,l) u_i)
   \label{fixed points}
\end{equation}
\footnote{ $\text{\footnotesize{gcd}}(k,l)$ denotes the greatest common divisor of $k,l$ with the convention $\text{\footnotesize{gcd}}(a,0)=\text{\footnotesize{gcd}}(0,a)=a$.}is the number of simultaneous \say{chiral} fixed points\footnote{The orbifolds that we consider have fixed points on the $T^4$. However, due to the shift on the circle, there are no points left invariant under the full orbifold action.} of $\theta^k$ and $\theta^l$. We note here that equation \eqref{fixed points} is valid for $ku_{3,4} \notin \mathbb{Z}$. If there exists $j\in [3,4]$ such that $ku_j \in \mathbb{Z}$\footnote{An example is the orbifold with $\tilde{u}=u=(0,0,0,\frac{1}{2})$. From \eqref{psum_u_even} it follows that $p=4$, such that $k=0,1,2,3$. For $k=2$, one then has $ku_4\in \mathbb{Z}$. The mass parameters in this case are $m_1=m_2=-m_3=-m_4=\pi/2.$}, $ \chi[\theta^k,\theta^l]$ should be divided by $2\sin(\pi l u_j)$ for $l\neq 0$, and replaced by $\prod_{i\neq j,ku_i \notin \mathbb{Z}}2 \sin(\pi k u_i)$ for $l=0$  (see \cite{katsuki1990zn} for a relevant discussion). Furthermore, under modular transformations \eqref{general torus partition} transforms as
\begin{equation}
    \begin{aligned}
     & \mathcal{Z}_{T^4}[\theta^k,\theta^l] \xrightarrow{\mathcal{T}} e^{-\frac{\pi i}{6}\lambda} \mathcal{Z}_{T^4}[\theta^k,\theta^{l-k}]\,,\\
     &\mathcal{Z}_{T^4}[\theta^k,\theta^l] \xrightarrow{\mathcal{S}}  e^{-\frac{\pi i}{2}\lambda}  \mathcal{Z}_{T^4}[\theta^l,\theta^{-k}]\,,
    \end{aligned}
    \label{bosonic modular transformation}
\end{equation}
where $\lambda$ is the number of $u_i\notin\mathbb{Z}$. The partition function for the left-movers is simply obtained by substituting $q\to \bar{q}, \eta\to \bar{\eta}, \vartheta\to \bar{\vartheta}$ and $u\to \tilde{u}$, and obeys the transformations \eqref{bosonic modular transformation} but with phases of opposite sign. Finally, notice that if the orbifold acts trivially on ${W}_{R}^{1,2}$, i.e. $u_3,u_4 \in \mathbb{Z}$, equation \eqref{torus one}, or equivalently \eqref{torus l}, simply becomes $\eta^{-4}$. 

For the zero modes 
there are  integer momenta and windings   and the partition function includes a lattice sum of left and right-moving momenta. In general, if we compactify the bosonic string on $T^4$, we obtain the Narain lattice $\Gamma_{4,4}$. On the other hand, for the orbifold compactifications we have to specify the sublattice $\Lambda \subset \Gamma_{4,4}$ that is invariant under the orbifold action. Moreover, this sublattice will contribute to the partition function in the twisted sectors with an overall multiplicative factor that is equal to its volume \cite{Narain:1986qm,narain1991asymmetric}. We will return later to this issue when we discuss explicit examples.

For the construction of the fermionic partition function we combine the non-compact and compact fermions in one expression and we consider left and right-movers separately. In the NS-sector the right-moving fermionic partition function in a sector labeled by $k$ reads 
\begin{equation}
    \mathcal{Z}_{\text{NS}}[\theta^k,\theta^l]=\frac{1}{2}e^{\pi i \sum\limits_{i=3}^4\left(k u_i l u_i\right)}\bigg[\left(\frac{\vartheta_3}{\eta}\right)^2\prod^4_{i=3}\frac{{\vartheta[\psymbol{ku_i}{ -lu_i}]}}{\eta}-e^{\pi i \sum\limits_{i=3}^4  k u_i}\left(\frac{\vartheta_4}{\eta}\right)^2\prod^4_{i=3}\frac{\vartheta[\psymbol{ku_i}{ -\frac{1}{2}-lu_i}]}{\eta}\bigg]\ .
    \label{partition NS}
\end{equation}
In the R-sector we have
\begin{equation}
    \mathcal{Z}_{\text{R}}[\theta^k,\theta^l]=\frac{1}{2}e^{\pi i \sum\limits_{i=3}^4\left(k u_i l u_i\right)}\bigg[\left(\frac{\vartheta_2}{\eta}\right)^2\prod^4_{i=3}\frac{{\vartheta[\psymbol{\frac{1}{2}+ku_i} {-lu_i}]}}{\eta}+e^{\pi i \sum\limits_{i=3}^4 k u_i}\left(\frac{\vartheta_1}{\eta}\right)^2\prod^4_{i=3}\frac{{\vartheta[\psymbol{\frac{1}{2}+ku_i} {-\frac{1}{2}-lu_i}]}}{\eta}\bigg]\ .
    \label{partition R}
\end{equation}
By combining the above, we find the right-moving fermionic partition function in a sector labeled by $k$, which reads
\begin{equation}
    \mathcal{Z}_{{F}}[\theta^k,\theta^l]=\mathcal{Z}_{\text{NS}}[\theta^k,\theta^l]-\mathcal{Z}_{\text{R}}[\theta^k,\theta^l]\,,
\end{equation}
and transforms under modular transformations as
\begin{equation}
    \begin{aligned}
      & \mathcal{Z}_F[\theta^k,\theta^l] \xrightarrow{\mathcal{T}} e^{\frac{4\pi i}{6}} \mathcal{Z}_F[\theta^k,\theta^{l-k}]\,,\\
       & \mathcal{Z}_F[\theta^k,\theta^l] \xrightarrow{\mathcal{S}}   \mathcal{Z}_F[\theta^l,\theta^{-k}]\,.
    \end{aligned}
    \label{fermionic modular transformation}
\end{equation}
For later convenience, we rewrite expressions \eqref{partition NS} and \eqref{partition R} in terms of infinite sums as\footnote{In the literature, this is usually referred to as \say{bosonization}. For an alternative construction of the partition function see \cite{condeescu2012asymmetric,Condeescu:2013yma}.} 
\begin{equation}
   \mathcal{Z}_{\text{NS,R}}[\theta^k,\theta^l]= \frac{1}{\eta^4}\, e^{\pi i \sum\limits_{i=3}^4\left(k u_i l u_i\right)}\sum_{{r}} q^{\frac{1}{2}({r}+k {u})^2}\,e^{-2\pi i l[({r}+k {u})\cdot{u}]}\,.
   \label{fermionic infinite sums}
\end{equation}
Here $r=(r_1,r_2,r_3,r_4)$ is an SO(8) weight vector with each component ${r}_i\in \mathbb{Z}$ in the NS-sector and ${r}\in \mathbb{Z}+\tfrac{1}{2}$ in the R-sector. The GSO projection is $\sum_{i=1}^4 r_i \in 2\mathbb{Z}+1$ in the NS-sector and $\sum_{i=1}^4 r_i \in 2\mathbb{Z}$ in the R-sector. Finally, the left-moving fermionic partition function is obtained by substituting $q\to \bar{q}, \eta\to \bar{\eta}, \vartheta\to \bar{\vartheta}, u\to \tilde{u}$  and $r \to \tilde{r}$, where $\tilde{r}=(\tilde{r}_1,\tilde{r}_2,\tilde{r}_3,\tilde{r}_4)$, and transforms as in \eqref{fermionic modular transformation} but with a phase of opposite sign. 

As a last comment here, we observe from \eqref{bosonic modular transformation} and \eqref{fermionic modular transformation} that if we consider only the right-movers, we do not obtain a modular invariant partition function. Of course, modular invariance can be achieved by taking also into account the contribution from the left-movers. This can be easily verified in the case of symmetric orbifolds because the expression for the left-moving partition function is essentially the complex conjugate of the right-moving one. Consequently, the constant phases cancel out, as the partition function is the tensor product of left and right-movers. However, this argument does not hold for asymmetric orbifolds. Therefore, one shall carefully examine modular invariance for each asymmetric orbifold construction. In the next, we will address this issue by discussing specific examples.

\section{Closed string spectrum}
\label{closed string spectrum}
In this section we will first present the general formalism that we use in order to obtain the closed string spectrum that arises from our orbifold constructions. Afterwards, we will give explicit examples of symmetric and asymmetric orbifolds, preserving $\mathcal{N}=6,4,2,$ or 0 supersymmetry. In general, we treat the untwisted and twisted sectors separately. In other words, we fix $k$ and then we sum over $l$ and divide by the orbifold rank $p$ in order to implement the orbifold projection. First, we consider  the untwisted sector, i.e. the sector with $k=0$ boundary conditions. Furthermore, we  focus on the lowest excited states, i.e. the states that are massless without the addition of momentum and/or winding modes. Consequently, we expand the $\vartheta$-functions coming from the bosonic contributions as well as all the $\eta$-functions and we keep only the lowest order terms. We expand then the partition function in the untwisted sector as (omitting for now overall factors of ${\tau_2}$ which are reinstated later)
\begin{equation}
    {Z}[0,l]= (q\bar{q})^{-\frac{1}{2}}\sum_{n,w \in \mathbb{Z}}e^{\frac{2\pi i n}{p}l}\, q^{\frac{\alpha'}{4}P_{R}^2(0)}\, (\bar{q})^{\frac{\alpha'}{4}P_{L}^2(0)}\sum_{{r},\tilde{{r}}} q^{\frac{1}{2}{r}^2}\,(\bar{q})^{\frac{1}{2}\tilde{r}^2}e^{2\pi il (\tilde{r}\cdot \tilde{u}-r\cdot u)}\, \left(1+\cdots\right)\,,
    \label{generic partition untwisted}
\end{equation}
where the dots denote contributions from higher excited oscillator states. We present in \autoref{tablemasslessstates} the NS and R-sector weight vectors for the states of the lowest level that survive the GSO projection (all of these are massless in the absence of momentum and/or winding modes). Furthermore, we table their representations under both the massless little group SO$(3)\approx\text{SU}(2)$ and the massive little group SO$(4)\approx\text{SU}(2)\times\text{SU}(2)$ in five dimensions. The latter is important when adding momenta or windings such that the states become massive.

\renewcommand{\arraystretch}{2}
\begin{table}[h!]
\centering
 \begin{tabular}{|c|c|c|c|}
    \hline
    Sector &  $\tilde{r}, {r}$  & SO(3) rep & SO(4) rep\\
    \hline
    \hline
   \multirow{3}{*}{NS}  & $(\underline{\pm 1,0},0,0)$ & $\textbf{3}\oplus \textbf{1}$ & $(\textbf{2},\textbf{2})$\\
    \cline{2-4}
    &$(0,0,\pm 1,0)$& 2$\,\times\,\textbf{1}$ & 2$\,\times\,(\textbf{1},\textbf{1})$\\
    \cline{2-4}
   & $(0,0,0,\pm 1)$ & 2$\,\times\,\textbf{1}$ & 2$\,\times\,(\textbf{1},\textbf{1})$\\
    \hline
    \hline
    \multirow{4}{*}{R}  & $(\pm\frac{1}{2},\pm\frac{1}{2},\frac{1}{2},\frac{1}{2})$ & $\textbf{2}$ & $(\textbf{2},\textbf{1})$\\
    \cline{2-4}
    & $(\pm\frac{1}{2},\pm\frac{1}{2},-\frac{1}{2},-\frac{1}{2})$ & $\textbf{2}$ & $(\textbf{2},\textbf{1})$\\
    \cline{2-4}
    & $(\underline{\frac{1}{2},-\frac{1}{2}},\frac{1}{2},-\frac{1}{2})$ & $\textbf{2}$ & $(\textbf{1},\textbf{2})$\\
    \cline{2-4}
    & $(\underline{\frac{1}{2},-\frac{1}{2}},-\frac{1}{2},\frac{1}{2})$ & $\textbf{2}$ & $(\textbf{1},\textbf{2})$\\
   \hline
    \end{tabular}
\captionsetup{width=.9\linewidth}
\caption{\textit{Here we write down all the weight vectors for states that are massless in the absence of momentum and/or winding modes, including their representations under the massless and massive little groups in 5D. We write down both left-moving and right-moving weight vectors, where underlining denotes permutations.}}
\label{tablemasslessstates}
\end{table}
\renewcommand{\arraystretch}{1}

We construct string states by tensoring the left and right-moving weight vectors from \autoref{tablemasslessstates}. In general, a state carries a non-trivial orbifold charge, given by the phase $e^{2\pi i l (\tilde{r}\cdot \tilde{u}-r\cdot u)}$, and its degeneracy is 
\begin{equation}
    D(k=0)=\frac{1}{p}\sum_{l=0}^{p-1}e^{2\pi il [(\tilde{r}\cdot \tilde{u}-r\cdot u)+\frac{n}{p}]}\,.
    \label{degeneracy untwisted}
\end{equation}
A charged state will be projected out of the orbifold spectrum when we perform the summation over $l$. However, we can fix this issue by adding appropriate momentum modes on the circle  to the state, such that the orbifold charge is cancelled. This also means that the state will become massive, since momentum modes contribute to the mass of a state; we will discuss this in detail later. Finally, if $(\tilde{r}\cdot \tilde{u}-r\cdot u)\in \mathbb{Z}$, the orbifold charge is trivial. States with trivial charge survive the orbifold projection and remain massless. As follows from \eqref{degeneracy untwisted}, the degeneracy of orbifold invariant states in the untwisted sector is 1.

For the construction of states, we use the rules
\begin{equation}
\begin{aligned}
\textbf{3}\otimes \textbf{3} = \textbf{5} \oplus \textbf{3} \oplus \textbf{1} \,, \qquad\quad \textbf{2}\otimes \textbf{2} = \textbf{3} \oplus \textbf{1} \,, \qquad\quad \textbf{3}\otimes \textbf{2} = \textbf{4} \oplus \textbf{2} \,,
\end{aligned}
\end{equation}
for tensoring SU$(2)$ representations.  In addition, we table the (massless and massive) representations that correspond to various supergravity fields in five dimensions in \autoref{table5Dfieldreps}.

\renewcommand{\arraystretch}{1.2}
\begin{table}[h!]
\centering
\begin{tabular}{|c|c|}
\hline
\;Massless field\; & \;SO$(3)$ rep\; \\ \hline\hline
$g_{\mu\nu}$ & \textbf{5} \\
$\psi_\mu$ & \textbf{4} \\
$A_\mu$ & \textbf{3} \\
$\chi$ & \textbf{2} \\
$\phi$ & \textbf{1} \\ \hline
\end{tabular}
\hspace{1.5cm}
\begin{tabular}{|c|c|}
\hline
\;\,Massive field\,\; & SO$(4)$ rep \\ \hline\hline
$B_{\mu\nu}^+$ / $B_{\mu\nu}^-$ & $(\textbf{3},\textbf{1})$ / $(\textbf{1},\textbf{3})$ \\
$\psi_\mu^+$ / $\psi_\mu^-$ & \;\:$(\textbf{2},\textbf{3})$ / $(\textbf{3},\textbf{2})$\:\; \\
$A_\mu$ & $(\textbf{2},\textbf{2})$ \\
$\chi^+$ / $\chi^-$ & $(\textbf{1},\textbf{2})$ / $(\textbf{2},\textbf{1})$ \\
$\phi$ & $(\textbf{1},\textbf{1})$ \\ \hline
\end{tabular}
\captionsetup{width=.83\linewidth}
\caption{\textit{Here we show the various massless and massive 5D fields and their representations under the appropriate little group.}}
\label{table5Dfieldreps}
\end{table}
\renewcommand{\arraystretch}{1}
Regarding the construction of states in twisted sectors, the procedure is similar with the untwisted sector. The expansion of the partition function in a twisted sector labeled by $k$ yields\footnote{Here we consider general twist vectors of the form $\tilde{u}=(0,0,\tilde{u}_3,\tilde{u}_4)$, $u=(0,0,u_3,u_4)$. As in the untwisted sector, we omit factors of $\tau_2$.}
\begin{equation}
    \begin{aligned}
    {Z}[k,l]=&\left|{\chi} [\theta^k,\theta^l]\,\tilde{\chi}[\tilde{\theta}^k,\tilde{\theta}^l]\right| (q\bar{q})^{-\frac{1}{2}}\,e^{i(\tilde{\varphi}-\varphi)}q^{E_k}\,(\bar{q})^{\tilde{E}_k}\sum_{n,w \in \mathbb{Z}}e^{ \frac{2\pi i n}{p}l}\, q^{\frac{\alpha'}{4}P_{R}^2(k)}\, (\bar{q})^{\frac{\alpha'}{4}P_{L}^2(k)}\,\times\\
    & \sum_{r,\tilde{r}} q^{\frac{1}{2}({r}+k u)^2}\, (\bar{q})^{\frac{1}{2}(\tilde{r}+k \tilde{u})^2}\,e^{2\pi il (\tilde{r}\cdot \tilde{u}-r\cdot u)}\, e^{2\pi il k (\tilde{u}^2-u^2)}\, \left(1+\cdots\right)\,,
    \end{aligned}
    \label{generic partition twisted}
\end{equation}
where
\begin{equation}
    \varphi= 2\pi  \sum_{u_i\notin \mathbb{Z}}\left(\frac{1}{2}-k |u_i|\right)l |u_i|
\end{equation}
is a phase arising from bosonic contributions and
\begin{equation}
    E_k = \sum_{u_i\notin \mathbb{Z}}\frac{1}{2}k|u_i| (1-k|u_i|)\,
    \label{shifted energy}
\end{equation}
is a shift to the zero point energy\footnote{The absolute values in \eqref{generic partition twisted}-\eqref{shifted energy} appear due to restriction on the allowed values of the characteristics of the $\vartheta$-functions in the product representation \eqref{product representation}. Recall that in general we consider $k,l\geq0$.}. The expressions for $\tilde{\chi}[\tilde{\theta}^k,\tilde{\theta}^l],\tilde{\varphi}, \tilde{E}_k$ are simply obtained by substituting $u\to\tilde{u}$. Finally, the degeneracy of a state in a $k$ twisted sector is given by (see also \cite{ibanez1988heterotic,font1990construction} for a relevant discussion)
\begin{equation}
   D(k)= \frac{1}{p}\sum_{l=0}^{p-1}\left|{\chi} [\theta^k,\theta^l]\,\tilde{\chi}[\tilde{\theta}^k,\tilde{\theta}^l]\right| e^{2\pi il [(\tilde{r}\cdot \tilde{u}-r\cdot u)+k(\tilde{u}^2-u^2)+i(\tilde{\varphi}-\varphi)+\frac{n}{p}]}\,.
   \label{degeneracy twisted}
\end{equation}
However, if the orbifold acts trivially on ${W}_{{L/R}}^{1}$ and/or ${W}_{{L/R}}^{2}$, this degeneracy should be modified because integer momentum and winding numbers can appear. In particular, one should also divide \eqref{degeneracy twisted} by the volume of the invariant momentum sublattice.

Now, we can proceed with the construction of explicit orbifold models. In particular, we treat supersymmetric models: two asymmetric $\mathbb{Z}_2$ orbifolds, an $\mathcal{N}=6$ and an $\mathcal{N}=2$, a symmetric $\mathbb{Z}_4$, $\mathcal{N}=4\,(0,2)$ and an asymmetric $\mathbb{Z}_2$, $\mathcal{N}=4\,(1,1)$ orbifold, as well as a non-supersymmetric, symmetric $\mathbb{Z}_3$ orbifold. We discuss the untwisted orbifold spectra and verify that they match exactly with the Scherk-Schwarz supergravity spectra obtained in \cite{Hull:2020byc}. In addition, we construct purely stringy states arising from the orbifold twisted sectors. Regarding the non-supersymmetric orbifolds, we focus on the twisted sectors where tachyons can appear and we find a critical value for the orbifold circle radius above which the spectrum is tachyon-free.

\subsection{$\mathcal{N}=6$}
\label{sec:N=6}
In this section we discuss models with $\mathcal{N}=6$ supersymmetry in five dimensions. As we have seen in the example of section \ref{n=6 allows only p=2 or 3}, these are asymmetric orbifolds of rank $p=2$ or 3. Such models are chiral orbifolds with the twist acting only on the left or the right-movers on the torus. Without loss of generality we consider twist vectors of the form $u=(0,0,0,0)$ and $\tilde{u}=(0,0,\tilde{u}_3,\tilde{u}_4)$, with $\pm \tilde{u}_3\pm \tilde{u}_4=0$ mod 2 for some choice of signs, which ensures that we break half of the left-moving supersymmetries. This choice of twist vectors also implies that the left-moving momenta are projected out, while there exists an invariant sublattice of right-moving momenta $\Lambda_R \subset \Gamma_{4,4}$ contributing to the partition function ${Z}_{T^4}[k,l]$ (for $k$ or $l\neq 0$). 
Following the examples of \cite{dabholkar1999string,bianchi2022perturbative}, we take the right-moving momenta to lie in the lattice $D_4$ for $p=2$ or $A_2 \oplus A_2$ for $p=3$. Once we have the twist vectors and the invariant sublattice, we can calculate the degeneracy of the ground state of the twisted sectors, which is given by the number of fixed points \eqref{fixed points} divided by the volume of the invariant sublattice \cite{Narain:1986qm,narain1991asymmetric}. For a consistent physical theory, the degeneracies of states should be integers. Orbifolds whose partition functions give non-integer degeneracies should be excluded. As shown in \cite{bianchi2022perturbative}, this only allows the cases $\mathbb{Z}_2$ and $\mathbb{Z}_3$, which precisely agrees with the T-duality approach that we use to construct our orbifold models, as we saw in the example of \ref{n=6 allows only p=2 or 3}.

Now, let us present an example of a $\mathbb{Z}_2$ orbifold, breaking 8 left-moving supersymmetries, with twist vectors $\tilde{u}=\left(0,0,\tfrac{1}{2},\tfrac{1}{2}\right)$ and $u=(0,0,0,0)$, or in terms of the mass parameters $\vec{m}=(\pi,0,0,0)$\footnote{Here we employ the notation $\vec{m}=(m_1,m_2,m_3,m_4)$.}. This orbifold acts trivially on the right-movers, while left-movers obtain a non-zero phase under the orbifold action. The right-moving momenta lie in the $D_4$ lattice. The associated lattice sum is\footnote{A thorough discussion on lattices and theta functions can be found in \cite{conway2013sphere}.} 
\begin{equation}
    \Theta_{D_4}(\tau)= \sum_{P\in D_4}q^{\frac{1}{2}P^2}= \frac{1}{2}\left(\vartheta_3(\tau)^4+\vartheta_4(\tau)^4\right)\,.
\end{equation}
The remaining parts of the partition function can be obtained as described in section \ref{sec:Partition function}. We find
\begin{equation}
    \begin{aligned}
       Z[0,1]=Z_{\mathbb{R}^{1,4}}{Z}_{S^1}[0,1]& \left(\frac{\bar{\eta}}{\bar{\vartheta_2}}\right)^2\frac{1}{\bar{\eta}^4} [(\bar{\vartheta_3}\bar{\vartheta_4})^2-(\bar{\vartheta_4}\bar{\vartheta_3})^2-(\bar{\vartheta_2}\bar{\vartheta_1})^2-(\bar{\vartheta_1}\bar{\vartheta_2)^2}] \,\times\\
      &  \frac{1}{{\eta}^4}\Theta_{D_4}(\tau)\frac{1}{{\eta}^4}
      [(\vartheta_3)^4-(\vartheta_4)^4-(\vartheta_2)^4-(\vartheta_1)^4]\,.
    \end{aligned}
\end{equation}

\begin{equation}
    \begin{aligned}
     Z[1,0]=Z_{\mathbb{R}^{1,4}}{Z}_{S^1}[1,0]& \left(\frac{\bar{\eta}}{\bar{\vartheta_4}}\right)^2\frac{1}{\bar{\eta}^4}  [(\bar{\vartheta_3}\bar{\vartheta_2})^2+(\bar{\vartheta_4}\bar{\vartheta_1})^2-(\bar{\vartheta_2}\bar{\vartheta_3})^2+(\bar{\vartheta_1}\bar{\vartheta_4)^2}] \,\times\\
    & \frac{1}{2{\eta}^4}\Theta_{D_4^*}(\tau)\frac{1}{{\eta}^4}
      [(\vartheta_3)^4-(\vartheta_4)^4-(\vartheta_2)^4-(\vartheta_1)^4]\,.
    \end{aligned}
\end{equation}

\begin{equation}
    \begin{aligned}
     Z[1,1]=Z_{\mathbb{R}^{1,4}}{Z}_{S^1}[1,1]& \left(\frac{\bar{\eta}}{\bar{\vartheta_3}}\right)^2\frac{1}{\bar{\eta}^4}  [(\bar{\vartheta_3}\bar{\vartheta_1})^2+(\bar{\vartheta_4}\bar{\vartheta_2})^2-(\bar{\vartheta_2}\bar{\vartheta_4})^2+(\bar{\vartheta_1}\bar{\vartheta_3)^2}] \,\times\\
    & \frac{1}{2{\eta}^4}\Theta_{D_4^*}(\tau+1)\frac{1}{{\eta}^4}
      [(\vartheta_3)^4-(\vartheta_4)^4-(\vartheta_2)^4-(\vartheta_1)^4]\,.
    \end{aligned}
\end{equation}
The above pieces of the partition function satisfy the following modular transformations
\begin{equation}
     {{\mathcal{T}}} \lcirclearrowright Z[0,1]      \xleftrightarrow{\mathcal{S}} Z[1,0]  \xleftrightarrow{\mathcal{T}} Z[1,1] \rcirclearrowleft  {\mathcal{S}}\,,
     \label{explicit modular orbits}
\end{equation}
which ensure modular invariance. It is worth mentioning here that the above $\mathbb{Z}_2$ chiral twist is a symmetry of the $(A_1)^4$ lattice as well\footnote{$(A_1)^4$ is a shorthand notation for $A_1 \oplus A_1 \oplus A_1 \oplus A_1$. }. However, choosing the $(A_1)^4$ instead of the $D_4$ lattice does not lead to a modular invariant partition function. In particular, for the model based on the $(A_1)^4$ lattice, performing a modular $\mathcal{S}$ transformation on $Z[1,1]$ does not give back $Z[1,1]$, due to a sign difference in the lattice sum. Furthermore, under two consecutive $\mathcal{T}$ transformations neither $Z[1,0]$ nor $Z[1,1]$ get back to themselves, which in turn means that level-matching is not satisfied \cite{vafa1986modular}. The fact that $(A_1)^4$ fails level-matching was also discussed in \cite{blumenhagen1999orientifolds}. In addition, the model constructed in \cite{sen1995dual} based on the $(A_1)^4$ lattice suffers from this problem, as  was pointed out in \cite{dabholkar1999string}.

We now move on to the construction of the closed string spectrum. First, we work out the massless spectrum in the untwisted sector. Recall that for doing so we do not add momentum or winding modes on the circle. We use \eqref{generic partition untwisted} with $n=w=0$, $p=2$, $\tilde{u}=\left(0,0,\tfrac{1}{2},\tfrac{1}{2}\right)$ and $u=(0,0,0,0)$. Thus, we obtain
\begin{equation}
    {Z}[0,l]= (q\bar{q})^{-\frac{1}{2}}\sum_{{r},\tilde{{r}}} q^{\frac{1}{2}{r}^2}\,(\bar{q})^{\frac{1}{2}\tilde{r}^2}e^{\pi il (\tilde{r}_3+\tilde{r}_4)}\, \left(1+\cdots\right)\,.
    \label{N=6 massless untwisted}
\end{equation}
We observe that orbifold invariant states have to satisfy $\tilde{r}_3+\tilde{r}_4=0$ mod 2. By examining \autoref{tablemasslessstates} we find the following invariant states

NS-NS sector:
\begin{equation}
    \begin{aligned}
     (\underline{\pm 1,0},0,0) &\otimes (\underline{\pm 1,0},0,0)=\textbf{5}\oplus3\times\textbf{3}\oplus2\times\textbf{1}\\
     (\underline{\pm 1,0},0,0) &\otimes (0,0,\underline{\pm 1,0})=4\times\textbf{3}\oplus4\times\textbf{1}
    \end{aligned}
\end{equation}
NS-R sector:
\begin{equation}
    \begin{aligned}
        (\underline{\pm 1,0},0,0) &\otimes \pm (\pm\tfrac{1}{2},\pm\tfrac{1}{2},\tfrac{1}{2},\tfrac{1}{2})= 2\times \textbf{4}\oplus 4\times \textbf{2}  \\
        (\underline{\pm 1,0},0,0) &\otimes \pm(\underline{\tfrac{1}{2},-\tfrac{1}{2}},\tfrac{1}{2},-\tfrac{1}{2})=2\times \textbf{4}\oplus 4\times \textbf{2}
       \end{aligned}
\end{equation}
R-NS sector:
\begin{equation}
    \begin{aligned}
      \pm (\underline{\tfrac{1}{2},-\tfrac{1}{2}},\tfrac{1}{2},-\tfrac{1}{2})&\otimes (\underline{\pm 1,0},0,0) = 2\times \textbf{4}\oplus 4\times \textbf{2}\\
      \pm (\underline{\tfrac{1}{2},-\tfrac{1}{2}},\tfrac{1}{2},-\tfrac{1}{2}) &\otimes (0,0,\underline{\pm 1,0})=8\times \textbf{2}
    \end{aligned}
\end{equation}
R-R sector
\begin{equation}
    \begin{aligned}
      \pm (\underline{\tfrac{1}{2},-\tfrac{1}{2}},\tfrac{1}{2},-\tfrac{1}{2})&\otimes \pm (\pm\tfrac{1}{2},\pm\tfrac{1}{2},\tfrac{1}{2},\tfrac{1}{2})=4\times \textbf{3}\oplus 4 \times \textbf{1}\\
     \pm  (\underline{\tfrac{1}{2},-\tfrac{1}{2}},\tfrac{1}{2},-\tfrac{1}{2})&\otimes \pm (\underline{\tfrac{1}{2},-\tfrac{1}{2}},\tfrac{1}{2},-\tfrac{1}{2})=4 \times \textbf{3}\oplus 4\times \textbf{1}
     \end{aligned}
\end{equation}
All together, we find the graviton, 15 vectors, 14 scalars, 6 gravitini and 20 dilatini. These massless fields fit into the $\mathcal{N}=6$ gravity multiplet. Regarding the notation, underlining denotes permutations, e.g. $(0,0,\underline{\pm 1,0})$ corresponds to $(0,0,1,0), (0,0,0,1), (0,0,-1,0)$ and $(0,0,0,-1)$. Now, we move on to the massive spectrum. For the construction of massive states, we take combinations from table \autoref{tablemasslessstates} that obtain a non-zero phase under the orbifold action and we cancel this phase by adding momentum modes on the circle. Whenever we add momentum and/or winding to a state, we denote it only on the left-movers by $(\tilde{r}_1,\tilde{r}_2,\tilde{r}_3,\tilde{r}_4;n,w)$. Once again, we use \eqref{generic partition untwisted} with the same values as in \eqref{N=6 massless untwisted}, with the exception that we allow $n \neq 0$. We obtain
\begin{equation}
     {Z}[0,l]=(q\,\bar{q})^{-\frac{1}{2}}\sum_{n \in \mathbb{Z}}e^{ {\pi i n}l}\, (q\,\bar{q})^{\frac{\alpha'n^2}{4\mathcal{R}^2}}\,\sum_{{r},\tilde{r}}  q^{\frac{1}{2}{r}^2} (\bar{q})^{\frac{1}{2}\tilde{r}^2} e^{\pi il (\tilde{r}_3+\tilde{r}_4)}\, \left(1+\cdots\right)\,.
\end{equation}
We find the following massive states

NS-NS sector:
\begin{equation}
    \begin{aligned}
     \pm(0,0,\underline{1,0};-1)&\otimes  (\underline{\pm 1,0},0,0)= 4\times(\textbf{2},\textbf{2})\\
    \pm (0,0,\underline{1,0};-1)&\otimes (0,0,\underline{\pm1,0})= 16\times(\textbf{1},\textbf{1})
    \end{aligned}
\end{equation}
NS-R sector:
\begin{equation}
    \begin{aligned}
     \pm (0,0,\underline{1,0};-1)&\otimes\pm(\pm\tfrac{1}{2},\pm\tfrac{1}{2},\tfrac{1}{2},\tfrac{1}{2})=8\times (\textbf{2},\textbf{1})\\
      \pm  (0,0,\underline{1,0};-1)&\otimes\pm(\underline{\tfrac{1}{2},-\tfrac{1}{2}},\tfrac{1}{2},-\tfrac{1}{2})=8\times (\textbf{1},\textbf{2})
    \end{aligned}
\end{equation}
R-NS sector:
\begin{equation}
    \begin{aligned}
    \pm   (\pm\tfrac{1}{2},\pm\tfrac{1}{2},\tfrac{1}{2},\tfrac{1}{2};-1)&\otimes(\underline{\pm 1,0},0,0)=2 \times (\textbf{3},\textbf{2})\oplus 2 \times (\textbf{1},\textbf{2})\\
      \pm (\pm\tfrac{1}{2},\pm\tfrac{1}{2},\tfrac{1}{2},\tfrac{1}{2};-1)&\otimes(0,0,\underline{\pm1,0})=8\times (\textbf{2},\textbf{1})
    \end{aligned}
\end{equation}
R-R sector:
\begin{equation}
    \begin{aligned}
     \pm (\pm\tfrac{1}{2},\pm\tfrac{1}{2},\tfrac{1}{2},\tfrac{1}{2};-1) &\otimes\pm(\pm\tfrac{1}{2},\pm\tfrac{1}{2},\tfrac{1}{2},\tfrac{1}{2})=4\times(\textbf{3},\textbf{1})\oplus 4 \times (\textbf{1},\textbf{1})\\
      \pm  (\pm\tfrac{1}{2},\pm\tfrac{1}{2},\tfrac{1}{2},\tfrac{1}{2};-1)&\otimes\pm (\underline{\tfrac{1}{2},-\tfrac{1}{2}},\tfrac{1}{2},-\tfrac{1}{2})=4\times(\textbf{2},\textbf{2})
    \end{aligned}
\end{equation}
In total, we find 2 gravitini (\textbf{3},\textbf{2}), 4 tensors (\textbf{3},\textbf{1}), 8 vectors (\textbf{2},\textbf{2}), 26 dilatini, $16\times(\textbf{2},\textbf{1})$ and $10\times(\textbf{1},\textbf{2})$, and 20 scalars (\textbf{1},\textbf{1}). All these fields have mass
\begin{equation}
    \frac{\alpha' m^2_{L}}{2}=\frac{\alpha' m^2_{R}}{2}=\frac{\alpha'}{4\mathcal{R}^2} \quad\Rightarrow \quad m= \left|\frac{1}{\mathcal{R}}\right|\,,
\end{equation}
due to the contribution of the $n=\pm1$ momentum mode on the circle, and fit into a complex (1,2) BPS supermultiplet with the representations
\begin{equation}
(3,2)\oplus2\times(3,1)\oplus4\times(2,2)\oplus5\times(1,2)\oplus8\times(2,1)\oplus10\times(1,1)\,.
\label{bps reps n=6}
\end{equation}
All massive and massless states are constructed such that the combination between left and right-movers ensures zero phase. However, we can add to all these states
a trivial phase $e^{(\pi i l)2\mathbb{Z}}$ simply by adding $2\mathbb{Z}$ momentum modes along the circle. In this way we can construct Kaluza-Klein towers on the circle, where the contribution from each (even) momentum mode to the mass of the state is $\left|{2\mathbb{Z}}/{\mathcal{R}}\right|$. Furthermore, we can generally identify the orbifold radius $\mathcal{R}$, with the Scherk-Schwarz radius $R$, by $\mathcal{R}=p{R}$, where $p$ is the orbifold rank. In our $\mathbbm{Z}_2$ example, this means that we can write the mass of the BPS supermultiplet and the contribution from the Kaluza-Klein towers as $|{1}/{2R}|$ and $|{\mathbb{Z}}/{R}|$ respectively. This entire spectrum arising from our orbifold construction in the untwisted sector matches exactly with the one found in \cite{Hull:2020byc} from the Scherk-Schwarz reduction on the level of supergravity.

Besides the untwisted spectrum, we are also interested in finding the lightest states in the twisted sectors (in a $\mathbb{Z}_2$ orbifold there is only one such sector for $k=1$). In order to construct the twisted spectrum we use \eqref{generic partition twisted}-\eqref{shifted energy} (with $p=2$, $k=1$, $\tilde{u}=\left(0,0,\tfrac{1}{2},\tfrac{1}{2}\right)$ and $u=(0,0,0,0)$). We obtain
\begin{equation}
     Z[1,l]=2q^{-\frac{1}{2}}(\bar{q})^{-\frac{1}{4}}\,\sum_{{n,w\in \mathbb{Z}}}e^{{\pi i n}l} q^{\frac{\alpha'}{4}P_{R}^2(1)} (\bar{q})^{\frac{\alpha'}{4}P_{L}^2(1)}\,\sum_{{r},\tilde{r}}  q^{\frac{1}{2}{r}^2}(\bar{q})^{\frac{1}{2}(\tilde{r}+\tilde{u})^2} e^{{\pi}il(\tilde{r}_3+\tilde{r}_4+1)}(1+\cdots)\,.
\end{equation}
Note that all states come with a multiplicative factor of 2, which is precisely the number of chiral fixed points\footnote{Note that in the $k=1$ sector the expression $\prod_{i}2\sin(\pi \text{\footnotesize{gcd}}(1,l)\tilde{u}_i)$ simply becomes $\prod_{i}2\sin(\pi \tilde{u}_i)$.} $4 \sin^2({\pi}/{2})=4$, divided by the volume of the invariant sublattice ${\text{Vol}(D_4)}=2$. The weight vectors for the lightest right-moving states are, again, given in \autoref{tablemasslessstates} because the orbifold acts trivially on the right-movers. On the contrary, the weight vectors for the lightest left-moving states are listed in \autoref{k=1 N=6 states}. Since the twisted states are in general massive, we only write down the representation of these states under the massive little group in five dimensions.
\renewcommand{\arraystretch}{2}
\begin{table}[h!]
\centering
 \begin{tabular}{|c|c|c|}
    \hline
    Sector &  $\tilde{r}$ & SO(4) rep\\
    \hline
    \hline
  NS & $(0,0,\underline{-1,0})$ & 2$\,\times\,(\textbf{1},\textbf{1})$\\
  \hline
   R  & $(\pm\frac{1}{2},\pm\frac{1}{2},-\frac{1}{2},-\frac{1}{2})$ & $(\textbf{2},\textbf{1})$\\
   \hline
    \end{tabular}
\captionsetup{width=.9\linewidth}
\caption{\textit{Here we list the weight vectors of the lightest left-moving states in the $k=1$ twisted sector of the $\mathbb{Z}_2$, $\mathcal{N}=6$ orbifold and their representations under the massive little group in 5D.}}
\label{k=1 N=6 states}
\end{table}
\renewcommand{\arraystretch}{1}

In the twisted sector, orbifold invariant states have to satisfy $\tilde{r}_3+\tilde{r}_4+1=0$ mod 2. We list below the states that we find in each sector

NS-NS sector:
\begin{equation}
\begin{aligned}
 (0,0,\underline{-1,0}) &\otimes (\underline{\pm 1,0},0,0)= 2 \times (\textbf{2},\textbf{2})\\
     (0,0,\underline{-1,0}) &\otimes  (0,0,\underline{\pm1,0})= 8\times (\textbf{1},\textbf{1})
\end{aligned}
\end{equation}
NS-R sector:
\begin{equation}
    \begin{aligned}
       (0,0,\underline{-1,0})&\otimes\pm (\pm\tfrac{1}{2},\pm\tfrac{1}{2},\tfrac{1}{2},\tfrac{1}{2})=4\times (\textbf{2},\textbf{1})\\
         (0,0,\underline{-1,0})&\otimes\pm(\underline{\tfrac{1}{2},-\tfrac{1}{2}},\tfrac{1}{2},-\tfrac{1}{2})=4\times (\textbf{1},\textbf{2})
    \end{aligned}
\end{equation}
R-NS sector:
\begin{equation}
    \begin{aligned}
        (\pm\tfrac{1}{2},\pm\tfrac{1}{2},-\tfrac{1}{2},-\tfrac{1}{2})&\otimes(\underline{\pm 1,0},0,0)=(\textbf{3},\textbf{2})\oplus (\textbf{1},\textbf{2})\\
       (\pm\tfrac{1}{2},\pm\tfrac{1}{2},-\tfrac{1}{2},-\tfrac{1}{2})&\otimes(0,0,\underline{\pm1,0})=4\times (\textbf{2},\textbf{1})
    \end{aligned}
\end{equation}
R-R sector:
\begin{equation}
    \begin{aligned}
            (\pm\tfrac{1}{2},\pm\tfrac{1}{2},-\tfrac{1}{2},-\tfrac{1}{2}) &\otimes\pm (\pm\tfrac{1}{2},\pm\tfrac{1}{2},\tfrac{1}{2},\tfrac{1}{2})=2\times (\textbf{3},\textbf{1})\oplus 2 \times (\textbf{1},\textbf{1})\\
        (\pm\tfrac{1}{2},\pm\tfrac{1}{2},-\tfrac{1}{2},-\tfrac{1}{2})&\otimes \pm (\underline{\tfrac{1}{2},-\tfrac{1}{2}},\tfrac{1}{2},-\tfrac{1}{2})=2\times (\textbf{2},\textbf{2})
    \end{aligned}
\end{equation}
Remember that all these states have degeneracy 2. In total we find 2 gravitini (\textbf{3},\textbf{2}), 4 tensors (\textbf{3},\textbf{1}), 8 vectors (\textbf{2},\textbf{2}), 26 dilatini, $16\times(\textbf{2},\textbf{1})$ and $10\times(\textbf{1},\textbf{2})$, and 20 scalars (\textbf{1},\textbf{1}). All these fields have mass $\left|{\mathcal{R}}/{2\alpha'}\right|$ due to the $\tfrac{1}{2}$-winding on the circle, and fit into a complex (1,2) BPS supermultiplet, as in \eqref{bps reps n=6}.

At this point it is interesting to discuss the low energy spectrum that is obtained without the shift on the circle, i.e. from the non-freely acting orbifold $T^4/\mathbb{Z}_2 \times S^1$. In this case, there are no massive fields in the untwisted sector (these are projected out of the spectrum instead) and all fields in the twisted sector are massless. In addition to the 6 massless gravitini from the untwisted sector, there are another 2 massless gravitini coming from the twisted sector. This leads to a supersymmetry enhancement from 24 to 32 supersymmetries and in essence, one simply recovers the $\mathcal{N}=8, D=5$ theory. Furthermore, in the de-compactification limit $\mathcal{R}\to \infty$, which corresponds to an asymmetric $T^4/\mathbb{Z}_2$ orbifold compactification, one obtains the $\mathcal{N}=8, D=6$ theory. 

The same result can be obtained from the lowest lying spectrum of our freely acting orbifold in the limit $\mathcal{R}\to \infty$. Recall that the mass of all fields in the untwisted sector is equal to $|1/\mathcal{R}|$. In the de-compactification limit these fields become massless and we obtain 2 additional massless gravitini. On the other hand, the mass of the fields in the twisted sector is proportional to $\mathcal{R}$. These become infinitely massive and decouple. Consequently, we find 8 massless gravitini and we retrieve the $\mathcal{N}=8, D=6$ theory. In fact, in the de-compactification limit all our orbifolds reduce to the $\mathcal{N}=8, D=6$ theory because, as we will also see in the next examples, the masses of all fields in the untwisted sector are proportional to $1/\mathcal{R}$, while in the twisted sectors they are proportional to $\mathcal{R}$ (for large $\mathcal{R}$).

Similar arguments also hold for the $\mathbb{Z}_3$, $\mathcal{N}=6$ orbifold. Finally, it is worth noting that both freely and non-freely acting orbifolds prevent the appearance of an $\mathcal{N}=6, D=6$ theory. Such a theory can be defined classically but in the quantum level is inconsistent since it suffers from gravitational anomalies (for a discussion on the $\mathcal{N}=6, D=6$ supergravity see also \cite{d1998n}).

\subsection{$\mathcal{N}=4$}
\label{sec:N=4 symmetric and asymmetric}
In this section we discuss orbifolds with $\mathcal{N}=4$ supersymmetry in five dimensions. These models can be realized by symmetric or asymmetric constructions. Here we will present a symmetric $\mathbb{Z}_4$,  $\mathcal{N}=4\,(0,2)$ and an asymmetric  $\mathbb{Z}_2$,  $\mathcal{N}=4\,(1,1)$  orbifold.

\subsubsection*{A symmetric $\mathbb{Z}_4,$ $\mathcal{N}=4\,(0,2)$ orbifold}

As a first example, we consider a non-chiral symmetric $\mathbb{Z}_4$ orbifold (cf. example \ref{non-chiral N=4}) with twist vectors $\tilde{u}=u=(0,0,\tfrac{1}{4},\tfrac{1}{4})$, or equivalently $\vec{m}=(\tfrac{\pi}{2},\tfrac{\pi}{2},0,0)$, breaking half of the left and right-moving supersymmetries. We choose the torus lattice to be the $(A_1)^4$ root lattice with basis vectors $R_1(1,0,0,0)$, $R_1(0,1,0,0)$, $R_2(0,0,1,0)$ and $R_2(0,0,0,1)$, and we set the $B$-field to zero. This $\mathbb{Z}_4$ orbifold acts non-trivially on all toroidal dimensions. Consequently, there will be no invariant sublattices of left or right-moving momenta contributing to ${Z}_{T^4}[k,l]$ (for $k$ or $l\neq 0$). As was discussed in section \ref{sec:Partition function}, the partition functions for the symmetric orbifolds that we consider is always modular invariant and we will not  present them here. 

Regarding the orbifold spectrum, we construct closed string states following the same procedure as in section \ref{sec:N=6}. We begin with the massless spectrum in the untwisted sector where invariant states satisfy $\tilde{r}_3+\tilde{r_4}-r_3-r_4=0$ mod 4. We find the following states

NS-NS sector:
\begin{equation}
    \begin{aligned}
      (\underline{\pm 1,0},0,0) &\otimes (\underline{\pm 1,0},0,0)=\textbf{5}\oplus3\times\textbf{3}\oplus2\times\textbf{1}\\
    \pm[  (0,0,\underline{ 1,0}) &\otimes (0,0,\underline{ 1,0})] = 8\times\textbf{1}
    \end{aligned}
\end{equation}
NS-R sector:
\begin{equation}
    \begin{aligned}
      (\underline{\pm 1,0},0,0) &\otimes \pm(\underline{\tfrac{1}{2},-\tfrac{1}{2}},\tfrac{1}{2},-\tfrac{1}{2})=2\times \textbf{4}\oplus 4\times \textbf{2}\\
      \pm[ (0,0,\underline{ 1,0}) &\otimes (\pm\tfrac{1}{2},\pm\tfrac{1}{2},\tfrac{1}{2},\tfrac{1}{2})]=   4\times \textbf{2}
    \end{aligned}
\end{equation}
R-R sector:
\begin{equation}
    \begin{aligned}
    \pm[ (\pm\tfrac{1}{2},\pm\tfrac{1}{2},\tfrac{1}{2},\tfrac{1}{2})&\otimes (\pm\tfrac{1}{2},\pm\tfrac{1}{2},\tfrac{1}{2},\tfrac{1}{2})] = 2\times \textbf{3} \oplus 2\times \textbf{1}\\
    \pm (\underline{\tfrac{1}{2},-\tfrac{1}{2}},\tfrac{1}{2},-\tfrac{1}{2}) &\otimes \pm (\underline{\tfrac{1}{2},-\tfrac{1}{2}},\tfrac{1}{2},-\tfrac{1}{2}) = 4\times \textbf{3} \oplus 4 \times \textbf{1} 
    \end{aligned}
\end{equation}
The spectrum in the R-NS sector is identical with the one found in the NS-R sector. This is expected because we are treating a symmetric orbifold.  Collecting together our results from the four sectors, we find the graviton, 9 vectors, 4 gravitini, 16 dilatini and 16 scalars. These form the $\mathcal{N}=4$ gravity multiplet, consisting of the graviton, 4 gravitini, 6 vectors, 4 dilatini and 1 scalar, coupled to three vector multiplets, each made up from 1 vector, 4 dilatini and 5 scalars. We continue with the massive states

NS-NS sector:
\begin{equation}
    \begin{aligned}
   \pm[   (\underline{\pm 1,0},0,0;1) &\otimes (0,0,\underline{ 1,0})]= 4\times (\textbf{2},\textbf{2})\\
       \pm (0,0,\underline{1,0};-1)&\otimes  (\underline{\pm 1,0},0,0)= 4 \times (\textbf{2},\textbf{2})\\
  \pm[   (0,0,\underline{1,0};-2)&\otimes (0,0,\underline{-1,0})] = 8\times (\textbf{1},\textbf{1})
    \end{aligned}
\end{equation}
NS-R sector:
\begin{equation}
\begin{aligned}
  \pm[  (\underline{\pm 1,0},0,0;1) &\otimes (\pm\tfrac{1}{2},\pm\tfrac{1}{2},\tfrac{1}{2},\tfrac{1}{2})]= 2\times (\textbf{3},\textbf{2}) \oplus 2\times (\textbf{1},\textbf{2})\\
  \pm[ (0,0,\underline{1,0};-2) &\otimes (\pm\tfrac{1}{2},\pm\tfrac{1}{2},-\tfrac{1}{2},-\tfrac{1}{2})]=4\times (\textbf{2},\textbf{1})\\
   \pm (0,0,\underline{1,0};-1) &\otimes \pm(\underline{\tfrac{1}{2},-\tfrac{1}{2}},\tfrac{1}{2},-\tfrac{1}{2})  =8\times (\textbf{1},\textbf{2})
   \end{aligned}
\end{equation}
R-R sector: 
\begin{equation}
    \begin{aligned}
  \pm[  (\pm\tfrac{1}{2},\pm\tfrac{1}{2},\tfrac{1}{2},\tfrac{1}{2};-2)  &\otimes (\pm\tfrac{1}{2},\pm\tfrac{1}{2},-\tfrac{1}{2},-\tfrac{1}{2})]= 2\times (\textbf{3},\textbf{1}) \oplus 2\times (\textbf{1},\textbf{1})\\
 \pm(\pm\tfrac{1}{2},\pm\tfrac{1}{2},\tfrac{1}{2},\tfrac{1}{2};-1)  &\otimes \pm(\underline{\tfrac{1}{2},-\tfrac{1}{2}},\tfrac{1}{2},-\tfrac{1}{2}) = 4\times (\textbf{2},\textbf{2})\\
\pm(\underline{\tfrac{1}{2},-\tfrac{1}{2}},\tfrac{1}{2},-\tfrac{1}{2}) & \otimes \pm(\pm\tfrac{1}{2},\pm\tfrac{1}{2},\tfrac{1}{2},\tfrac{1}{2};1) = 4\times (\textbf{2},\textbf{2})
  \end{aligned}
\end{equation}
\footnote{In the last line of the R-R sector, for clearer notation we denoted the momentum of the states on the right-movers.}Again, in the R-NS sector we find the same spectrum as in the NS-R sector. In total, we find 16 vectors $(\textbf{2},\textbf{2})$ with mass $|{1}/{\mathcal{R}}|$, 2 tensors $(\textbf{3},\textbf{1})$ and 10 scalars $(\textbf{1},\textbf{1})$ with mass $|{2}/{\mathcal{R}}|$, 4 gravitini $(\textbf{3},\textbf{2})$ and 20 dilatini  $(\textbf{1},\textbf{2})$ with mass $|{1}/{\mathcal{R}}|$ and 8 dilatini $(\textbf{2},\textbf{1})$ with mass $|{2}/{\mathcal{R}}|$. These fields fit into two complex $(0,2)$ spin-$\tfrac{3}{2}$ multiplets of the form $(\textbf{3},\textbf{2}) \oplus 4 \times (\textbf{2},\textbf{2}) \oplus 5\times (\textbf{1},\textbf{2})$ with mass $|1/\mathcal{R}|$, and one complex $(0,2)$ tensor multiplet $(\textbf{3},\textbf{1}) \oplus 4 \times (\textbf{2},\textbf{1}) \oplus 5\times (\textbf{1},\textbf{1})$ with mass $|2/\mathcal{R}|$. Finally, we construct Kaluza-Klein towers by adding a trivial phase $e^{(\frac{\pi i l}{2})4\mathbb{Z}}$ to all states and we identify $\mathcal{R}=4R$. In this way, we can verify that the orbifold untwisted spectrum matches exactly with the Scherk-Schwarz supergravity one found in \cite{Hull:2020byc}. Now, we move on to the twisted sectors and we start our analysis with the $k=1$ sector. The weight vectors for the lightest left and right-moving states are the same, and coincide with those listed in  \autoref{k=1 N=6 states}. Orbifold invariant states satisfy $\tilde{r}_3+\tilde{r_4}-r_3-r_4=0$ mod 4. We list below the states that we find in each sector

NS-NS sector:
\begin{equation}
     (0,0,\underline{-1,0})\otimes   (0,0,\underline{-1,0}) = 4\times (\textbf{1},\textbf{1})
\end{equation}
NS-R sector:
\begin{equation}
     (0,0,\underline{-1,0})\otimes (\pm\tfrac{1}{2},\pm\tfrac{1}{2},-\tfrac{1}{2},-\tfrac{1}{2})= 2\times   (\textbf{2},\textbf{1})
\end{equation}
R-NS sector
\begin{equation}
    (\pm\tfrac{1}{2},\pm\tfrac{1}{2},-\tfrac{1}{2},-\tfrac{1}{2}) \otimes (0,0,\underline{-1,0}) = 2\times   (\textbf{2},\textbf{1})
\end{equation}
R-R sector:
\begin{equation}
     (\pm\tfrac{1}{2},\pm\tfrac{1}{2},-\tfrac{1}{2},-\tfrac{1}{2})\otimes  (\pm\tfrac{1}{2},\pm\tfrac{1}{2},-\tfrac{1}{2},-\tfrac{1}{2}) = (\textbf{3},\textbf{1}) \oplus (\textbf{1},\textbf{1})
\end{equation}
The degeneracy of the above states is 4. In total, we find 4 tensors $(\textbf{3},\textbf{1})$ 16 dilatini $(\textbf{2},\textbf{1})$ and 20 scalars $(\textbf{1},\textbf{1})$. These fields have mass $|{\mathcal{R}}/{4\alpha'}|$, due to the $\tfrac{1}{4}$-winding on the circle, and they fit into two complex $(0,2)$ tensor multiplets of the form $(\textbf{3},\textbf{1}) \oplus 4 \times (\textbf{2},\textbf{1}) \oplus 5\times (\textbf{1},\textbf{1})$. Regarding the $k=2$ twisted sector, the weight vectors for the lightest states coincide with those of the $k=1$ sector. Hence, we find the same states. However, the degeneracy of the states in the $k=2$ twisted sector is 10 and their mass is $|{\mathcal{R}}/{2\alpha'}|$. Finally, the $k=3$ twisted sector is equivalent to the $k=1$ sector. Note that the lightest states in the $k=3$ sector are those with winding $w=-1$\footnote{The same result can be obtained by considering $k=-1$ instead of $k=3$.}. Concluding, in the twisted sectors of the symmetric $\mathbb{Z}_4$, $\mathcal{N}=4\,(0,2)$ orbifold we find 9 complex tensor multiplets, 4 with mass $|{\mathcal{R}}/{4\alpha'}|$ and 5 with mass $|{\mathcal{R}}/{2\alpha'}|$.

Finally, we would like to briefly discuss here the spectrum obtained by the corresponding non-freely acting orbifold $T^4/\mathbb{Z}_4\times S^1$. In the untwisted sector the massless spectrum consists of the $\mathcal{N}=4$ gravity multiplet coupled to 3 vector multiplets. In the twisted sectors one finds 18 massless vector multiplets. In total, there exist 27 massless vectors.
The resulting number of vectors is consistent with the anomaly cancellation condition in type IIB $\mathcal{N}=4\,(0,2)$ theory in 6$D$, where 21 massless tensor multiplets are required for the anomalies to cancel \cite{townsend1984new}. Then compactification of the 6$D$ theory on a circle yields the 5$D$ theory with exactly 27 vectors, upon dualizing the tensors into vectors (see also \cite{bonetti2013exploring}). For the $T^4/\mathbb{Z}_{3,6}\times S^1$ orbifolds the result is exactly the same and for the $T^4/\mathbb{Z}_2\times S^1$ orbifold, 11 vectors come from the untwisted sector and 16 from the untwisted sector, giving again a total of 27 vectors\footnote{Type IIB on $K3\times S^1$  also gives 27 vectors in $D=5$.}.

\subsubsection*{An asymmetric $\mathbb{Z}_2,$ $\mathcal{N}=4\,(1,1)$ orbifold} 
\label{subsec N=4 1,1}

Here we consider an asymmetric $\mathbb{Z}_2$ orbifold with twist vectors $\tilde{u}=(0,0,0,0)$ and $u=(0,0,0,1)$, i.e. $\vec{m}=(0,\pi,0,-\pi)$. An interesting characteristic of this orbifold is the form of the right-moving twist vector $u$, which generates a $(-1)^{F_R}$ action. As we saw in the example of the fermionic monodromies in section \ref{fermionic monodromies}, this orbifold acts trivially on the torus coordinates, but spacetime fermions do feel the twist. Consequently, the torus bosonic partition is given by
\begin{equation}
    {Z}_{T^4} = \frac{1}{(\eta\,\bar{\eta})^4} \Gamma_{4,4}\,,
\end{equation}
which simply corresponds to a $T^4$ compactification and is invariant under modular transformations. For the remaining pieces of the partition function we find

\begin{equation}
\begin{aligned}
     Z[0,1]=Z_{\mathbb{R}^{1,4}}{Z}_{S^1}[0,1]{Z}_{T^4}\frac{1}{4(\eta\,\bar{\eta})^4} &[(\bar{\vartheta}_3)^4-(\bar{\vartheta}_4)^4-(\bar{\vartheta}_2)^4-(\bar{\vartheta}_1)^4] \,\times\\
&[(\vartheta_3)^4-(\vartheta_4)^4+(\vartheta_2)^4+(\vartheta_1)^4]\,.
\end{aligned}
\end{equation}
\begin{equation}
\begin{aligned}
     Z[1,0]=Z_{\mathbb{R}^{1,4}}{Z}_{S^1}[1,0]{Z}_{T^4}\frac{1}{4(\eta\,\bar{\eta})^4} &[(\bar{\vartheta}_3)^4-(\bar{\vartheta}_4)^4-(\bar{\vartheta}_2)^4-(\bar{\vartheta}_1)^4] \,\times\\
&[(\vartheta_3)^4+(\vartheta_4)^4-(\vartheta_2)^4+(\vartheta_1)^4]\,.
\end{aligned}
\end{equation}
\begin{equation}
\begin{aligned}
     Z[1,1]=Z_{\mathbb{R}^{1,4}}{Z}_{S^1}[1,1]{Z}_{T^4}\frac{1}{4(\eta\,\bar{\eta})^4} &[(\bar{\vartheta}_3)^4-(\bar{\vartheta}_4)^4-(\bar{\vartheta}_2)^4-(\bar{\vartheta}_1)^4] \,\times\\
&[-(\vartheta_3)^4-(\vartheta_4)^4-(\vartheta_2)^4+(\vartheta_1)^4]\,.
\end{aligned}
\end{equation}
Under modular transformations, the above pieces of the partition function transform as in \eqref{explicit modular orbits} and this guarantees modular invariance. Concerning the orbifold spectrum, in the untwisted sector orbifold invariant states satisfy $r_4\in \mathbb{Z}$. This means that all NS-NS and R-NS states survive the orbifold projection and remain massless. On the other hand, all NS-R and R-R states are charged under the orbifold action and survive the orbifold projection only with the addition of $n=\pm 1$ momentum modes. In this case, there is no need to write down explicitly the states that we find in each sector. The massless spectrum consists of the graviton, 11 vectors, 26 scalars, 4 gravitini and 24 dilatini. These fields make up the $\mathcal{N}=4$ gravity multiplet coupled to five vector multiplets. The massive spectrum consists of 8 vectors $(\textbf{2},\textbf{2})$, 8 tensors, $4\times(\textbf{3},\textbf{1})$ and $4\times(\textbf{1},\textbf{3})$, 8 scalars $(\textbf{1},\textbf{1})$, 4 gravitini, $2\times(\textbf{3},\textbf{2})$ and $2\times(\textbf{2},\textbf{3})$ and 20 dilatini,  $10\times(\textbf{1},\textbf{2})$ and $10\times(\textbf{2},\textbf{1})$. All these fields have mass $|{1}/{\mathcal{R}}|$ and form two complex $(1,1)$ spin-$\tfrac{3}{2}$ multiplets in the representations
\begin{equation}
    \begin{aligned}
       &(3,2)\oplus2\times(3,1)\oplus2\times(2,2)\oplus(1,2)\oplus4\times(2,1)\oplus2\times(1,1)\,,\\
       &(2,3)\oplus2\times(1,3)\oplus2\times(2,2)\oplus(2,1)\oplus4\times(1,2)\oplus2\times(1,1)\,.
    \end{aligned}
    \label{n=4 1,1 massive reps}
\end{equation}
Finally, for the construction of the Kaluza-Klein towers one works exactly as in section \ref{sec:N=6}. Once again, we can verify that the untwisted orbifold spectrum matches exactly the one found in \cite{Hull:2020byc} from the Scherk-Schwarz reduction on the level of supergravity. 

Now, consider the $k=1$ twisted sector where we use\footnote{Here there is an additional phase $e^{\pi i l}$ coming from the fermionic partition function \eqref{fermionic infinite sums} which is not cancelled by the torus bosonic partition function due to the form of the twist vector $u$.} 
\begin{equation}
     Z[1,l]=(q\,\bar{q})^{-\frac{1}{2}}\,\sum_{ {n,w\in \mathbb{Z}}}e^{ {\pi i n}l} q^{\frac{\alpha'}{4}P_{R}^2(1)} (\bar{q})^{\frac{\alpha'}{4}P_{L}^2(1)}\,\sum_{{r},\tilde{r}}  q^{\frac{1}{2}{(r+u)^2}}(\bar{q})^{\frac{1}{2}\tilde{r}^2} e^{{-\pi}il(2r_4+1)}(1+\cdots)\,.
\end{equation}
Since the orbifold acts trivially on the left-movers, the weight vectors for the lightest left-moving states in the absence of momentum and/or winding modes are given in \autoref{tablemasslessstates}. The weight vectors for the lightest right-moving states are listed in \autoref{k=1 N=4 asymmetric states}. Note here that the state with $r=(0,0,0,-1)$ is tachyonic. However, as we shall demonstrate in what follows, tachyonic states do not survive the orbifold projection, as expected in a supersymmetric model.
\renewcommand{\arraystretch}{2}
\begin{table}[h!]
\centering
 \begin{tabular}{|c|c|c|}
    \hline
    Sector &  r & SO(4) rep \\
    \hline
    \hline
  NS  & $(0,0,0,-1)$ &  $(\textbf{1},\textbf{1})$   \\
  \hline
   \multirow{4}{*}{R}   &$(\pm\frac{1}{2},\pm\frac{1}{2},-\frac{1}{2},-\frac{1}{2})$&$(\textbf{2},\textbf{1})$ \\  
   \cline{2-3}
  & $(\pm\frac{1}{2},\pm\frac{1}{2},\frac{1}{2},-\frac{3}{2})$ & $(\textbf{2},\textbf{1})$\\
   \cline{2-3}
  &  $(\underline{\frac{1}{2},-\frac{1}{2}},\frac{1}{2},-\frac{1}{2})$ & $(\textbf{1},\textbf{2})$\\
   \cline{2-3}
  &  $(\underline{\frac{1}{2},-\frac{1}{2}},-\frac{1}{2},-\frac{3}{2})$ & $(\textbf{1},\textbf{2})$\\
    \hline
    \end{tabular}
\captionsetup{width=.9\linewidth}
\caption{\textit{Here we list the weight vectors of the lightest right-moving states in the $k=1$ twisted sector of the asymmetric $\mathbb{Z}_2$, $\mathcal{N}=4$ orbifold and their representations under the massive little group in 5D.}}
\label{k=1 N=4 asymmetric states}
\end{table}
\renewcommand{\arraystretch}{1}

Orbifold invariant states satisfy $2r_4+1=0$ mod 2. As it follows from \autoref{k=1 N=4 asymmetric states}, all states in the NS-R and R-R sectors are invariant under the orbifold action. These are 8 vectors $(\textbf{2},\textbf{2})$, 8 tensors, $4\times(\textbf{3},\textbf{1})$ and $4\times(\textbf{1},\textbf{3})$, 8 scalars $(\textbf{1},\textbf{1})$, 4 gravitini, $2\times(\textbf{3},\textbf{2})$, and $2\times(\textbf{2},\textbf{3})$ and 20 dilatini,  $10\times(\textbf{1},\textbf{2})$ and $10\times(\textbf{2},\textbf{1})$. All these fields have mass $|{\mathcal{R}}/{2\alpha'}|$, due to the $\tfrac{1}{2}$-winding on the circle, and form two complex $(1,1)$ spin-$\tfrac{3}{2}$ multiplets, exactly as in \eqref{n=4 1,1 massive reps}. Moving on to the NS-NS and R-NS sectors, we notice that states are neither invariant under the orbifold action nor level-matched. However we can fix both issues either by adding $n=-1,w=0$ or $n=+1, w=-1$ momentum and winding modes respectively on the circle. We find the following states

NS-NS sector:
\begin{equation}
    \begin{aligned}
        (\underline{\pm1,0},0,0;-1,0)&\otimes (0,0,0,-1) = (\textbf{2},\textbf{2})\\
          (\underline{\pm1,0},0,0;1,-1)&\otimes (0,0,0,-1) = (\textbf{2},\textbf{2})\\
          (0,0,\underline{\pm1,0};-1,0)&\otimes (0,0,0,-1) = 4\times (\textbf{1},\textbf{1})\\
          (0,0,\underline{\pm1,0};1,-1)&\otimes (0,0,0,-1) = 4\times (\textbf{1},\textbf{1})
    \end{aligned}
\end{equation}
R-NS sector:
\begin{equation}
    \begin{aligned}
         (\pm(\pm\tfrac{1}{2},\pm\tfrac{1}{2},\tfrac{1}{2},\tfrac{1}{2});-1,0)&\otimes (0,0,0,-1) = 2\times (\textbf{2},\textbf{1})\\
         (\pm(\pm\tfrac{1}{2},\pm\tfrac{1}{2},\tfrac{1}{2},\tfrac{1}{2});1,-1)&\otimes (0,0,0,-1) = 2\times (\textbf{2},\textbf{1})\\
         (\pm(\underline{\tfrac{1}{2},-\tfrac{1}{2}},\tfrac{1}{2},-\tfrac{1}{2});-1,0) &\otimes (0,0,0,-1) = 2\times (\textbf{1},\textbf{2})\\
          (\pm(\underline{\tfrac{1}{2},-\tfrac{1}{2}},\tfrac{1}{2},-\tfrac{1}{2});1,-1) &\otimes (0,0,0,-1) = 2\times (\textbf{1},\textbf{2})
    \end{aligned}
\end{equation}
In total we find 2 vectors $(\textbf{2},\textbf{2})$, 8 scalars $(\textbf{1},\textbf{1})$ and 8 dilatini,  $4\times(\textbf{1},\textbf{2})$ and $4\times(\textbf{2},\textbf{1})$. These fields form one complex $(1,1)$ vector multiplet with mass $ m = \left|{1}/{\mathcal{R}}-{\mathcal{R}}/{2\alpha'}\right|$. As we have seen so far, states in the twisted sectors are generally massive. However, some states can become massless at special points of the moduli space. In the specific example that we discuss here, this can be achieved at circle radius $\mathcal{R}=\sqrt{2\alpha'}$. In this massless limit the massive complex vector multiplet gives two massless real vector multiplets (this can be seen by using the appropriate representations of the states under the massless little group in five dimensions). This is a generic feature of $\mathcal{N}=4\,(1,1)$ theories, where we break all the left, or right-moving supersymmetries, and we will revisit it when we discuss the moduli spaces of the orbifolds in section \ref{moduli spaces}.

\subsection{$\mathcal{N}=2$}
\label{sec:N=2 asymmetric orbifold}
In this section we discuss orbifolds with $\mathcal{N}=2$ supersymmetry in five dimensions. These are non-chiral asymmetric constructions (cf. example \ref{non-chiral N=2}) which break half of the left-moving and all the right-moving supersymmetries (or the other way around). As an example, consider the $\mathbb{Z}_2$ orbifold with twist vectors $\tilde{u}=(0,0,\tfrac{1}{2},\tfrac{1}{2})$ and $u=(0,0,0,1)$, i.e. $\vec{m}=(\pi,\pi,0,-\pi)$. Regarding the bosonic contribution to the partition function, notice that the twist vector $\tilde{u}$ is the same as in section \ref{sec:N=6}. Moreover, the twist vector $u$ acts trivially on the torus coordinates, exactly as the twist vector of section \ref{sec:N=6}. Therefore, the bosonic partition function of this $\mathcal{N}=2$ model coincides with the partition function of the $\mathcal{N}=6$ model obtained in section \ref{sec:N=6}. However, the fermionic partition function is different because the twist vector $u$ generates a $(-1)^{F_R}$ action. Putting all together we find 
\begin{equation}
    \begin{aligned}
       Z[0,1]=Z_{\mathbb{R}^{1,4}}{Z}_{S^1}[0,1]& \left(\frac{\bar{\eta}}{\bar{\vartheta_2}}\right)^2\frac{1}{\bar{\eta}^4} [(\bar{\vartheta_3}\bar{\vartheta_4})^2-(\bar{\vartheta_4}\bar{\vartheta_3})^2-(\bar{\vartheta_2}\bar{\vartheta_1})^2-(\bar{\vartheta_1}\bar{\vartheta_2)^2}] \,\times\\
      &  \frac{1}{{\eta}^4}\Theta_{D_4}(\tau)\frac{1}{{\eta}^4}
      [(\vartheta_3)^4-(\vartheta_4)^4+(\vartheta_2)^4+(\vartheta_1)^4]\,.
    \end{aligned}
\end{equation}
\begin{equation}
    \begin{aligned}
     Z[1,0]=Z_{\mathbb{R}^{1,4}}{Z}_{S^1}[1,0]& \left(\frac{\bar{\eta}}{\bar{\vartheta_4}}\right)^2\frac{1}{\bar{\eta}^4}  [(\bar{\vartheta_3}\bar{\vartheta_2})^2+(\bar{\vartheta_4}\bar{\vartheta_1})^2-(\bar{\vartheta_2}\bar{\vartheta_3})^2+(\bar{\vartheta_1}\bar{\vartheta_4)^2}] \,\times\\
    & \frac{1}{2{\eta}^4}\Theta_{D_4^*}(\tau)\frac{1}{{\eta}^4}
      [(\vartheta_3)^4+(\vartheta_4)^4-(\vartheta_2)^4+(\vartheta_1)^4]\,.
    \end{aligned}
\end{equation}
\begin{equation}
    \begin{aligned}
     Z[1,1]=Z_{\mathbb{R}^{1,4}}{Z}_{S^1}[1,1]& \left(\frac{\bar{\eta}}{\bar{\vartheta_3}}\right)^2\frac{1}{\bar{\eta}^4}  [(\bar{\vartheta_3}\bar{\vartheta_1})^2+(\bar{\vartheta_4}\bar{\vartheta_2})^2-(\bar{\vartheta_2}\bar{\vartheta_4})^2+(\bar{\vartheta_1}\bar{\vartheta_3)^2}] \,\times\\
    & \frac{1}{2{\eta}^4}\Theta_{D_4^*}(\tau+1)\frac{1}{{\eta}^4}
      [-(\vartheta_3)^4-(\vartheta_4)^4-(\vartheta_2)^4+(\vartheta_1)^4]\,.
    \end{aligned}
\end{equation}
These pieces of the partition function satisfy the transformations \eqref{explicit modular orbits}. This guarantees modular invariance\footnote{As in the $\mathcal{N}=6$ example of section \ref{sec:N=6}, the model based on the $(A_1)^4$ lattice, which is also discussed in \cite{sen1995dual}, does not give a modular invariant partition function.}. Now, let us proceed with the construction of the orbifold spectrum. As usual, we start with the massless states in the untwisted sector, which obey $\tilde{r}_3+\tilde{r}_4-2r_4=0$ mod 2 

NS-NS sector:
\begin{equation}
    \begin{aligned}
       (\underline{\pm 1,0},0,0) &\otimes (\underline{\pm 1,0},0,0)=\textbf{5}\oplus3\times\textbf{3}\oplus2\times\textbf{1}\\
     (\underline{\pm 1,0},0,0) &\otimes (0,0,\underline{\pm 1,0})=4\times\textbf{3}\oplus4\times\textbf{1}
    \end{aligned}
\end{equation}
NS-R sector:
\begin{equation}
    \begin{aligned}
      (0,0,\underline{\pm 1,0}) &\otimes \pm (\pm\tfrac{1}{2},\pm\tfrac{1}{2},\tfrac{1}{2},\tfrac{1}{2})=   8\times \textbf{2}  \\
       (0,0,\underline{\pm 1,0}) &\otimes \pm (\underline{\tfrac{1}{2},-\tfrac{1}{2}},\tfrac{1}{2},-\tfrac{1}{2})=8\times \textbf{2}
    \end{aligned}
\end{equation}
R-NS sector:
\begin{equation}
    \begin{aligned}
    \pm (\underline{\tfrac{1}{2},-\tfrac{1}{2}},\tfrac{1}{2},-\tfrac{1}{2}) & \otimes  (\underline{\pm 1,0},0,0) = 2\times \textbf{4} \oplus 4\times \textbf{2}\\
      \pm(\underline{\tfrac{1}{2},-\tfrac{1}{2}},\tfrac{1}{2},-\tfrac{1}{2}) & \otimes   (0,0,\underline{\pm 1,0}) = 8\times \textbf{2}
    \end{aligned}
\end{equation}
R-R sector:
\begin{equation}
    \begin{aligned}
    \pm (\pm\tfrac{1}{2},\pm\tfrac{1}{2},\tfrac{1}{2},\tfrac{1}{2})&\otimes\pm  (\pm\tfrac{1}{2},\pm\tfrac{1}{2},\tfrac{1}{2},\tfrac{1}{2})=4\times \textbf{3}\oplus 4\times \textbf{1}\\
     \pm (\pm\tfrac{1}{2},\pm\tfrac{1}{2},\tfrac{1}{2},\tfrac{1}{2})&\otimes \pm (\underline{\tfrac{1}{2},-\tfrac{1}{2}},\tfrac{1}{2},-\tfrac{1}{2})=4\times \textbf{3}\oplus 4\times \textbf{1}
    \end{aligned}
\end{equation}
We find the graviton, 15 vectors, 14 scalars, 2 gravitini and 28 dilatini. These fields form the $\mathcal{N}=2$ gravity multiplet consisting of the graviton, 2 gravitini and 1 vector, and 14 vector multiplets each consisting of 1 vector 2 dilatini and 1 scalar. We continue with the construction of the massive spectrum

NS-NS sector:
\begin{equation}
    \begin{aligned}
    \pm(0,0,\underline{1,0};-1) & \otimes (\underline{\pm1,0},0,0)=  4 \times (\textbf{2},\textbf{2})\\
     \pm (0,0,\underline{1,0};-1) & \otimes  (0,0,{\pm 1,0}) = 8 \times (\textbf{1},\textbf{1})\\
     \pm [(0,0,\underline{1,0};1) & \otimes (0,0,0,1) = 4 \times (\textbf{1},\textbf{1})\\
     \pm [(0,0,\underline{1,0};-3) & \otimes (0,0,0,-1) = 4 \times (\textbf{1},\textbf{1})
        \end{aligned}
\end{equation}
NS-R sector:
\begin{equation}
    \begin{aligned}
   \pm[  (\underline{\pm1,0},0,0;1) &\otimes (\pm\tfrac{1}{2},\pm\tfrac{1}{2},\tfrac{1}{2},\tfrac{1}{2})] = 2\times (\textbf{3},\textbf{2}) \oplus 2\times (\textbf{1},\textbf{2})\\
   \pm[   (\underline{\pm1,0},0,0;-1) &\otimes  (\underline{\tfrac{1}{2},-\tfrac{1}{2}},\tfrac{1}{2},-\tfrac{1}{2})] = 2\times (\textbf{2},\textbf{3}) \oplus 2\times (\textbf{2},\textbf{1})
    \end{aligned}
\end{equation}
R-NS sector:
\begin{equation}
    \begin{aligned}
    \pm (\pm\tfrac{1}{2},\pm\tfrac{1}{2},\tfrac{1}{2},\tfrac{1}{2};-1) &\otimes  (\underline{\pm1,0},0,0) = 2 \times (\textbf{3},\textbf{2}) \oplus 2 \times (\textbf{1},\textbf{2})\\
     \pm (\pm\tfrac{1}{2},\pm\tfrac{1}{2},\tfrac{1}{2},\tfrac{1}{2};-1) &\otimes (0,0,\pm1,0) = 4 \times   (\textbf{2},\textbf{1})\\
    \pm [(\pm\tfrac{1}{2},\pm\tfrac{1}{2},\tfrac{1}{2},\tfrac{1}{2};1) &\otimes (0,0,0,1)] = 2 \times (\textbf{2},\textbf{1})\\
     \pm [(\pm\tfrac{1}{2},\pm\tfrac{1}{2},\tfrac{1}{2},\tfrac{1}{2};-3) &\otimes (0,0,0,-1)] = 2 \times (\textbf{2},\textbf{1})
      \end{aligned}
\end{equation}
R-R sector:
\begin{equation}
    \begin{aligned}
    \pm (\underline{\tfrac{1}{2},-\tfrac{1}{2}},\tfrac{1}{2},-\tfrac{1}{2}) & \otimes \pm(\pm\tfrac{1}{2},\pm\tfrac{1}{2},\tfrac{1}{2},\tfrac{1}{2};1)= 4 \times (\textbf{2},\textbf{2})\\
   \pm (\underline{\tfrac{1}{2},-\tfrac{1}{2}},\tfrac{1}{2},-\tfrac{1}{2}) & \otimes \pm(\underline{\tfrac{1}{2},-\tfrac{1}{2}},\tfrac{1}{2},-\tfrac{1}{2};-1) =  4 \times (\textbf{1},\textbf{3}) \oplus  4 \times (\textbf{1},\textbf{1})
    \end{aligned}
\end{equation}
\footnote{In the R-R sector, for clearer notation we denoted the momentum of the states on the right-movers.}In total, we find 6 gravitini, $4\times(\textbf{3},\textbf{2})$ and $2\times(\textbf{2},\textbf{3})$, 4 tensors (\textbf{1},\textbf{3}), 8 vectors (\textbf{2},\textbf{2}), 12 dilatini, $8\times(\textbf{2},\textbf{1})$ and $4\times(\textbf{1},\textbf{2})$, and 16 scalars (\textbf{1},\textbf{1}) with mass $|1/\mathcal{R}|$, and in addition, 2 dilatini (\textbf{2},\textbf{1}) and 4 scalars (\textbf{1},\textbf{1}) with mass $|3/\mathcal{R}|$. These fields fit into the following complex multiplets: Two spin-$\tfrac{3}{2}$ multiplets $(\textbf{3},\textbf{2}) \oplus 2 \times (\textbf{2},\textbf{2}) \oplus (\textbf{1},\textbf{2})$, another multiplet containing a spin-$\tfrac{3}{2}$ particle; $ (\textbf{2},\textbf{3}) \oplus 2 \times (\textbf{1},\textbf{3})$, and four hypermultiplets $(\textbf{2},\textbf{1})\oplus 2 \times (\textbf{1},\textbf{1})$ with mass $|1/\mathcal{R}|$, as well as one hypermultiplet with mass $|3/\mathcal{R}|$. Finally, the construction of the Kaluza-Klein towers proceeds in exactly the same way as in section \ref{sec:N=6}. Once again, we confirm that the orbifold spectrum in the untwisted sector matches exactly the one found in \cite{Hull:2020byc} from the Scherk-Schwarz reduction on the level of supergravity. Now, we move on to the $k=1$ twisted sector where we use\footnote{As in the asymmetric $\mathbb{Z}_2, \mathcal{N}=4$ orbifold, here there is an additional phase $e^{\pi i l}$ coming from the fermionic partition function \eqref{fermionic infinite sums} due to the form of the twist vector $u$.} 
\begin{equation}
     Z[1,l]=2q^{-\frac{1}{2}}(\bar{q})^{-\frac{1}{4}}\,\sum_{ {n,w\in \mathbb{Z}}}e^{ {\pi i n}l} q^{\frac{\alpha'}{4}P_{R}^2(1)} (\bar{q})^{\frac{\alpha'}{4}P_{L}^2(1)}\,\sum_{{r},\tilde{r}}  q^{\frac{1}{2}{(r+u)}^2}(\bar{q})^{\frac{1}{2}(\tilde{r}+\tilde{u})^2} e^{{\pi}il(\tilde{r}_3+\tilde{r}_4-2r_4)}(1+\cdots)\,.
\end{equation}
The weight vectors for the lightest states in the absence of momentum and/or winding modes are given in \autoref{k=1 N=2 states}. 
\renewcommand{\arraystretch}{2}
\begin{table}[h!]
\centering
 \begin{tabular}{|c|c|c|c|c|}
    \hline
    Sector &  $\tilde{r}$ & SO(4) rep & r & SO(4) rep \\
    \hline
    \hline
  NS & $(0,0,\underline{-1,0})$ & 2$\,\times\,(\textbf{1},\textbf{1})$ & $(0,0,0,-1)$ &  $(\textbf{1},\textbf{1})$   \\
  \hline
   \multirow{4}{*}{R}   &  \multirow{4}{*}{$(\pm\frac{1}{2},\pm\frac{1}{2},-\frac{1}{2},-\frac{1}{2})$}  & \multirow{4}{*}{$(\textbf{2},\textbf{1})$} &$(\pm\frac{1}{2},\pm\frac{1}{2},-\frac{1}{2},-\frac{1}{2})$&$(\textbf{2},\textbf{1})$ \\  
   \cline{4-5}
  & & & $(\pm\frac{1}{2},\pm\frac{1}{2},\frac{1}{2},-\frac{3}{2})$ & $(\textbf{2},\textbf{1})$\\
   \cline{4-5}
  & & & $(\underline{\frac{1}{2},-\frac{1}{2}},\frac{1}{2},-\frac{1}{2})$ & $(\textbf{1},\textbf{2})$\\
   \cline{4-5}
  & & & $(\underline{\frac{1}{2},-\frac{1}{2}},-\frac{1}{2},-\frac{3}{2})$ & $(\textbf{1},\textbf{2})$\\
    \hline
    \end{tabular}
\captionsetup{width=.9\linewidth}
\caption{\textit{Here we list the weight vectors of the lightest left and right-moving states in the $k=1$ twisted sector of the $\mathbb{Z}_2$, $\mathcal{N}=2$ orbifold and their representations under the massive little group in 5D.}}
\label{k=1 N=2 states}
\end{table}
\renewcommand{\arraystretch}{1}

Orbifold invariant states satisfy $\tilde{r}_3+\tilde{r}_4-2r_4=0$ mod 2 and they have degeneracy 2. All the states in the NS-R and R-R sectors survive the orbifold projection. In these sectors we find 16 dilatini, $8\times(\textbf{2},\textbf{1})$ and $8\times(\textbf{1},\textbf{2})$, 4 tensors (\textbf{3},\textbf{1}), 4 vectors (\textbf{2},\textbf{2}) and 4 scalars (\textbf{1},\textbf{1}). All these fields have mass $\left|{\mathcal{R}}/{2\alpha'}\right|$, due to the $\tfrac{1}{2}$-winding on the circle, and they fit into two complex tensor multiplets $(\textbf{3},\textbf{1})\oplus 2\times (\textbf{2},\textbf{1})\oplus (\textbf{1},\textbf{1})$ and two complex vector multiplets $(\textbf{2},\textbf{2})\oplus 2\times (\textbf{1},\textbf{2})$. Regarding the NS-NS and R-NS sectors, we observe that, as in the $\mathcal{N}=4\,(1,1)$ case, the state $r=(0,0,0,-1)$ is tachyonic. In these sectors states are not level-matched and do not survive the orbifold projection. However, with the addition of $n=-1,w=0$ or $n=+1, w=-1$ momentum and winding modes respectively on the circle, states become level-matched and survive the orbifold. Thereby, we find the following states

NS-NS sector:
\begin{equation}
\begin{aligned}
    (0,0,\underline{-1,0};-1,0)&\otimes (0,0,0,-1) = 2\times (\textbf{1},\textbf{1})\\
    (0,0,\underline{-1,0};1,-1)&\otimes (0,0,0,-1) = 2\times (\textbf{1},\textbf{1})
    \end{aligned}
\end{equation}
R-NS sector:
\begin{equation}
\begin{aligned}
     (\pm\tfrac{1}{2},\pm\tfrac{1}{2},-\tfrac{1}{2},-\tfrac{1}{2};-1,0)&\otimes (0,0,0,-1)= (\textbf{2},\textbf{1})\\
     (\pm\tfrac{1}{2},\pm\tfrac{1}{2},-\tfrac{1}{2},-\tfrac{1}{2};1,-1)&\otimes (0,0,0,-1)= (\textbf{2},\textbf{1})
     \end{aligned}
\end{equation}
These states come with a degeneracy factor of 2. We find 8 scalars (\textbf{1},\textbf{1}) and 4 dilatini (\textbf{2},\textbf{1}) which form two massive complex hypermultiplets with mass $|{1}/{\mathcal{R}}-{\mathcal{R}}/{2\alpha'}|$. As a final remark, note that it is possible to make these hypermultiplets massless by fixing the circle radius at $\mathcal{R}=\sqrt{2\alpha'}$. We will return to this issue in section \ref{moduli spaces}, where we discuss the moduli spaces of the orbifolds.

\subsection{$\mathcal{N}=0$}
\label{sec:N=0}
In this section we discuss non-supersymmetric symmetric orbifolds (see also \cite{font2002non,kawazu2004non}). These are models with $\tilde{u}=u=(0,0,u_3,u_4)$ which do not satisfy $\pm u_3\pm u_4 = 0$ mod 2 for any choice of signs, such that all supersymmetries are broken. In general, for the construction of states in all sectors we follow the same procedure as in section \ref{sec:N=6}. The interesting feature of non-supersymmetric orbifolds is the appearance of tachyons in the twisted sectors. As we will demonstrate, if the radius of the orbifold circle is large enough compared to the string scale, there will be no tachyons in the spectrum, see e.g. \cite{Dabholkar:2002sy,Angelantonj:2006ut,Scrucca:2001ni,acharya2021stringy} for more examples. As an example, consider a symmetric $\mathbb{Z}_3$ orbifold breaking all supersymmetry, with twist vectors $\tilde{u}=u=(0,0,0,\tfrac{2}{3})$, i.e. $\vec{m}=(\tfrac{2\pi}{3},\tfrac{2\pi}{3},-\tfrac{2\pi}{3},-\tfrac{2\pi}{3})$. For this orbifold is it more convenient to decompose the $T^4$ as $T^2\times T^2$. In this way we can see that the orbifold action leaves one $T^2$ intact. Therefore, from this $T^2$, there will be an invariant lattice of left and right-moving momenta contributing to ${Z}_{T^4}[k,l]$. As usual in toroidal compactification, this lattice is a 4-dimensional even, self-dual Lorentzian lattice $\Gamma_{2,2}$, with volume equal to 1. As a consequence, there are no inconsistencies regarding the degeneracy number of the twisted states. For the other $T^2$, in order for the orbifold to be well-defined, we choose the $A_2$ root lattice with basis vectors $R_1(\sqrt{2},0)$ and $R_1(-\frac{1}{\sqrt{2}},\sqrt{\frac{3}{2}})$. We set the $B$-field to zero.

We start with the spectrum in the untwisted sector. As we are mostly interested in the twisted sectors, we do not write down explicitly all the states but we simply state the results. Orbifold invariant states satisfy $2(\tilde{r}_4-r_4)=0$ mod 3. The massless spectrum consists of the graviton, 15 vectors, 20 scalars and 8 dilatini. Note that there are no massless gravitini in the spectrum, hence the spectrum is non-supersymmetric. The massive spectrum consists of 8 vectors (\textbf{2},\textbf{2)}, 4 tensors, $2\times(\textbf{3},\textbf{1})$ and $2\times(\textbf{1},\textbf{3})$, and 12 scalars (\textbf{1},\textbf{1)} with mass $|{2}/{\mathcal{R}}|$, 2 scalars (\textbf{1},\textbf{1)} with mass $|{4}/{\mathcal{R}}|$ , and 8 gravitini, $4\times(\textbf{3},\textbf{2})$ and $4\times(\textbf{2},\textbf{3})$, and 32 dilatini, $16\times(\textbf{2},\textbf{1})$ and $16\times(\textbf{1},\textbf{2})$, with mass $|{1}/{\mathcal{R}}|$. In addition, we build the Kaluza-Klein towers by adding a trivial phase $e^{(\frac{2\pi i l}{3})3\mathbb{Z}}$ to all states. Finally, we identify the orbifold radius $\mathcal{R}$, with the Scherk-Schwarz radius $R$, by $\mathcal{R}=3{R}$ and we confirm that the entire untwisted spectrum matches with the one found in \cite{Hull:2020byc} from the Scherk-Schwarz reduction on the level of supergravity.

Let us now discuss the spectrum in the twisted sectors. As usual, we are interested in finding the lightest states in the absence of momentum and winding modes. We start from the $k=1$ twisted sector and we use\footnote{Note that in the $k=1$ sector the expression $\prod_{i}2\sin(\pi \text{\footnotesize{gcd}}(1,l)\tilde{u}_i)$ simply becomes $\prod_{i}2\sin(\pi \tilde{u}_i)$, and similarly for ${u}$.} 
\begin{equation}
    {Z}[1,l]= 3(q\bar{q})^{-\frac{7}{18}}\sum_{{n,w\in \mathbb{Z}}}e^{\frac{2\pi i n}{3}l} q^{\frac{\alpha'}{4}P_{R}^2(1)} (\bar{q})^{\frac{\alpha'}{4}P_{L}^2(1)}\,\sum_{r,\tilde{r}} q^{\frac{1}{2}({r}+ u)^2} (\bar{q})^{\frac{1}{2}(\tilde{r}+{u})^2}\,e^{\frac{4 \pi i l}{3}(\tilde{r}_4-r_4)}\, \left(1+\cdots\right)\,.
    \label{k=1 N=0}
\end{equation}
The weight vectors for the lightest states are given in \autoref{k=1 N=0 states}.
\renewcommand{\arraystretch}{2}
\begin{table}[h!]
\centering
 \begin{tabular}{|c|c|c|}
    \hline
    Sector &  $\tilde{r},r$ & SO(4) rep\\
    \hline
    \hline
  NS & $(0,0,0,-1)$ & $(\textbf{1},\textbf{1})$\\
  \hline
  \multirow{2}{*}{R}  & $(\pm\frac{1}{2},\pm\frac{1}{2},-\frac{1}{2},-\frac{1}{2})$ &(\textbf{2},\textbf{1})\\
    \cline{2-3}
     & $(\underline{\frac{1}{2},-\frac{1}{2}},\frac{1}{2},-\frac{1}{2})$ & (\textbf{1},\textbf{2})\\
   \hline
    \end{tabular}
\captionsetup{width=.9\linewidth}
\caption{\textit{Here we list the weight vectors of the lightest states in the $k=1$ twisted sector of the symmetric $\mathbb{Z}_3$, $\mathcal{N}=0$ orbifold and their representations under the massive little group in 5D.}}
\label{k=1 N=0 states}
\end{table}
\renewcommand{\arraystretch}{1}

Orbifold invariant states satisfy $2(\tilde{r}_4-r_4)=0$ mod 3. We list below the states that we find in each sector (all the states below come with a multiplicity 3, which we will omit writing down explicitly)

NS-NS sector:
\begin{equation}
    (0,0,0,-1) \otimes (0,0,0,-1) = (\textbf{1},\textbf{1})
\end{equation}
This state is a scalar with mass
\begin{equation}
    \alpha' m^2= \frac{\mathcal{R}^2}{9\alpha'}-\frac{4}{3}\,.
\end{equation}
We note that this state is tachyonic if 
\begin{equation}
    {\mathcal{R}}<2\sqrt{3\alpha'}\,.
\end{equation}
Taking the circle radius to be above this tachyon bound ensures that the spectrum is tachyon-free. We move on to the R-R sector:
\begin{equation}
    \begin{aligned}
      (\pm\tfrac{1}{2},\pm\tfrac{1}{2},-\tfrac{1}{2},-\tfrac{1}{2}) & \otimes (\pm\tfrac{1}{2},\pm\tfrac{1}{2},-\tfrac{1}{2},-\tfrac{1}{2}) =  (\textbf{3},\textbf{1}) \oplus (\textbf{1},\textbf{1})\\
      (\pm\tfrac{1}{2},\pm\tfrac{1}{2},-\tfrac{1}{2},-\tfrac{1}{2}) & \otimes (\underline{\tfrac{1}{2},-\tfrac{1}{2}},\tfrac{1}{2},-\tfrac{1}{2}) = (\textbf{2},\textbf{2})\\
      (\underline{\tfrac{1}{2},-\tfrac{1}{2}},\tfrac{1}{2},-\tfrac{1}{2}) & \otimes   (\pm\tfrac{1}{2},\pm\tfrac{1}{2},-\tfrac{1}{2},-\tfrac{1}{2})= (\textbf{2},\textbf{2})\\
      (\underline{\tfrac{1}{2},-\tfrac{1}{2}},\tfrac{1}{2},-\tfrac{1}{2}) & \otimes  (\underline{\tfrac{1}{2},-\tfrac{1}{2}},\tfrac{1}{2},-\tfrac{1}{2})  =  (\textbf{1},\textbf{3}) \oplus (\textbf{1},\textbf{1})
    \end{aligned}
\end{equation}
In this sector we find 2 tensors, 1$\times$(\textbf{3},\textbf{1}) and 1$\times$(\textbf{1},\textbf{3}), 2 vectors (\textbf{2},\textbf{2}) and 2 scalars (\textbf{1},\textbf{1}) with mass $|{\mathcal{R}}/{3\alpha'}|$. In the NS-R/R-NS sectors, states are not level-matched and do not survive the orbifold projection. However, we can fix both issues by adding $n=+1/-1$ momentum modes on the circle:

NS-R sector:
\begin{equation}
    \begin{aligned}
        (0,0,0,-1;1) &\otimes (\pm\tfrac{1}{2},\pm\tfrac{1}{2},-\tfrac{1}{2},-\tfrac{1}{2}) = (\textbf{2},\textbf{1})\\
        (0,0,0,-1;1) &\otimes  (\underline{\tfrac{1}{2},-\tfrac{1}{2}},\tfrac{1}{2},-\tfrac{1}{2})  = (\textbf{1},\textbf{2})
    \end{aligned}
\end{equation}
R-NS sector:
\begin{equation}
    \begin{aligned}
       (\pm\tfrac{1}{2},\pm\tfrac{1}{2},-\tfrac{1}{2},-\tfrac{1}{2};-1) &\otimes (0,0,0,-1) = (\textbf{2},\textbf{1})\\
       (\underline{\tfrac{1}{2},-\tfrac{1}{2}},\tfrac{1}{2},-\tfrac{1}{2};-1)&\otimes (0,0,0,-1) = (\textbf{1},\textbf{2})
    \end{aligned}
\end{equation}
In these sectors we find 4 dilatini, 2$\times$(\textbf{2},\textbf{1}) and 2$\times$(\textbf{1},\textbf{2}), with mass $|{1}/{\mathcal{R}}-{\mathcal{R}}/{3\alpha'}|$. We note that the states in the NS-R/R-NS sectors become massless for $\mathcal{R}=\sqrt{3\alpha'}$. However, this value is under the tachyon bound. Finally, the spectrum in the $k=2$ twisted sector is identical to the spectrum in the $k=1$ sector.

\subsubsection{Tachyon bounds}
In general, the spectrum of non-supersymmetric orbifolds contains tachyons coming from the twisted sectors. Nevertheless, we can repeat the same analysis as in the $\mathbb{Z}_3$ orbifold discussed above, in order to find the tachyon bounds for other non-supersymmetric, symmetric orbifolds. For each model, there is a critical circle radius $\mathcal{R}_*$ above which the string spectrum renders tachyon-free. The critical radius can be determined by examining the $k=1$ twisted sector\footnote{As $k$ becomes bigger, the critical radius becomes smaller. The strongest constraint comes from the $k=1$ sector.}. For $\mathbb{Z}_p$ orbifolds with twist vectors of the form $(0,0,u_3,u_4)$ with $|u_3|-|u_4|={n}/{p}$, $n\in \mathbb{Z}$, we find 
\begin{equation}
   \frac{\mathcal{R}_*}{\sqrt{\alpha'}}=\sqrt{2 |n| p}\ .
    \label{critical radious non-susy}
\end{equation}
The twist vectors that we used to obtain the above result are listed in appendix \ref{Ap C}, \autoref{tab breaking all}.

\section{Low-energy limit}
\label{Supergravity}
In this section we  give details of the relation between the compactification of type IIB string theory on freely acting orbifolds and the Scherk-Schwarz reduction of type IIB supergravity presented in \cite{Hull:2020byc}. We then discuss the moduli spaces, the classical Scherk-Schwarz potential and the supertrace formulae.

\subsection{Lowest lying orbifold states and Scherk-Schwarz spectrum}
\label{states}

In this subsection, we will explicitly construct the untwisted orbifold sector in terms of oscillator states. As in section \ref{closed string spectrum}, we will only focus on the lowest excited states, i.e the states that are massless without the addition of momentum and/or winding modes. This will manifest the correspondence between freely acting orbifolds and Scherk-Schwarz mechanism. 

As we discussed in section \ref{The orbifold action and partition function}, in the untwisted sector the NS-vacuum is a spacetime scalar and the R-vacuum is a spacetime spinor in all target space dimensions. The NS-vacua are invariant under the orbifold action, while the R-vacua transform as in \eqref{transformation of ramond vacua}. Having these at hand, we can discuss the resulting spectrum.

First, we present in \autoref{tablemasslessstates2} the lightest NS and R-sector states that survive the GSO projection. We write down general states that appear both in a left-moving and in a right-moving version, and we write down the orbifold charges that both of these versions carry. Furthermore, we table the representations of these states under both the massless little group SO$(3)\approx\text{SU}(2)$ and the massive little group SO$(4)\approx\text{SU}(2)\times\text{SU}(2)$ in five dimensions. 

\renewcommand{\arraystretch}{1.4}
\begin{table}[h!]
\centering
\begin{tabular}{|c|c|c|c|c|c|}
\hline
\;Sector\; & State & $L$ orbifold charge & $R$ orbifold charge & SO$(3)$ rep & SO$(4)$ rep \\ \hline\hline
NS & \;\;${b}^{\hat{\mu}}_{-1/2}\ket{0}$\;\; & $1$ & $1$ & ${\textbf{3}\oplus\textbf{1}}$ & {$({\textbf{2}},{\textbf{2}})$} \\ \cline{2-6}
& ${b}^i_{-1/2}\ket{0}$ & $e^{i(m_1\pm m_3)}$ & $e^{i(m_2\pm m_4)}$ & {$2\times\textbf{1}$} & $2\times({\textbf{1}},{\textbf{1}})$ \\ \cline{2-6}
& $\bar{{b}}^i_{-1/2}\ket{0}$ & $e^{-i(m_1\pm m_3)}$ & $e^{-i(m_2\pm m_4)}$ & {$2\times\textbf{1}$} & $2\times({\textbf{1}},{\textbf{1}})$ \\ \hline\hline
R & $|a_{1,2}\rangle$ & $e^{\pm im_1}$ & $e^{\pm im_2}$ & {$2\times\textbf{2}$} & $2\times({\textbf{2}},{\textbf{1}})$ \\ \cline{2-6}
& $|a_{3,4}\rangle$ & $e^{\pm im_3}$ & $e^{\pm im_4}$ & {$2\times\textbf{2}$} & $2\times({\textbf{1}},{\textbf{2}})$ \\ \hline
\end{tabular}
\captionsetup{width=.9\linewidth}
\caption{\textit{Here we write down all states that are massless in the absence of momentum and/or winding modes, including their charges under the orbifold action and their representations under the massless and massive little groups in 5D. We write down general states that appear both left-moving and right-moving. The tildes on the oscillators in the left-moving sector, and the subscripts $L$ and $R$ on the vacua are omitted. The index $i$ on the oscillators takes the values 1 and 2. The $+$ sign in the ${L/R}$ orbifold charge corresponds to $i=1$ and the $-$ sign to $i=2$.}}
\label{tablemasslessstates2}
\end{table}
\renewcommand{\arraystretch}{1}

String states are constructed by tensoring the left and right-moving states from \autoref{tablemasslessstates2}. If a state carries a non-trivial orbifold charge, we compensate for this by adding momentum modes on the circle. In \autoref{huiberttable} we give the spectrum of lowest excited string states, including their orbifold charge and little group representations.

\renewcommand{\arraystretch}{1.4}
\begin{table}[h!]
\centering
\begin{tabular}{|c|c|c|c|c|}
\hline
\;Sector\; & State & Orbifold charge & SO$(3)$ rep & SO$(4)$ rep \\ \hline\hline
NS-NS & \;\;$\tilde{b}^{\hat{\mu}}_{-1/2}\ket{0}_L \otimes {b}^{\hat{\nu}}_{-1/2}\ket{0}_R$\;\; & $1$ & $\textbf{5} \oplus 3\times\textbf{3}\oplus2\times \textbf{1}$ & $(\textbf{3}\oplus\textbf{1},\textbf{3}\oplus\textbf{1})$ \\ \cline{2-5}
& $\tilde{b}^{\hat{\mu}}_{-1/2}\ket{0}_L \otimes {b}^i_{-1/2}\ket{0}_R$ & $e^{i(m_2\pm m_4)}$ & {$2\times{\textbf{3}\oplus2\times\textbf{1}}$} & $2\times({\textbf{2}},{\textbf{2}})$ \\ \cline{2-5}
& $\tilde{b}^{\hat{\mu}}_{-1/2}\ket{0}_L \otimes \bar{{b}}^i_{-1/2}\ket{0}_R$ & $e^{-i(m_2\pm m_4)}$ & {$2\times{\textbf{3}\oplus2\times\textbf{1}}$} & $2\times({\textbf{2}},{\textbf{2}})$ \\ \cline{2-5}
& $\tilde{b}^{i}_{-1/2}\ket{0}_L \otimes {b}^{\hat{\mu}}_{-1/2}\ket{0}_R$ & $e^{i(m_1\pm m_3)}$ & {$2\times{\textbf{3}\oplus2\times\textbf{1}}$} & $2\times({\textbf{2}},{\textbf{2}})$ \\ \cline{2-5}
& $\bar{\tilde{b}}^{i}_{-1/2}\ket{0}_L \otimes {b}^{\hat{\mu}}_{-1/2}\ket{0}_R$ & $e^{-i(m_1\pm m_3)}$ & {$2\times{\textbf{3}\oplus2\times\textbf{1}}$} & $2\times({\textbf{2}},{\textbf{2}})$ \\ \cline{2-5}
& $\tilde{b}^{i}_{-1/2}\ket{0}_L \otimes {b}^j_{-1/2}\ket{0}_R$ & $e^{i(m_1\pm m_3)+i(m_2\pm m_4)}$ & {$4\times\textbf{1}$} & $4\times({\textbf{1}},{\textbf{1}})$ \\ \cline{2-5}
& $\tilde{b}^{i}_{-1/2}\ket{0}_L \otimes \bar{{b}}^j_{-1/2}\ket{0}_R$ & $e^{i(m_1\pm m_3)-i(m_2\pm m_4)}$ & {$4\times\textbf{1}$} & $4\times({\textbf{1}},{\textbf{1}})$ \\ \cline{2-5}
& $\bar{\tilde{b}}^{i}_{-1/2}\ket{0}_L \otimes {b}^j_{-1/2}\ket{0}_R$ & $e^{-i(m_1\pm m_3)+i(m_2\pm m_4)}$ & {$4\times\textbf{1}$} & $4\times({\textbf{1}},{\textbf{1}})$ \\ \cline{2-5}
& $\bar{\tilde{b}}^{i}_{-1/2}\ket{0}_L \otimes \bar{{b}}^j_{-1/2}\ket{0}_R$ & $e^{-i(m_1\pm m_3)-i(m_2\pm m_4)}$ & {$4\times\textbf{1}$} & $4\times({\textbf{1}},{\textbf{1}})$ \\ \hline\hline

R-R & $|a_{1,2}\rangle_L \otimes |a_{1,2}\rangle_R$ & $e^{\pm im_1\pm im_2}$ & {$4\times{\textbf{3}\oplus4\times\textbf{1}}$} & $4\times({\textbf{3}\oplus\textbf{1}},{\textbf{1}})$ \\ \cline{2-5}
& $|a_{1,2}\rangle_L \otimes |a_{3,4}\rangle_R$ & $e^{\pm im_1\pm im_4}$ & {$4\times{\textbf{3}\oplus4\times\textbf{1}}$} & $4\times({\textbf{2}},{\textbf{2}})$ \\ \cline{2-5}
& $|a_{3,4}\rangle_L \otimes |a_{1,2}\rangle_R$ & $e^{\pm im_3\pm im_2}$ & {$4\times{\textbf{3}\oplus4\times\textbf{1}}$} & $4\times({\textbf{2}},{\textbf{2}})$ \\ \cline{2-5}
& $|a_{3,4}\rangle_L \otimes |a_{3,4}\rangle_R$ & $e^{\pm im_3\pm im_4}$ & {$4\times{\textbf{3}\oplus4\times\textbf{1}}$} & $4\times({{\textbf{1},\textbf{3}\oplus\textbf{1}}})$ \\ \hline\hline

NS-R & $\tilde{b}^{\hat{\mu}}_{-1/2}\ket{0}_L \otimes |a_{1,2}\rangle_R$ & $e^{\pm im_2}$ & {$2\times{\textbf{4}\oplus4\times\textbf{2}}$} & $2\times({\textbf{3}\oplus\textbf{1}},{\textbf{2}})$ \\ \cline{2-5}
& $\tilde{b}^{\hat{\mu}}_{-1/2}\ket{0}_L \otimes |a_{3,4}\rangle_R$ & $e^{\pm im_4}$ & {$2\times{\textbf{4}\oplus4\times\textbf{2}}$} & $2\times({\textbf{2}},{\textbf{3}\oplus\textbf{1}})$ \\ \cline{2-5}
& $\tilde{b}^{i}_{-1/2}\ket{0}_L \otimes |a_{1,2}\rangle_R$ & $e^{i(m_1\pm m_3)\pm im_2}$ & {$4\times\textbf{2}$} & $4\times({\textbf{2}},{\textbf{1}})$ \\ \cline{2-5}
& $\tilde{b}^{i}_{-1/2}\ket{0}_L \otimes |a_{3,4}\rangle_R$ & $e^{i(m_1\pm m_3)\pm im_4}$ & {$4\times\textbf{2}$} & $4\times({\textbf{1}},{\textbf{2}})$ \\ \cline{2-5}
& $\bar{\tilde{b}}^{i}_{-1/2}\ket{0}_L \otimes |a_{1,2}\rangle_R$ & $e^{-i(m_1\pm m_3)\pm im_2}$ & $4\times\textbf{2}$ & $4\times({\textbf{2}},{\textbf{1}})$ \\ \cline{2-5}
& $\bar{\tilde{b}}^{i}_{-1/2}\ket{0}_L \otimes |a_{3,4}\rangle_R$ & $e^{-i(m_1\pm m_3)\pm im_4}$ & {$4\times\textbf{2}$} & $4\times({\textbf{1}},{\textbf{2}})$ \\ \hline\hline

R-NS & $|a_{1,2}\rangle_L \otimes {b}^{\hat{\mu}}_{-1/2}\ket{0}_R$ & $e^{\pm im_1}$ & {$2\times{\textbf{4}\oplus4\times\textbf{2}}$} & $2\times({\textbf{3}\oplus\textbf{1}},{\textbf{2}})$ \\ \cline{2-5}
& $|a_{3,4}\rangle_L \otimes {b}^{\hat{\mu}}_{-1/2}\ket{0}_R$ & $e^{\pm im_3}$ & {$2\times{\textbf{4}\oplus4\times\textbf{2}}$} & $2\times({\textbf{2}},{\textbf{3}\oplus\textbf{1}})$ \\ \cline{2-5}
& $|a_{1,2}\rangle_L \otimes {b}^i_{-1/2}\ket{0}_R$ & $e^{\pm im_1+i(m_2\pm m_4)}$ & {$4\times\textbf{2}$} & $4\times({\textbf{2}},{\textbf{1}})$ \\ \cline{2-5}
& $|a_{3,4}\rangle_L \otimes {b}^i_{-1/2}\ket{0}_R$ & $e^{\pm im_3+i(m_2\pm m_4)}$ & {$4\times\textbf{2}$} & $4\times({\textbf{1}},{\textbf{2}})$ \\ \cline{2-5}
& $|a_{1,2}\rangle_L \otimes \bar{{b}}^i_{-1/2}\ket{0}_R$ & $e^{\pm im_1-i(m_2\pm m_4)}$ & {$4\times\textbf{2}$} & $4\times({\textbf{2}},{\textbf{1}})$ \\ \cline{2-5}
& $|a_{3,4}\rangle_L \otimes \bar{{b}}^i_{-1/2}\ket{0}_R$ & $e^{\pm im_3-i(m_2\pm m_4)}$ & {$4\times\textbf{2}$} & $4\times({\textbf{1}},{\textbf{2}})$ \\ \hline
\end{tabular}
\captionsetup{width=.9\linewidth}
\caption{\textit{The spectrum of lowest excited string states including their orbifold charge and representations under the massless little group} SO$(3)\approx\text{SU}(2)$ \textit{and the massive little group} SO$(4)\approx\text{SU}(2)\times\text{SU}(2)$ \textit{in five dimensions.}}
\label{huiberttable}
\end{table}
\renewcommand{\arraystretch}{1}

We can now find the field content of our orbifold construction from \autoref{huiberttable}. As an example, take the string state $|a_{1}\rangle_L \otimes {b}^{\hat{\mu}}_{-1/2}\ket{0}_R$, which has orbifold charge $e^{im_1}$. First recall that we can always write $m_1 = {2\pi N_1}/{p}$ where $N_1$ is an integer and $p$ is the rank of the orbifold. To make the state invariant under the orbifold action, we add momentum along the $S^1$. States with momentum on the circle obtain a phase $e^{2\pi i n / p}$ with $n$ the number of modes. If we now choose $n=-N_1$, this phase becomes $e^{-2\pi i N_1 / p}$ which cancels exactly against the phase that the string state had before the addition of momentum. In other words, the state $|a_{1};-N_1,0\rangle_L \otimes {b}^{\hat{\mu}}_{-1/2}\ket{0}_R$ is invariant under the orbifold action and therefore survives in the spectrum. Here we use the notational convention to denote the momentum and winding numbers on the $S^1$ as $\ket{\;\;\;\:;n,w}$ on the left-moving vacuum.

At this point, we would like to mention that for the states in \autoref{huiberttable} it is always possible to find an integer-valued momentum number that cancels the phase due to the orbifold action. All mass parameters can be written as $m_i = 2\pi N_i/p$ with $N_i\in\mathbb{Z}$; any sum or difference of mass parameters can thus also be written as $2\pi/p$ times an integer. If we take this integer with the sign flipped as the momentum number, the total phase cancels.

Next, we can use the little group representations in \autoref{huiberttable} to determine what kind of fields the spectrum consists of. We return to the example state $|a_{1};-N_1,0\rangle_L \otimes {b}^{\hat{\mu}}_{-1/2}\ket{0}_R$. Due to the addition of momentum, the state has become massive with mass $|{N_1}/{\mathcal{R}}|$. From the table we then read off the representation as $({\textbf{3},\textbf{2})\oplus(\textbf{1}},{\textbf{2}})$, i.e. it corresponds to a massive gravitino and a massive dilatino.

We can rewrite the mass $|N_1/\mathcal{R}|$ slightly differently in order to make contact with the Scherk-Schwarz supergravity spectrum. We know that $N_1 = p\,m_1/2\pi$, and we know that the radius of the orbifold circle $\mathcal{R}$ and the radius of the Scherk-Schwarz circle $R$ are related by $\mathcal{R}=pR$. The mass of the state is therefore equal to $|m_1/2\pi R|$. Masses of this form are precisely what was found in \cite{Hull:2020byc}. We have included the relevant table of masses of the supergravity fields in appendix \ref{Ap D} (see also section 3.3 of \cite{Hull:2020byc}).

Some further care must be taken when comparing the representation of the monodromy matrix acting on the world-sheet fields or on the supergravity fields. For instance, the two gravitini from the NS-R sector (the $(\textbf{3},\textbf{2})$ and the $(\textbf{2},\textbf{3})$ in \autoref{huiberttable}) only pick up a Spin(4)$_R$ monodromy (with mass parameters $m_2, m_4$), but their spacetime chirality is opposite in six dimensions. Similarly the ones from the R-NS sector transform only under Spin(4)$_L$ and again have opposite $6D$ chirality. As supergravity fields, the $6D$ gravitini sit in the $(\textbf{4},\textbf{1})$ and $(\textbf{1},\textbf{4})$ representations of the R-symmetry $\text{USp}(4)_L\times\text{USp}(4)_R$, where $L,R$ now indicates the $6D$ chirality. Therefore, the monodromy matrix appearing in \cite{Hull:2020byc} (see e.g. eq. (3.39) in that paper) has grouped together $m_1$ and $m_2$, and similarly $m_3$ and $m_4$.

This systematic approach can be used to construct the entire field content coming from the lowest excited string states. Each of the states in \autoref{huiberttable} gives fields whose mass can be read off from their orbifold charge (it is always the absolute value of this linear combination of $m_i$'s times $p/2\pi \mathcal{R}$). The type of fields that this state gives can then be found from its massless or massive little group representation, depending on whether the aforementioned mass is zero (mod $2\pi$) or not. In this way, the entire supergravity spectrum from \cite{Hull:2020byc} can be reproduced. It is important to note however that the Scherk-Schwarz reduction can easily be carried out for any monodromy in Spin(5,5), but only for choices in Spin(4,4) we can compare to the orbifold picture. 

Finally, recall that we can build the Kaluza-Klein towers on the circle by adding $p\,\mathbb{Z}$ momentum modes to the ones that were added following the procedure above. This addition doesn't change the orbifold charge, so all of these states survive as well. The masses shift by $p\,\mathbb{Z}/\mathcal{R} = \mathbb{Z}/R$, e.g. the masses of the KK-tower on our example state become $|m_1/2\pi R+\mathbb{Z}/R|$. Again, this agrees with the supergravity calculation.

\subsection{Moduli spaces}
\label{moduli spaces}

We discussed in section \ref{closed string spectrum} the different patterns of supersymmetry breaking and gave examples of residual $\mathcal{N}=6,4,2$ or 0 supersymmetry. The resulting theories have massless modes and the massless scalars parametrize the moduli space of the orbifold. We illustrate this here for the examples given in the previous section, and we also provide some additional examples. In the absence of any twist, the moduli space in $D=6$ is $\text{SO}(5,5)/\text{SO}(5)\times \text{SO}(5)$ and has dimension 25. In $D=5$ with maximal supersymmetry, so in the absence of any twist, it is $\text{E}_{6(6)}/\text{USp}(8)$ and is 42-dimensional. Some of these scalars will become massive after the twist and the moduli space will become smaller. For generic values of the Scherk-Schwarz circle, it suffices to look only at the scalars in the untwisted sector to determine the moduli space. At special values of the radius, additional scalars and vectors can become massless, and we will discuss an example of this too.

\subsubsection{$\mathcal{N}=6$}

These theories can be realized by asymmetric orbifolds and the example we discussed was a $\mathbb{Z}_2$ orbifold with $\vec{m}=(\pi,0,0,0)$. The massless fields constitute the $\mathcal{N}=6, D=5$ supergravity multiplet \cite{Cremmer:1980gs}. It contains the graviton and 6 massless gravitini, with $\text{USp}(6)_R=\text{Sp}(3)$ R-symmetry. Next to the 15 vectors and 20 dilatini, we have 14 scalars that define the residual moduli space
\begin{equation}
{\cal M}_{\mathcal{N}=6}=\frac{\text{SU}^*(6)}{\text{USp}(6)}\ .
\end{equation}
6 of the 14 massless scalars descend from the NS-NS sector and 8 come from the R-R sector. 28 scalars have become massive in the Scherk-Schwarz reduction. 

Similarly, one can construct the moduli space of the $\mathcal{N}=6, \mathbb{Z}_3$ orbifold, with mass parameters $\vec{m}=(\tfrac{2\pi}{3},0,0,0)$, corresponding to $u=(0,0,0,0)$ and $\tilde u=(0,0,\frac{1}{3},\frac{1}{3})$. It has the same massless spectrum as the $p=2$ orbifold and hence the same moduli space, dictated by $\mathcal{N}=6$ supersymmetry. Asymmetric orbifolds with $\mathcal{N}=6$, in four dimensions, have also been constructed in \cite{Ferrara:1989nm} and more recently in \cite{bianchi2022perturbative,Bianchi:2008cj,Bianchi:2010aw}.

\subsubsection{$\mathcal{N}=4$}

For the $\mathbb{Z}_4$ (0,2) symmetric orbifold that we discussed in section \ref{sec:N=4 symmetric and asymmetric}, we have mass parameters $\vec{m}=(\tfrac{\pi}{2},\tfrac{\pi}{2},0,0)$. We saw that the massless spectrum consisted of the $D=5, \mathcal{N}=4$ supergravity multiplet, containing 1 real scalar and 6 vectors, together with three massless vector multiplets, each containing 5 scalars. Out of the 15 scalars from the vector multiplets, 9 come from the NS-NS sector, and are remaining geometric moduli, and 6 from the R-R sector. The total scalar manifold with the $16=1+15$ scalars including those of the three vector multiplets is
\begin{equation}
{\cal M}_{\mathcal{N}=4}^{\mathbb{Z}_4}=\text{SO}(1,1)\times \frac{\text{SO}(5,3)}{\text{SO}(5)\times \text{SO}(3)}\ .
\end{equation}
One gets the same moduli space for the symmetric $\mathbb{Z}_3$ and $\mathbb{Z}_6$ orbifolds, but the masses are different. The moduli space is fixed by supersymmetry and fits into the general structure of $\mathcal{N}=4, D=5$ supergravity coupled to $n$ $\mathcal{N}=4$ vector multiplets, where the scalars parametrize \cite{Awada:1985ep}
\begin{equation}
{\cal M}_{\mathcal{N}=4}=\text{SO}(1,1)\times \frac{\text{SO}(5,n)}{\text{SO}(5)\times \text{SO}(n)}\ .
\end{equation}
For the $\mathbb{Z}_4$ symmetric orbifold from above, we had $n=3$, but it is easy to get other values for $n$.
Take for instance the symmetric $\mathbb{Z}_2$ (0,2) orbifold with $\vec{m}=(\pi,\pi,0,0)$, corresponding to $\vec \alpha=(\pi,\pi,0,0)$ (see example \ref{non-chiral N=4}). We find from \autoref{huiberttable} that there are $26=1+25$ massless scalars with moduli space
\begin{equation}\label{nV=5}
{\cal M}_{\mathcal{N}=4}^{\mathbb{Z}_2}=\text{SO}(1,1)\times \frac{\text{SO}(5,5)}{\text{SO}(5)\times \text{SO}(5)}\ .
\end{equation}
Hence, for $\mathbb{Z}_2$ we get four accidental vector multiplets, and for $\mathbb{Z}_{3,4,6}$ we get two accidental vector multiplets. For the $\mathbb{Z}_2$ case, 3 vectors come from the NS-NS sector, and 8 from the R-R sector. As the gravity multiplet contains 6 vectors, we are left over with 5 vectors to form vector multiplets indeed. 

There is another way to get the moduli space \eqref{nV=5}, namely from the asymmetric $\mathbb{Z}_2$ (1,1) orbifold that we discussed in section \ref{subsec N=4 1,1} with $\vec{m}=(0,\pi,0,-\pi)$. The bosonic massless field content is the same, except that the fields have a different ten-dimensional origin. In this case, all the vectors and scalars come from the NS-NS sectors. Furthermore, for the asymmetric $\mathbb{Z}_2$ orbifold, we saw that for a certain value of the orbifold circle radius, we can obtain two extra massless vector multiplets coming from the twisted sector. In this case, the moduli space becomes
\begin{equation}\label{nV=7}
{\cal M}_{\mathcal{N}=4}^{\mathbb{Z}_2}=\text{SO}(1,1)\times \frac{\text{SO}(5,7)}{\text{SO}(5)\times \text{SO}(7)}\ .
\end{equation}

In general, for the $\mathcal{N}=4, D=5$ theories, if there are no accidental massless modes, the moduli space reduces to
\begin{equation}
{\cal M}_{\mathcal{N}=4}=\text{SO}(1,1)\times \frac{\text{SO}(5,1)}{\text{SO}(5)}\ .
\end{equation}
This seems only possible for asymmetric orbifolds, for instance an asymmetric $\mathbb{Z}_6\,(0,2)$  orbifold with $\vec{m}=(\tfrac{\pi}{3},\pi,0,0)$, corresponding to $\vec \alpha=(\frac{2\pi}{3},\frac{2\pi}{3},-\frac{\pi}{3},-\frac{\pi}{3})$, and with twist vectors $u=(0,0,\frac{1}{2},\frac{1}{2})$ and $\tilde u=(0,0,\frac{1}{6},\frac{1}{6})$. As we can see from \autoref{huiberttable}, there are 6 massless scalars, 2 from the NS-NS and 4 from the R-R sector. There is only one vector multiplet. 3 vectors come from the NS-NS sector, and 4 from the R-R sector. Another interesting case is the asymmetric $\mathbb{Z}_6\,(1,1)$  orbifold with $\vec{m}=(\tfrac{\pi}{3},0,\pi,0)$. It has the same $D=5$ massless field content and moduli space, but the ten-dimensional origin is different as all the bosonic fields, both scalars and vectors come from the NS-NS sector, and the theory contains no massless R-R fields.

The $\mathcal{N}=4$ orbifolds we discussed yield an odd number $n$ of massless vector multiplets, with $n\leq 21$ (all massless tensors are dualized into vectors), and the corresponding unbroken gauge group (at generic points in the moduli space) is U$(1)^n$. First of all, this result agrees with the upper bound on the rank of the gauge group $r_G=26-D$ given in \cite{montero2021cobordism,kim2020four}. Secondly, it should be pointed out that in all the examples we gave $n\in 2\mathbb{Z}+1$. Furthermore, it is easy to derive from \autoref{huiberttable} that all orbifolds give an odd number of untwisted vector multiplets. Regarding the twisted sectors, we saw that in $\mathcal{N}=4\,(1,1)$ theories (where we break all the left, or right-moving supersymmetries) it is possible to tune the circle radius such that some of the twisted vector multiplets become massless. In this massless limit each complex vector multiplet will give two real vector multiplets. As a result, regardless of the specific orbifold, we can only get an even number of real vector multiplets coming from the twisted sectors. Therefore, combining both untwisted and twisted sectors we can see that $n\in 2\mathbb{Z}+1$ is a generic feature of our orbifolds.

The appearance of only an odd number of vector multiplets seems to be a characteristic of a broader class of string constructions with 16 supersymmetries in $5D$. Theories with 16 supersymmetries in $D>6$ have been studied extensively (see e.g. \cite{de2001triples,bedroya2022compactness} and references therein). In addition, $\mathcal{N}=(1,1)$ theories in $6D$ were recently classified in \cite{fraiman2023unifying}.
Dimensional reduction of the various higher dimensional theories given in these references always yields an odd number of vector multiplets in $5D$, which supports our findings\footnote{Similar results were found in the context of four dimensional asymmetric orbifolds in \cite{Blumenhagen:2016rof}. In four dimensions, the number of vector multiplets in $\mathcal{N}=4$ was always found to be even, consistent with odd number in five dimensions.}. Consequently, cases with $n \in 2\mathbb{Z}$ appear not to be part of the string landscape of $\mathcal{N}=4, D=5$ Minkowski vacua\footnote{
In \cite{dabholkar1999string}, an asymmetric orbifold was proposed which appeared to give pure supergravity with no vector multiplets. However, the monodromy in this example involves a single T-duality and so the construction is not a quotient by a symmetry of the IIB string theory: the bosonic action of the monodromy is in  $\text{O}(d,d)$, not  in $\text{SO}(d,d)$. Such a quotient seems problematic and is outside the class of constructions we discuss here.}
They do however seem to appear as moduli spaces in AdS$_5$ vacua in the context of holography, see e.g. section 7 of \cite{Freedman:1999gp} (and earlier work obtained in \cite{Khavaev:1998fb}) for the case of two vector multiplets, $n=2$.

\subsubsection{$\mathcal{N}=2$}

$\mathcal{N}=2$ in five dimensions is the minimum amount of supersymmetry, namely eight real supersymmetries. In this case, only one mass parameter is set to zero, and therefore, all these orbifolds are necessarily asymmetric. In the absence of any accidental massless modes, one obtains the gravity multiplet (containing the metric and a vector), coupled to two vector multiplets. For the $\mathbb{Z}_2$ orbifold discussed in section \ref{sec:N=2 asymmetric orbifold} we have $\vec{m}=(\pi,\pi,0,-\pi)$. This orbifold gives twelve additional vector multiplets. In total there are 14 real scalars, one per each massless vector multiplet. Notice that the massless bosonic field content is exactly the same as in the asymmetric $\mathcal{N}=6, \mathbb{Z}_2$ orbifold, consisting of 6 NS-NS scalars and 8 R-R scalars. The moduli space is again
\begin{equation}
{\cal M}_{\mathcal{N}=2}^{\mathbb{Z}_2}=\frac{\text{SU}^*(6)}{\text{USp}(6)}\ .
\end{equation}
This is one of the magic square supergravities of \cite{Gunaydin:1983rk}.

We can also easily determine the moduli space of a $\mathbb{Z}_4$ orbifold with $\vec{m}=(\tfrac{\pi}{2},\tfrac{\pi}{2},\tfrac{\pi}{2},0)$. The corresponding twist vectors are given by $u=(0,0,\frac{1}{4},\frac{1}{4})$ and $\tilde{u}=(0,0,\frac{1}{2},0)$ and satisfy the quantization conditions \eqref{psum_u_even}. The $\alpha$'s are given by 
$\vec \alpha=(\frac{3\pi}{4},\frac{\pi}{4},\frac{\pi}{4},-\frac{\pi}{4})$ (cf. example \ref{non-chiral N=2}) and do not satisfy \eqref{quantalpha}. This could indicate that this orbifold is not allowed and be part of the swampland. It remains to be seen whether this is due to the restricted conjugacy classes we did consider or otherwise it would have to show up that the integrality condition or modularity of the partition function is not satisfied. We leave this for future research. If for now we assume it is a consistent orbifold, it would give a massless spectrum consisting of eight vector multiplets coupled to supergravity, and no hypermultiplets. The most likely candidate for the (classical) moduli space is another of the magic square supergravities,
\begin{equation}
    {\cal M}_{\mathcal{N}=2}^{\mathbb{Z}_4}=\frac{\text{SL}(3,\mathbb{C})}{\text{SU}(3)}\ ,
\end{equation}
which is known to be a truncation of $\mathcal{N}=8, D=5$ supergravity if we break the $\mathcal{N}=8$ supermultiplet into $\mathcal{N}=2$ supermultiplets and truncate the spin-$\tfrac{3}{2}$ and matter multiplets  \cite{Gunaydin:1983rk}. In Scherk-Schwarz, we do not truncate these multiplets; instead, these become lifted and massive. 

In both of the above cases, there are no hypermultiplets and the dilaton sits in a vector multiplet. Hence, the classical moduli space can be quantum corrected. Of course, as we saw in section \ref{sec:N=2 asymmetric orbifold} it is possible to obtain massless hypermultiplets coming from the twisted sectors by tuning the orbifold circle radius. However, at a generic point of the moduli space the orbifolds discussed above are free of massless hypermultiplets. Similar models of hyper-free superstrings were constructed in \cite{Dolivet:2007sz}. It is interesting that we can produce two out of the four magical supergravity theories of \cite{Gunaydin:1983rk,Gunaydin:1983bi}, but apparently not the other two (with 5 and 26 vector multiplets). 

Another example is the $\mathbb{Z}_6$ orbifold defined by  $\vec{m}=(\pi,\tfrac{2\pi}{3},\pi,0)$. This corresponds to $\vec \alpha=(\frac{4\pi}{3},\frac{\pi}{3},\frac{2\pi}{3},-\frac{\pi}{3})$ and twist vectors $u=(0,0,\frac{1}{3},\frac{1}{3})$ and $\tilde u=(0,0,1,0)$. This gives six vector multiplets coupled to $\mathcal{N}=2$ supergravity, and again no hypers. The moduli space does not belong to the magic square. All the scalars come from the NS-NS sector and the most likely candidate is the factorizable moduli space
\begin{equation}
\frac{\text{SO}(5,1)}{\text{SO}(5)}\times \text{SO}(1,1)\ ,
\end{equation}
as this can be obtained from a truncation of the $\mathcal{N}=6$ theory \cite{Gunaydin:1983rk}.

Hypermultiplets in the untwisted sector can be produced by other types of orbifolds, namely when three of the non-vanishing mass parameters add up to 0 mod 2$\pi$. For instance, consider the $\mathbb{Z}_4$ orbifold with $\vec{m}=(\pi,\tfrac{\pi}{2},\tfrac{\pi}{2},0)$. It leads to four vector multiplets and two hypermultiplets. In general, one always gets an even number of hypermultiplets as one can easily derive from \autoref{huiberttable}.

Related interesting work on asymmetric orbifolds and non-geometric string backgrounds with few moduli in four dimensions, preserving $\mathcal{N}=2$ supersymmetry, can be found e.g. in
\cite{Anastasopoulos:2009kj,Condeescu:2013yma,Hull:2017llx,Gautier:2019qiq,Israel:2013wwa}.

What all the models seem to have in common is that the total number of vectors $r$ is always odd. In $\mathcal{N}=6$ we have $r=15$. In $\mathcal{N}=4$, the number of vectors in the gravity multiplet is six, so this means always an odd number of vector fields. In $\mathcal{N}=2$, there is one vector in the gravity multiplet, and we find always an even number of vector multiplets. It would be nice to relate these observations to the work in higher dimensions in $D\geq 7$ and the string lamppost principle \cite{montero2021cobordism}.

\subsubsection{$\mathcal{N}=0$ and supertraces}
\label{n=0 and supertraces}
 
For $\mathcal{N}=0$, all four mass parameters are switched on and the complete R-symmetry group $\text{USp}(4)_L\times \text{USp}(4)_R=\text{Spin}(5)_L\times \text{Spin}(5)_R$ is broken. The states neutral under the Scherk-Schwarz twist are the singlets, i.e. with zero orbifold charge. In the absence of accidental modes these are 3 vectors, 2 scalars, and the graviton, all coming from the NS-NS sector. All fermions become massive, and hence all supersymmetry is broken. So we get the Einstein-Hilbert action for gravity coupled to 3 vectors and 2 scalars. 

A particular model breaking all supersymmetry is the $\mathbb{Z}_2$ symmetric orbifold with all masses turned on and equal to $\pi$. The twist vectors are given by $u=\tilde{u}=(0,0,1,0)$ and $\vec{\alpha}=(2\pi,0,0,0)$. There are now many accidental massless modes and in fact all bosons become massless, while all fermions become massive (cf. \autoref{huiberttable}). In fact, this $\mathbb{Z}_2$ orbifold is generated by 
$(-1)^{F_s}$ combined with a half-shift on $S^1$.
The bosonic field content is therefore the same as the one from $\mathcal{N}=8, D=5$ supergravity, and the classical moduli space is $\text{E}_{6(6)}/\text{USp}(8)$. The orbifold is tachyon-free above the critical radius $\mathcal{R}_*=2{\sqrt{2\alpha'}}$. The fact that we combined $(-1)^{F_s}$ with a shift on the circle makes this model different from type 0B string theory compactified on $T^4\times S^1$, which suffers from a tachyon for any value of the radius of $S^1$.

The Scherk-Schwarz reductions we considered generate positive definite scalar potentials at the classical level. In the presence of partial supersymmetry, the corresponding Minkowski vacua are perturbatively stable, and hence the cosmological constant vanishes. When all of the supersymmetries are spontaneously broken, a one-loop cosmological constant, which is determined by the minimum of the one-loop effective potential, may be non-vanishing. 

The effective potential can be expressed in terms of various {supertraces}, as it was first shown in \cite{coleman1973radiative}. Supertraces are defined as weighted sums over the masses of all fields in the spectrum of the theory, that is
\begin{equation}
  \text{Str}M^{2\beta}=   N_{\phi}(M_{\phi})^{2\beta}-2N_{\chi}(M_{\chi})^{2\beta}+3N_{B_{\mu \nu}}(M_{B_{\mu \nu}})^{2\beta}
    +4N_{A_{\mu}}(M_{A_{\mu}})^{2\beta}-6N_{\psi_{\mu}}(M_{\psi_{\mu}})^{2\beta},
    \label{explicit}
\end{equation}
where $\beta > 0$ is an integer\footnote{$\text{Str}M^0$ is equal to bosonic minus fermionic degrees of freedom and is always zero.} and we denote by $N_{\text{field}}$ the number of fields with mass $M_{\text{field}}$. Each field is multiplied by the corresponding (massive) degrees of freedom in five dimensions and fermion fields appear with a minus sign.

Regarding supersymmetric theories, all supertraces vanish on Minkowski vacua \cite{zumino1975supersymmetry}. This is a result of bose-fermi degeneracy. In supersymmetric theories all fields fit in supermultiplets and as a consequence, all supertraces are identically zero. For non-supersymmetric theories the situation is more complicated, due to the absence of bose-fermi degeneracy. However, it was noticed in \cite{ferrara1979mass,cremmer1979spontaneously} that for Scherk-Schwarz reductions of $\mathcal{N}=8$ supergravities with fully broken supersymmetry
    \begin{equation}
   \text{Str}M^2=\text{Str}M^4=\text{Str}M^6=0\,,\qquad \text{Str}M^8\neq 0\,.
   \label{first supertraces}
\end{equation}
The same result was later obtained in \cite{Dall'Agata:2012cp} for any gauging of $\mathcal{N}=8$ supergravity.

Here we also perform the supertrace calculation for an $\mathcal{N}=0$ supergravity theory with all mass parameters non-zero. Our result is consistent with \eqref{first supertraces}. Furthermore, we explicitly find the value of $\text{Str}M^8$ to be
\begin{equation}
    \text{Str}M^8=40320\left(m_1\,m_2\, m_3\, m_4\right)^2\,,
    \label{str8}
\end{equation}
which is positive definite. As shown in \cite{Dall'Agata:2012cp}, $\text{Str}M^2=\text{Str}M^4=\text{Str}M^6=0$ and $\text{Str}M^8>0$ imply that the one-loop effective potential is negative definite. 

It would be interesting to see how this calculation extends to the full string theory spectrum. For this one needs to compute the full partition function which is generating the cosmological constant. It can happen that for specific orbifolds, one can still obtain a vanishing cosmological constant at one-loop, see e.g. \cite{kachru1999vacuum,kachru1998self,kachru1999vanishing,harvey1998string,shiu1999bose,angelantonj1999non,Satoh:2015nlc,groot2020tension}, but it is not certain if the cosmological constant vanishes at higher loops. 

\section{Swampland examples}
\label{sec:bad examples}

The swampland programme deals with effective actions that cannot be consistently lifted to string theory or quantum gravity in general \cite{vafa2005string}. In our case, the effective supergravity actions are Scherk-Schwarz reductions to five spacetime dimensions. They yield $\mathcal{N}=8$ gauged supergravity in five dimensions with vacua that spontaneously break supersymmetry. The $\mathcal{N}=8$ supergravity multiplet in five dimensions contains as bosonic fields a graviton, 27 vector fields and 42 real scalars. The gauge group was discussed in detail in \cite{Hull:2020byc}, where it was shown how the structure constants of the gauge algebra are determined by the twist matrix. In the class of twists we considered in this paper (with the twist matrix (conjugate to) an element in the R-symmetry group  $\text{Spin}(5)_L\times \text{Spin}(5)_R$), the potential and the structure constants depend on four mass parameters $m_i$. The potential can be written as (see e.g. eq. (3.49) in \cite{Hull:2020byc})
\begin{equation}
    V(\mathcal{H})=\frac{1}{4}e^{-{\sqrt{8/3}}\phi_5}\Tr\Big[M^2+M^T\mathcal{H}^{-1}M\mathcal{H}\Big]\ ,
\end{equation}
where $\phi_5$ is the KK scalar coming from the metric in six dimensions, $\mathcal{H}\in \text{Spin}(5,5)$ parametrizes the 42 scalar fields, and $M\in \mathfrak{so}(5,5)$ is the mass matrix containing the mass parameters $m_i$ (see \cite{Hull:2020byc} for more details).

For this potential to belong to the string landscape, in particular the landscape of freely acting orbifolds, a number of conditions need to be satisfied. First of all, the monodromy matrix  should be inside the T-duality group, otherwise there is no obvious CFT description on the worldsheet in terms of an orbifold, as we discussed in detail in section \ref{sec:Orbifold constructions}.

Secondly, in supergravity the mass parameters $m_i$ are continuous parameters, while for a well defined orbifold the mass parameters must be quantized (see section \ref{sec:Orbifold constructions}) and possible accidental massless modes arise when some of the masses add up to a multiple of $2\pi$. In the case of symmetric orbifolds the quantization of the mass parameters is the only constraint for a consistent uplift, since modular invariance of the partition function is ensured. For asymmetric orbifolds, however, modular invariance is not guaranteed and this   restricts the possible supergravity theories that can be consistently uplifted to string theory (as freely acting orbifolds). Furthermore, even if modular invariance is achieved, one should carefully examine the degeneracy of states in the twisted sectors. For a sensible theory this degeneracy must be an integer number. As shown in \cite{bianchi2022perturbative}, this additional integrality constraint can exclude a seemingly consistent model.  

For example, consider the asymmetric $\mathbb{Z}_4$, $\mathcal{N}=6$ orbifold with twist vectors $\tilde{u}=\left(0,0,\tfrac{1}{4},-\tfrac{1}{4}\right)$ and $u=(0,0,0,0)$, or in terms of the mass parameters $\vec{m}=(0,0,\tfrac{\pi}{2},0)$. For this orbifold, the invariant momentum sublattice is $\Lambda_R=(A_1)^4$ and the degeneracy of states in the $k=1$ sector is $D(1)=\tfrac{1}{2}$ \cite{bianchi2022perturbative}. Consequently, this model is not allowed. Note that this orbifold is also excluded by the T-duality arguments presented in section \ref{sec:quantization conditions}. Specifically, for this orbifold the values of the $\alpha$'s are $\vec{\alpha}=(\tfrac{\pi}{4},-\tfrac{\pi}{4},\tfrac{\pi}{4},-\tfrac{\pi}{4})$, which are not valid (cf. example \ref{n=6 allows only p=2 or 3}). Therefore, the supergravity potential with this choice of mass matrix is in the swampland.

Now, let us take a chiral $\mathbb{Z}_4$, $\mathcal{N}=4\,(1,1)$ orbifold with twist vectors $\tilde{u}=\left(0,0,\tfrac{1}{4},\tfrac{3}{4}\right)$ and $u=(0,0,0,0)$, or $\vec{m}=(\pi,0,-\tfrac{\pi}{2},0)$. Although this orbifold breaks 16 supersymmetries, its action on the torus coordinates is the same as in the previous $\mathbb{Z}_4$, $\mathcal{N}=6$ example, i.e. the SO$(4,4)$ element is unchanged. Then the background is the same and the degeneracy of states in the $k=1$ sector is once again $D(1)=\tfrac{1}{2}$. This is another example of an apparently consistent orbifold that is excluded by the integrality constraint. Once more, the values of the $\alpha$'s for this orbifold are invalid, as we find $\vec{\alpha}=(\tfrac{\pi}{4},\tfrac{3\pi}{4},\tfrac{\pi}{4},\tfrac{3\pi}{4})$ (cf. example \ref{n=4 chiral orbis}). The chosen mass parameters therefore again yield a scalar potential that is in the swampland. 

Another interesting example is the $\mathbb{Z}_4$, $\mathcal{N}=2$ orbifold with twist vectors $\tilde{u}=\left(0,0,\tfrac{1}{4},-\tfrac{1}{4}\right)$ and $u=(0,0,0,1)$, or $\vec{m}=(0,\pi,\tfrac{\pi}{2},-\pi)$. The only difference of this orbifold compared to the $\mathbb{Z}_4$, $\mathcal{N}=6$ that we discussed previously, is that the twist vector $u$ generates an action $(-1)^{F_R}$, and one can show that this does not affect modular invariance\footnote{We have explicitly illustrated this in the $\mathbb{Z}_2$ examples of sections \ref{sec:N=4 symmetric and asymmetric} and \ref{sec:N=2 asymmetric orbifold}.}. Again, this orbifold suffers from the same integrality problem as the two models above. This is yet another example of a seemingly consistent orbifold which is not allowed by the integrality condition. In addition, this orbifold is not allowed by the quantization of the $\alpha$'s, since we find the invalid values $\vec{\alpha}=(\tfrac{\pi}{4},\tfrac{3\pi}{4},\tfrac{\pi}{4},-\tfrac{5\pi}{4})$ (see example \ref{non-chiral N=2}).

Similarly, one can start with the naively consistent $\mathbb{Z}_6$, $\mathcal{N}=6$ orbifold with twist vectors $\tilde{u}=\left(0,0,\tfrac{1}{6},-\tfrac{1}{6}\right)$ and $u=(0,0,0,0)$, which is also excluded due to the integrality constraint\cite{bianchi2022perturbative}. By repeating the same steps as above, one can show that the $\mathbb{Z}_6$, $\mathcal{N}=4$ orbifold with $\tilde{u}=\left(0,0,\tfrac{1}{6},\tfrac{5}{6}\right)$, $u=(0,0,0,0)$ and the $\mathbb{Z}_6$, $\mathcal{N}=2$ orbifold with $\tilde{u}=\left(0,0,\tfrac{1}{6},-\tfrac{1}{6}\right)$, $u=(0,0,0,1)$ are also not allowed by the integrality condition. Moreover, all these orbifolds are not allowed by the quantization of the $\alpha$'s (for the above orbifolds the $\alpha$'s are proportional to $\pi/6$). 

In all of the above examples, the particular choices of mass parameters are perfectly well-defined at the level of supergravity actions but they cannot be lifted to string theory orbifolds. Unless there are other stringy constructions than orbifolds, all these examples belong to the swampland. It remains to be seen if these other stringy constructions then are consistent with the chosen mass parameters.

\section{Conclusions and outlook}
\label{conclusion}

In this paper we have explored a landscape of freely acting orbifolds of type IIB on $(T^4\times S^1)/\mathbb{Z}_p$. Our techniques can be easily carried over to type IIA string theory and straightforwardly extended to higher dimensional tori. Freely acting orbifolds on $(T^4\times S^1)/\mathbb{Z}_p$ lead to five-dimensional effective actions with spontaneous (partial) supersymmetry breaking in Minkowski vacua. They are significantly  different from orbifolds by the same $T^4$ symmetry but without the shift on the extra circle, so that the orbifolds without shifts
have fixed points. It is nevertheless interesting to compare the two, as we did in in some examples. 
The main difference is the appearance of a scalar potential on the moduli space in the freely acting case, and the properties of this potential are interesting to study in the context of the landscape and moduli stabilization. In the supergravity description, these models correspond to Scherk-Schwarz reductions and the potential contains mass parameters $m_i$ that are continuous. In the orbifold picture, the masses are quantized and rational, $\frac{m_i}{2\pi}=\frac{N_i}{p}$, but not all integers $N_i$ and $p$ are allowed. In the space of string theories parametrized by these mass parameters, only a discrete and finite set arises and defines a landscape of freely acting orbifolds. 
In the models we considered, the highest possible value for $p$ we found is $p=24$. Here we have  only studied duality twists with monodromy matrices that are within the T-duality group Spin(4,4;$\mathbb{Z}$) and conjugate to the R-symmetry group. We
  have not discussed shifts on the dual circle or on multiple circles. It would be interesting to generalize to these cases and to     U-duality twists 
   to learn more about  the structure of this discrete, finite set of string compactifications and to investigate whether there is a   relationship  to the tamed structure found in \cite{Grimm:2021vpn}.

There are many directions in which to generalize and extend our work. One direction is to understand better the origin of the integrality constraints arising in the partition function of asymmetric orbifolds from the point of view of the duality twists and the possible monodromies. We discussed examples of $\mathbb{Z}_p$ orbifolds with $\mathcal{N}=6,4$ and 2 which have modular invariant partition functions but where for $p=4,6$ the integrality constraints are not satisfied. Another direction is to investigate the properties of the string orbifolds with completely broken supersymmetry. We have shown that they can be made free of tachyons, but it would be interesting to see if these models are perturbatively stable.

In the massless sector, we found some interesting restrictions on the number of vector multiplets arising in the class of orbifolds we consider.
For instance, for orbifolds with sixteen supersymmetries ($\mathcal{N}=4, D=5$) we find that only an odd number of vector multiplets arises, whereas generic $\mathcal{N}=4, D=5$ supergravities allow any number of vector multiplets. We are not aware of any anomaly conditions that could explain this. Moreover, the lift to six dimensions yields back the anomaly free $\mathcal{N}=8$ theory. It would be interesting to see if these observations can be promoted to a concrete conjecture on the full string landscape with sixteen supersymmetries.

We have limited ourselves to perturbative aspects of string theories with T-duality twists. Ongoing work will deal with the BPS spectrum of D-brane states in these T-folds. A particularly interesting case is to understand the fate of the D1-D5 brane system in the various orbifold theories and its applications to black holes and dual CFT. This was our original motivation laid out in \cite{Hull:2020byc} and will be studied further in \cite{GHSV}.

\section*{Acknowledgements}

It is a pleasure to acknowlegde discussions with Massimo Bianchi and Thomas Grimm. We also thank Eric Marcus for his collaboration in the initial stages of the project.
CMH was supported by the STFC Consolidated Grant ST/T000791/1 and a Royal
Society Leverhulme Trust Senior Research Fellowship.

\appendix

\section{Group theory}\label{app: group theory}

Here we collect some (known) group theoretical information and some conventions that are relevant for the main text.

\subsection{Two frames for SO(\textit{d,d})}\label{app: eta and tau frame}

Canonically, groups of the form SO$(d,d)$ consist of $2d$-dimensional matrices $A$ that satisfy the relation
\begin{equation}\label{etaframe}
    A\,\eta\,A^T = \eta \,,
\end{equation}
where $\eta$ is the indefinite metric
\begin{equation}
    \eta = \begin{pmatrix}
    \mathbbm{1} & 0 \\
    0 & -\mathbbm{1}
    \end{pmatrix} \,.
\end{equation}
Here and everywhere else in this subsection we adopt the notation that $\mathbbm{1}$ and $0$ denote $d$-dimensional blocks that make up a $2d$-dimensional matrix.

There is a different way of constructing SO$(d,d)$, namely as the group of matrices $\tilde{A}$ satisfying
\begin{equation}\label{tauframe}
    \tilde{A}\,\tau\,\tilde{A}^T = \tau \,,
\end{equation}
where
\begin{equation}
    \tau = \begin{pmatrix}
    0 & \mathbbm{1} \\
    \mathbbm{1} & 0
    \end{pmatrix} \,.
\end{equation}
It is straightforward to see that these two ways of constructing SO$(d,d)$ are equivalent, or, one might say, that the groups consisting of the matrices $A$ and $\tilde{A}$ are isomorphic. For example, the map
\begin{equation}\label{conjugation by X}
    A \;\;\rightarrow\;\; \tilde{A} = X\,A\,X^{-1} \,,
\end{equation}
where
\begin{equation}
    X = \frac{1}{\sqrt{2}}\,\begin{pmatrix}
    \mathbbm{1} & \mathbbm{1} \\
    \mathbbm{1} & -\mathbbm{1}
    \end{pmatrix} \,,
\end{equation}
takes matrices satisfying \eqref{etaframe} to matrices satisfying \eqref{tauframe}. It can be checked that this map is a proper isomorphism between what we call \say{SO$(d,d)$ in $\eta$-frame} and \say{SO$(d,d)$ in $\tau$-frame}.

We encounter both $\eta$ and $\tau$ frame in this work, and we will refer to them by these names.

\subsection{The isomorphism $\mathfrak{so}(4)\cong\mathfrak{su}(2)\oplus\mathfrak{su}(2)$}

We first discuss the isomorphism between compact Lie algebras before moving on to the non-compact version. The algebra $\mathfrak{so}(4)$ is spanned by six $4d$ anti-symmetric matrices. We denote the generators of this algebra by $M_{ij}$ where the $i,j$ denote which entries in the matrix are non-zero. We have
\begin{equation}
    M_{12} = \begin{pmatrix}
    0 & -1 & 0 & 0 \\
    1 & 0 & 0 & 0 \\
    0 & 0 & 0 & 0 \\
    0 & 0 & 0 & 0
    \end{pmatrix} \;,\qquad
    M_{13} = \begin{pmatrix}
    0 & 0 & -1 & 0 \\
    0 & 0 & 0 & 0 \\
    1 & 0 & 0 & 0 \\
    0 & 0 & 0 & 0
    \end{pmatrix} \;,\qquad \ldots
\end{equation}
Similarly we construct $M_{14},M_{23},M_{24},M_{34}$. We can compactly write the commutators between these generators as
\begin{equation}\label{commutation relations so4}
    [M_{ij},M_{kl}] = \delta_{ik} M_{jl} - \delta_{il} M_{jk} - \delta_{jk} M_{il} + \delta_{jl} M_{ik} \,,
\end{equation}
where it is understood that $M_{ii}=0$ and $M_{ij}=-M_{ji}$.

The algebra $\mathfrak{su}(2)$ is spanned by three $2d$ anti-Hermitian traceless matrices. We choose the generators
\begin{equation}\label{generators su2}
    N_1 = \frac{1}{2}\begin{pmatrix}
    0 & -1 \\
    1 & 0
    \end{pmatrix} \;, \qquad
    N_2 = \frac{i}{2}\begin{pmatrix}
    0 & 1 \\
    1 & 0
    \end{pmatrix} \;, \qquad
    N_3 = \frac{i}{2}\begin{pmatrix}
    1 & 0 \\
    0 & -1
    \end{pmatrix} \;, \qquad
\end{equation}
for this algebra. This yields the commutator
\begin{equation}\label{commutation relations su2}
    [N_I,N_J] = - \,\varepsilon_{IJK} \,N_K \;,
\end{equation}
where we choose $\varepsilon_{IJK}$ such that $\varepsilon_{123} = 1$.

We have now established sufficient notation to write down the isomorphism $\mathfrak{so}(4)\cong\mathfrak{su}(2)\oplus\mathfrak{su}(2)$. We choose this to be the map
\begin{equation}\label{compact isomorphism}
\begin{aligned}
    M_{12} \;\;&\rightarrow\;\; \big(N_1,\,N_1\big) \\
    M_{13} \;\;&\rightarrow\;\; \big(N_3,\,N_3\big) \\
    M_{14} \;\;&\rightarrow\;\; \big(N_2,\,-N_2\big) \\
    M_{23} \;\;&\rightarrow\;\; \big(N_2,\,N_2\big) \\
    M_{24} \;\;&\rightarrow\;\; \big(\!-\!N_3,\,N_3\big) \\
    M_{34} \;\;&\rightarrow\;\; \big(N_1,\,-N_1\big) \;,
\end{aligned}
\end{equation}
which can readily be checked to preserve the commutator, making it a proper isomorphism.

\subsection{The isomorphism $\mathfrak{so}(2,2)\cong\mathfrak{sl}(2)\oplus\mathfrak{sl}(2)$}\label{isomorphism so22}

Here we repeat the analysis of the previous section for the maximally non-compact version of the same isomorphism. On the $\mathfrak{so}(2,2)$ side, this means that we take four of the six generators to be symmetric rather than anti-symmetric. We choose the following generators to span the algebra\footnote{Note that these generators span the algebra $\mathfrak{so}(2,2)$ in $\eta$-frame, using the language from appendix \ref{app: eta and tau frame}.}:
\begin{equation}
\begin{alignedat}{6}
    \tilde{M}_{12} &= \begin{pmatrix}
    0 & -1 & 0 & 0 \\
    1 & 0 & 0 & 0 \\
    0 & 0 & 0 & 0 \\
    0 & 0 & 0 & 0
    \end{pmatrix} \;,\qquad
    &\tilde{M}_{13} &= \begin{pmatrix}
    0 & 0 & 1 & 0 \\
    0 & 0 & 0 & 0 \\
    1 & 0 & 0 & 0 \\
    0 & 0 & 0 & 0
    \end{pmatrix} \;,\qquad
    &\tilde{M}_{14} &= \begin{pmatrix}
    0 & 0 & 0 & 1 \\
    0 & 0 & 0 & 0 \\
    0 & 0 & 0 & 0 \\
    1 & 0 & 0 & 0
    \end{pmatrix} \;, \\
    \tilde{M}_{23} &= \begin{pmatrix}
    0 & 0 & 0 & 0 \\
    0 & 0 & 1 & 0 \\
    0 & 1 & 0 & 0 \\
    0 & 0 & 0 & 0
    \end{pmatrix} \;,\qquad
    &\tilde{M}_{24} &= \begin{pmatrix}
    0 & 0 & 0 & 0 \\
    0 & 0 & 0 & 1 \\
    0 & 0 & 0 & 0 \\
    0 & 1 & 0 & 0
    \end{pmatrix} \;,\qquad
    &\tilde{M}_{34} &= \begin{pmatrix}
    0 & 0 & 0 & 0 \\
    0 & 0 & 0 & 0 \\
    0 & 0 & 0 & -1 \\
    0 & 0 & 1 & 0
    \end{pmatrix} \;.
\end{alignedat}
\end{equation}
The algebra $\mathfrak{sl}(2)$ is spanned by traceless real matrices. The generators that we choose are equal to the ones we chose for the $\mathfrak{su}(2)$ algebra \eqref{generators su2}, but without the factors of $i$ in front. We define our three generators as
\begin{equation}
    \tilde{N}_1 = \frac{1}{2}\begin{pmatrix}
    0 & -1 \\
    1 & 0
    \end{pmatrix} \;, \qquad
    \tilde{N}_2 = \frac{1}{2}\begin{pmatrix}
    0 & 1 \\
    1 & 0
    \end{pmatrix} \;, \qquad
    \tilde{N}_3 = \frac{1}{2}\begin{pmatrix}
    1 & 0 \\
    0 & -1
    \end{pmatrix} \;. \qquad
\end{equation}

These algebras satisfy similar commutation relations as their compact counterparts, \eqref{commutation relations so4} and \eqref{commutation relations su2}, with some signs changed. In fact the signs are changed in such a way that the isomorphism $\mathfrak{so}(2,2)\cong\mathfrak{sl}(2)\oplus\mathfrak{sl}(2)$ can be written down in the exact same way as in \eqref{compact isomorphism}. We simply add the tildes to indicate that we consider the maximally non-compact generators:
\begin{equation}\label{noncompact isomorphism}
\begin{aligned}
    \tilde{M}_{12} \;\;&\rightarrow\;\; \big(\tilde{N}_1,\,\tilde{N}_1\big) \\
    \tilde{M}_{13} \;\;&\rightarrow\;\; \big(\tilde{N}_3,\,\tilde{N}_3\big) \\
    \tilde{M}_{14} \;\;&\rightarrow\;\; \big(\tilde{N}_2,\,-\tilde{N}_2\big) \\
    \tilde{M}_{23} \;\;&\rightarrow\;\; \big(\tilde{N}_2,\,\tilde{N}_2\big) \\
    \tilde{M}_{24} \;\;&\rightarrow\;\; \big(\!-\!\tilde{N}_3,\,\tilde{N}_3\big) \\
    \tilde{M}_{34} \;\;&\rightarrow\;\; \big(\tilde{N}_1,\,-\tilde{N}_1\big) \;.
\end{aligned}
\end{equation}
Again it is straightforward to check that this map preserves the commutator.

\subsection{Embedding rotations in Spin(4,4)}
\label{Ap:embedding}

One of the main motivations for adding in the previous sub-appendices was to be able to properly embed rotations in various subgroups of $\spinfourfour$ into an $8d$ matrix. Let us first consider how rotations map through the two isomorphisms that we just discussed. We start with rotations in
\begin{equation}
    \sotwo\times\sotwo\subset\sutwo\times\sutwo\cong\spinfour \,.
\end{equation}
If we take a rotation over an angle $m_1$ in the first $\sotwo$ and a rotation over an angle $m_3$ in the second $\sotwo$, we can use (the inverse of) the isomorphism \eqref{compact isomorphism} to map this to an $\sofour\subset\spinfour$ matrix:
\begin{align}
    &\left(
    \begin{pmatrix}
    \cos m_1 & -\sin m_1 \\
    \sin m_1 & \cos m_1
    \end{pmatrix}\;,\;
    \begin{pmatrix}
    \cos m_3 & -\sin m_3 \\
    \sin m_3 & \cos m_3
    \end{pmatrix}
    \right) \;=\; \left( e^{2m_1 N_1} , e^{2m_3 N_1} \right) \nonumber\\
    &\overset{\text{isomorphism}}{\longrightarrow} e^{(m_1+m_3)M_{12}+(m_1-m_3)M_{34}} = \scalebox{0.85}{$\begin{pmatrix}
    \cos (m_1+m_3) & -\sin (m_1+m_3) & 0 & 0 \\
    \sin (m_1+m_3) & \cos (m_1+m_3) & 0 & 0 \\
    0 & 0 & \cos (m_1-m_3) & -\sin (m_1-m_3) \\
    0 & 0 & \sin (m_1-m_3) & \cos (m_1-m_3)
    \end{pmatrix}$} \,.
    \label{embedding SO4}
\end{align}
We can repeat this for the maximally non-compact isomorphism, mapping rotations in the subgroup
\begin{equation}
\sotwo\times\sotwo\subset\sltwo\times
\sltwo\cong\spintwotwo \,,
\end{equation}
to an $\sotwotwo\subset\spintwotwo$ matrix. We find
\begin{align}
    &\left(
    \begin{pmatrix}
    \cos \alpha_1 & -\sin \alpha_1 \\
    \sin \alpha_1 & \cos \alpha_1
    \end{pmatrix} \;,\;
    \begin{pmatrix}
    \cos \alpha_3 & -\sin \alpha_3 \\
    \sin \alpha_3 & \cos \alpha_3
    \end{pmatrix}
    \right) \;=\; \left( e^{2\alpha_1 \tilde{N}_1} , e^{2\alpha_3 \tilde{N}_1} \right) \nonumber\\
    &\overset{\text{isomorphism}}{\longrightarrow} e^{(\alpha_1+\alpha_3)\tilde{M}_{12}+(\alpha_1-\alpha_3)\tilde{M}_{34}} = \scalebox{0.85}{$\begin{pmatrix}
    \cos (\alpha_1+\alpha_3) & -\sin (\alpha_1+\alpha_3) & 0 & 0 \\
    \sin (\alpha_1+\alpha_3) & \cos (\alpha_1+\alpha_3) & 0 & 0 \\
    0 & 0 & \cos (\alpha_1-\alpha_3) & -\sin (\alpha_1-\alpha_3) \\
    0 & 0 & \sin (\alpha_1-\alpha_3) & \cos (\alpha_1-\alpha_3)
    \end{pmatrix}$} \,.\label{embedding SO22}
\end{align}
The almost identical structure that we find through the compact and maximally non-compact isomorphisms should not be surprising, as the generators that appear are equal, e.g. $M_{12}=\tilde{M}_{12}$ and $N_1=\tilde{N}_1$. It is the other generators (the non-compact ones) that differ between these algebras.

Now we are ready to embed the four rotation parameters that are relevant in the main text (see section \ref{sec:Orbifold constructions}) in an $\sofourfour\subset\spinfourfour$ matrix.

We consider both the parameters $m_1,\ldots,m_4$ that rotate in the subgroup
\begin{equation}
    \sotwo^4\subset\sutwo^4\cong\spinfour^2\;,
\end{equation}
and the parameters $\alpha_1,\ldots,\alpha_4$ that rotate in the subgroup
\begin{equation}
    \sotwo^4\subset\sltwo^4\cong\spintwotwo^2 \;.
\end{equation}
Both are essentially a repetition of the embeddings shown in \eqref{embedding SO4} and \eqref{embedding SO22}.  The SO(4) matrices can be embedded into SO(4,4) in a block-diagonal way
\begin{equation}
    \begin{pmatrix}
    R(m_1+m_3) & 0 & 0 & 0 \\
    0 & R(m_1-m_3) & 0 & 0 \\
    0 & 0 & R(m_2+m_4) & 0 \\
    0 & 0 & 0 & R(m_2-m_4)
    \end{pmatrix} \in \sofourfour \;,
\end{equation}
while for the SO(2,2) matrices, the embeddding into SO(4,4) is given by
\begin{equation}
    \begin{pmatrix}
    R(\alpha_1+\alpha_3) & 0 & 0 & 0 \\
    0 & R(\alpha_2+\alpha_4) & 0 & 0 \\
    0 & 0 & R(\alpha_1-\alpha_3) & 0 \\
    0 & 0 & 0 & R(\alpha_2-\alpha_4)
    \end{pmatrix} \in \sofourfour \;.
\end{equation}
Here we use the shorthand notation $R(x)=\begin{psmallmatrix}\cos x & \,\,\,\,-\sin x \\ \sin x & \,\,\,\,\cos x \end{psmallmatrix}$ for a two by two rotation matrix. From the way that the two sets of rotation parameters are embedded in $\sofourfour$ we can deduce the relation between them. We find
\begin{equation}
\begin{aligned}
m_1&=\tfrac{1}{2}(\alpha_1+\alpha_2+\alpha_3+\alpha_4) \,,\qquad\quad
&m_2=\tfrac{1}{2}(\alpha_1+\alpha_2-\alpha_3-\alpha_4) \,,\\[4pt]
m_3&=\tfrac{1}{2}(\alpha_1-\alpha_2+\alpha_3-\alpha_4) \,,\qquad\quad
&m_4=\tfrac{1}{2}(\alpha_1-\alpha_2-\alpha_3+\alpha_4) \,.
\end{aligned}
\end{equation}
We use these relations in section \ref{sec:Orbifold constructions} to determine the allowed values for the $m$'s in terms of the allowed values for the $\alpha$'s.

Note that all $\sotwotwo$ and $\sofourfour$ matrices in this subsection are written down in $\eta$-frame. In order to obtain the relevant matrices in $\tau$-frame (which is the frame in which the monodromies are required to be integer valued), one would have to perform a conjugation \`a la \eqref{conjugation by X}.

\section{Modular functions and transformations}
\label{ap B}

The Dedekind $\eta$-function is defined as
\begin{equation}
   \eta(\tau)\equiv q^{\frac{1}{24}} \prod_{n=1}^{\infty}(1-q^n)\,,\hspace{0.5cm}q=e^{2\pi i \tau}\,.
\end{equation}
The Jacobi $\vartheta$-function with characteristics $\alpha,\beta$ is given by
\begin{equation}
     \vartheta[\psymbol{ \alpha}{ \beta}](\tau)=\sum_{n\in \mathbb{Z}}q^{\frac{1}{2}(n+\alpha)^2}e^{2\pi i(n+\alpha)\beta}\,.
     \label{14}
\end{equation}
For $-\tfrac{1}{2}\leq \alpha, \beta \leq \tfrac{1}{2}$ there is also a product representation of the $\vartheta$-function, which reads 
\begin{equation}
    \vartheta[\psymbol{ \alpha}{ \beta}](\tau)=\eta(\tau)e^{2\pi i\alpha\beta}q^{\frac{1}{2}\alpha^2-\frac{1}{24}}\prod_{n=1}^{\infty}(1+q^{n+\alpha-\frac{1}{2}}e^{2\pi i\beta})(1+q^{n-\alpha-\frac{1}{2}}e^{-2\pi i\beta})\,.
    \label{product representation}
\end{equation}
Particular $\vartheta$-functions that appear often are 
\begin{equation}
    \vartheta[\psymbol{ 0}{ 0}](\tau)\equiv\vartheta_3(\tau)\,,\hspace{0,2cm} \vartheta[\psymbol{ 0}{ \frac{1}{2}}](\tau)\equiv\vartheta_4(\tau)\,,\hspace{0,2cm} \vartheta[\psymbol{ \frac{1}{2}}{ 0}](\tau)\equiv\vartheta_2(\tau)\,,\hspace{0,2cm} \vartheta[\psymbol{ \frac{1}{2}}{ \frac{1}{2}}](\tau)\equiv\vartheta_1(\tau)\,.
\end{equation}
In addition, two useful $\vartheta$-function identities are\footnote{Various $\vartheta$-function identities can be found in \cite{gannon1992lattices,gannon1992lattices2}.}
\begin{equation}
    \vartheta[\psymbol{ -\alpha}{ -\beta}](\tau)=  \vartheta[\psymbol{ \alpha}{ \beta}](\tau)\,,\hspace{0,3 cm}  \vartheta[\psymbol{ \alpha+m}{ \beta+n}](\tau)= e^{2\pi i n \alpha}\,\vartheta[\psymbol{ \alpha}{ \beta}](\tau)\,,\,\,m,n \in \mathbb{Z}\,.
\end{equation}
Finally, the Poisson resummation formula is given by
\begin{equation}
    \sum_{n \in \mathbb{Z}}e^{-\pi a n^2+\pi b n}=\frac{1}{\sqrt{a}} \sum_{n \in \mathbb{Z}}e^{-\frac{\pi}{a}\left(n+i\frac{b}{2}\right)^2}\,.
\end{equation}
The modular transformations are defined as: $\mathcal{T}\equiv \tau\to \tau+1$ and $\mathcal{S}\equiv \tau \to -{1}/{\tau}$. The Dedekind $\eta$-function transforms as follows 
\begin{equation}
\begin{aligned}
    &\eta(\tau+1)= e^{{\pi i}/{12}}\,\eta(\tau)\,,\\
   & \eta(-{1}/{\tau}) = \sqrt{-i\tau}\,\eta(\tau)\,.
    \end{aligned}
\end{equation}
Note that under both $\mathcal{T}$ and $\mathcal{S}$ transformations the combination $\sqrt{\tau_2}\,\eta(\tau)\,\bar{\eta}(\bar{\tau})$ is invariant. Under modular transformations, the Jacobi $\theta$-function transforms as follows
\begin{equation}
    \begin{aligned}
      & \vartheta[\psymbol{ \alpha}{ \beta}] (\tau+1)=e^{-\pi i(\alpha^2-\alpha)} \, \vartheta[\psymbol{ \alpha}{\alpha+\beta-\frac{1}{2}}](\tau)\,,\\
       & \vartheta[\psymbol{ \alpha}{ \beta}] (-1/\tau) = \sqrt{-i\tau}\,  e^{2\pi i \alpha \beta}\, \vartheta[\psymbol{ -\beta}{ \alpha}](\tau)\,.
    \end{aligned}
\end{equation}
The theta series of the SO(2n) root lattice $D_n$ is
\begin{equation}
    \Theta_{D_n}(\tau)= \frac{1}{2}\left(\vartheta_3(\tau)^n+\vartheta_4(\tau)^n\right)\,.
\end{equation}
For the SU(3) root lattice $A_2$ and its dual $A_2^*$ we have
\begin{equation}
\begin{aligned}
 \Theta_{A_2}(\tau)= \vartheta_3(2\tau)\vartheta_3(6\tau) + \vartheta_2(2\tau)\vartheta_2(6\tau)\,,\\
 \Theta_{A_2^*}(\tau)= \vartheta_3(\tfrac{2\tau}{3})\vartheta_3(2\tau) + \vartheta_2(\tfrac{2\tau}{3})\vartheta_2(2\tau)\,.
 \end{aligned}
\end{equation}
Finally, for a $d$-dimensional lattice $\Lambda$ and its dual $\Lambda^*$ the following expression holds
\begin{equation}
    \Theta_{\Lambda}({-1}/{\tau})= \frac{(-i\tau)^{\frac{d}{2}}}{\text{Vol}(\Lambda)} \Theta_{\Lambda^*}({\tau})\,.
\end{equation}

\section{Twist vectors and lattices}
\label{Ap C}
Here we list in  \autoref{tab breaking all} and \autoref{tab breaking half} twist vectors $u$ (up to a sign and an exchange $u_3\leftrightarrow u_4$) breaking all and half of the right-moving supersymmetries  respectively. In addition, we compute the number of chiral fixed points for each twist vector, $\chi=\prod_{i}2 \sin(\pi u_i)$ and we list in \autoref{tab breaking half} only those $u$'s giving integer $\chi$. For symmetric orbifolds ($u=\tilde{u}$) we also write down examples of root lattices generated by root systems of simple Lie algebras, which represent the appropriate torus lattices admitting the corresponding $\mathbb{Z}_p$ symmetries (for symmetries of root lattices see e.g. \cite{baake2002guide,koca2014quasicrystallography} and references thereof). Furthermore, we evaluate the volume, Vol, of these lattices. The root lattices $A_n,B_n$ and $D_n$ are associated to the groups  $\text{SU}(n+1),\text{SO}(2n+1)$ and $\text{SO}(2n)$ respectively. Note that the lattice generated by the $A_2$ system is the same as the $G_2$ root lattice (hexagonal). Also, $A_1\oplus A_1\cong D_2$ and $B_2$ generate the same (square) lattice. Finally, we use a shorthand notation of the form $(A_1)^2\equiv A_1\oplus A_1$.
\newline

\begin{table}[h!]
    \centering
    {\renewcommand*{\arraystretch}{2}
    \begin{tabular}{|c|c|c|c|c|}
    \hline
      $\mathbb{Z}_p$ & $u$ & $\chi$ & Lattice & Vol  \\
        \hline
        \hline
       $\mathbb{Z}_2$  &  $(0,0,0,1)$ & - & & \\
       \hline
       $\mathbb{Z}_3$  &  $(0,0,0,\tfrac{2}{3})$ & ${\sqrt{3}}$ & $A_2$ & ${\sqrt{3}}$ \\
       \hline
      \multirow{2}{*}{ $\mathbb{\mathbb{Z}}_4$} &   $(0,0,0,\tfrac{1}{2})$ & 2 &   $(A_1)^2$ & 2\\
      \cline{2-5}
              &   $(0,0,\tfrac{1}{4},\tfrac{3}{4})$ & 2 & $(A_1)^4$ & 4\\
              \hline
               $\mathbb{Z}_5$ &  $(0,0,\frac{1}{5},\frac{3}{5})$  & $\sqrt{5}$ & $A_4$ & ${\sqrt{5}}$\\
               \hline
        \multirow{4}{*}{ $\mathbb{\mathbb{Z}}_6$} &   $(0,0,0,\tfrac{1}{3})$ & $\sqrt{3}$ & $A_2$ & ${\sqrt{3}}$\\
      \cline{2-5}
              & $(0,0,\tfrac{1}{3},\tfrac{2}{3})$ & 3 & $(A_2)^2$ & 3\\
              \cline{2-5}
              & $(0,0,\tfrac{1}{2},\tfrac{1}{6})$ & 2 & $(A_1)^2$ $\oplus$ $A_2$ &2${\sqrt{3}}$ \\
       \cline{2-5}
              &  $(0,0,\tfrac{1}{6},\tfrac{5}{6})$ & 1 & $(A_2)^2$ &3\\
    \hline
    $\mathbb{Z}_8$ & $(0,0,0,\frac{1}{4})$& $\sqrt{2}$ & $(A_1)^2$ & 2\\
    \hline
    \end{tabular}
    \begin{tabular}{|c|c|c|c|c|}
    \hline
      $\mathbb{Z}_p$ & $u$ & $\chi$ & Lattice & Vol  \\
        \hline
        \hline
  \multirow{2}{*}{$\mathbb{Z}_8$} 
  & $(0,0,\frac{1}{2},\frac{1}{4})$ & $2\sqrt{2}$ & $(A_1)^4$ &4\\
  \cline{2-5}
  & $(0,0,\frac{1}{8},\frac{3}{8})$  & $\sqrt{2}$ & $B_4,D_4$ & 2\\
    \hline
  \multirow{2}{*}{$\mathbb{Z}_{10}$} &   $(0,0,\frac{1}{5},\frac{2}{5})$  & $\sqrt{5}$ & $A_4$ & ${\sqrt{5}}$ \\
    \cline{2-5}
   & $(0,0,\frac{1}{10},\frac{3}{10})$  & ${1}$ &$A_4$ & ${\sqrt{5}}$ \\
    \hline
   \multirow{4}{*}{$\mathbb{Z}_{12}$}& $(0,0,0,\frac{1}{6})$ & 1  & $A_2$ & $\sqrt{3}$\\
    \cline{2-5}
    & $(0,0,\frac{1}{2},\frac{1}{3})$ & $2\sqrt{3}$ & $(A_1)^2\oplus A_2$ & $2\sqrt{3}$\\
    \cline{2-5}
    & $(0,0,\frac{1}{3},\frac{1}{6})$ & $\sqrt{3}$ & $(A_2)^2$ & 3\\
    \cline{2-5}
   & $(0,0,\frac{1}{12},\frac{5}{12})$  & 1  & $F_4,D_4$ & $\frac{1}{2},2$\\
   \hline
    \multirow{2}{*}{$\mathbb{Z}_{24}$}& $(0,0,\frac{1}{4},\frac{1}{3})$ & $\sqrt{6}$  & $(A_1)^2 \oplus A_2$ & $2\sqrt{3}$\\
    \cline{2-5}
    & $(0,0,\frac{1}{4},\frac{1}{6})$ & $\sqrt{2}$  & $(A_1)^2\oplus A_2$ & $2\sqrt{3}$\\
    \hline
   \end{tabular} 
    }
    \caption{}
    \label{tab breaking all}
\end{table}
\renewcommand{\arraystretch}{1}

\newpage
\begin{table}[h!]
    \centering
    {\renewcommand*{\arraystretch}{2}
    \begin{tabular}{|c|c|c|c|c|}
    \hline
      $\mathbb{Z}_p$ & $u$ & $\chi$ & Lattice & Vol\\
        \hline
        \hline
       $\mathbb{Z}_2$  &  $(0,0,\tfrac{1}{2},\tfrac{1}{2})$ & 4 & $D_4$, $(A_1)^4$ & 2, 4  \\
       \hline
       $\mathbb{Z}_3$  &  $(0,0,\tfrac{1}{3},\tfrac{1}{3})$ & 3 & $(A_2)^2$ &3\\
       \hline
           { $\mathbb{\mathbb{Z}}_4$}    &   $(0,0,\tfrac{1}{4},\tfrac{1}{4})$ & 2 & $(A_1)^4$ &4\\
              \hline
            { $\mathbb{\mathbb{Z}}_6$}    &   $(0,0,\tfrac{1}{6},\tfrac{1}{6})$ & 1 & $(A_2)^2$ &3\\
            \hline 
    \end{tabular}}
    \caption{}
    \label{tab breaking half}
\end{table}
\renewcommand{\arraystretch}{1}

\section{Scherk-Schwarz spectrum}
\label{Ap D}

Here we provide the table of masses of the various fields obtained by reduction of type IIB supergravity on $T^4$ and followed by a Scherk-Schwarz twist on $S^1$. The table is taken from \cite{Hull:2020byc}. The mass of each field is $|\mu(m_i)|/2\pi R$. The notation $m_{i,j}$ indicates that both indices $i$ and $j$ occur. There is no correlation between the $\pm$ signs and the $ij$
indices, so that e.g. ($\pm m_1\pm m_2$) denotes 4 different combinations of mass parameters,
and ($\pm m_{1,2}\pm m_{3,4}$) denotes 16 different combinations. For example, the 5 tensors in
the $(\textbf{5},\textbf{1})$ representation consist of two with mass $|m_1 + m_2|/2\pi R$, two with mass $|m_1 - m_2|/2\pi R$ and one with mass $0$. 

\begin{table}[h]
    \centering
     {\renewcommand*{\arraystretch}{1.5}
    \begin{tabular}{c|c|c}
    \hline
    Fields & Representation & $|\mu(m_i)|$ \\
    \hline
    \hline
    Scalars & $(\textbf{5},\textbf{5})$ & $|\pm m_1 \pm m_2 \pm m_3 \pm m_4|$   \\
    &  &  $|\pm m_1 \pm m_2 |$ \\ 
     &  &  $|\pm m_3 \pm m_4 |$\\
    &  &  0\\
    \hline
    Vectors & $(\textbf{4},\textbf{4})$ & $|\pm m_{1,2} \pm m_{3,4}|$   \\
    \hline
     Tensors  & $(\textbf{5},\textbf{1})$ & $|\pm m_1 \pm m_2|, 0$   \\
     & $(\textbf{1},\textbf{5})$ & $|\pm m_3 \pm m_4|, 0$   \\
     \hline
      Gravitini  & $(\textbf{4},\textbf{1})$ & $|\pm m_{1,2}|$   \\
     & $(\textbf{1},\textbf{4})$ & $|\pm  m_{3,4}|$   \\
     \hline
     Dilatini & $(\textbf{5},\textbf{4})$ & $|\pm m_1 \pm m_2 \pm m_{3,4}|$   \\
     & &  $|\pm m_{3,4}|$\\
     & $(\textbf{4},\textbf{5})$ & $|\pm m_{1,2}\pm m_3 \pm m_4|$   \\
     & &  $|\pm m_{1,2}|$
     \end{tabular}}
    \label{Table: SS reduction table}
\end{table}

\newpage

\bibliographystyle{unsrt}
\bibliography{bib}

\end{document}